\documentclass[10pt,journal,compsoc]{IEEEtran}

\AtBeginDocument{%
  \providecommand\BibTeX{{%
    \normalfont B\kern-0.5em{\scshape i\kern-0.25em b}\kern-0.8em\TeX}}}

\usepackage{CJKutf8}

\newcommand{\thoughts}[1]{\begin{CJK*}{UTF8}{gbsn}{\color{brown}[#1]}\end{CJK*}}

\renewcommand{\thoughts}[1]{} 

\usepackage{color}

\newcommand{\changed}[1]{#1}

\newcommand{\nbd}[0]{\nobreakdash}

\usepackage{stmaryrd} 
\usepackage{amsmath} 
\usepackage{multirow} 
\usepackage{subfigure}
\usepackage{rotating} 
\usepackage{amssymb}
\usepackage{booktabs}
\usepackage{hyperref}
\usepackage{comment}
\usepackage{cite}
\usepackage{caption}

\newcommand{\x}[0]{{\bf{x}}}
\newcommand{\y}[0]{{\bf{y}}}

\newcommand\numberthis{\addtocounter{equation}{1}\tag{\theequation}}

\begin{document}

\title{Hypothesis Testing for Progressive Kernel Estimation and VCM Framework}

\author{Zehui Lin, Chenxiao Hu, Jinzhu Jia, Sheng Li*,~\IEEEmembership{Member,~IEEE} 
\IEEEcompsocitemizethanks{
\IEEEcompsocthanksitem Zehui Lin, Chenxiao Hu, Sheng Li are with the School of Computer Science, Peking University, China.\\ E-mail: \{zehui\,$|$hineven\,$|$lisheng\}@pku.edu.cn. \\
Jinzhu Jia is with the Dept. of Biostatistics and Center for Statistical Science, Peking University, China. \\ Email: jzjia@math.pku.edu.cn.
\IEEEcompsocthanksitem Sheng Li is the corresponding author.
}
}


\IEEEtitleabstractindextext{
\begin{abstract}

Identifying an appropriate radius for unbiased kernel estimation is crucial for the efficiency of radiance estimation. However, determining both the radius and unbiasedness still faces big challenges. 
In this paper, we first propose a statistical model of photon samples and associated contributions for progressive kernel estimation, under which the kernel estimation is unbiased if the null hypothesis of this statistical model stands.
Then, we present a method to decide whether to reject the null hypothesis about the statistical population (i.e., photon samples) by the F-test in the Analysis of Variance. Hereby, we implement a progressive photon mapping (PPM) algorithm, wherein the kernel radius is determined by this hypothesis test for unbiased radiance estimation.
Secondly, we propose VCM+, a reinforcement of Vertex Connection and Merging (VCM), and derive its theoretically unbiased formulation.
VCM+ combines hypothesis testing-based PPM with bidirectional path tracing (BDPT) via multiple importance sampling (MIS), wherein our kernel radius can leverage the contributions from PPM and BDPT. We test our new algorithms, improved PPM and VCM+, on diverse scenarios with different lighting settings.
The experimental results demonstrate that our method can alleviate light leaks and visual blur artifacts of prior radiance estimate algorithms.
We also evaluate the asymptotic performance of our approach and observe an overall improvement over the baseline in all testing scenarios. 
\end{abstract}

}

\maketitle

\IEEEpeerreviewmaketitle

\IEEEraisesectionheading{\section{Introduction}}

\begin{figure*}
    \centering
     \includegraphics[width=\textwidth]{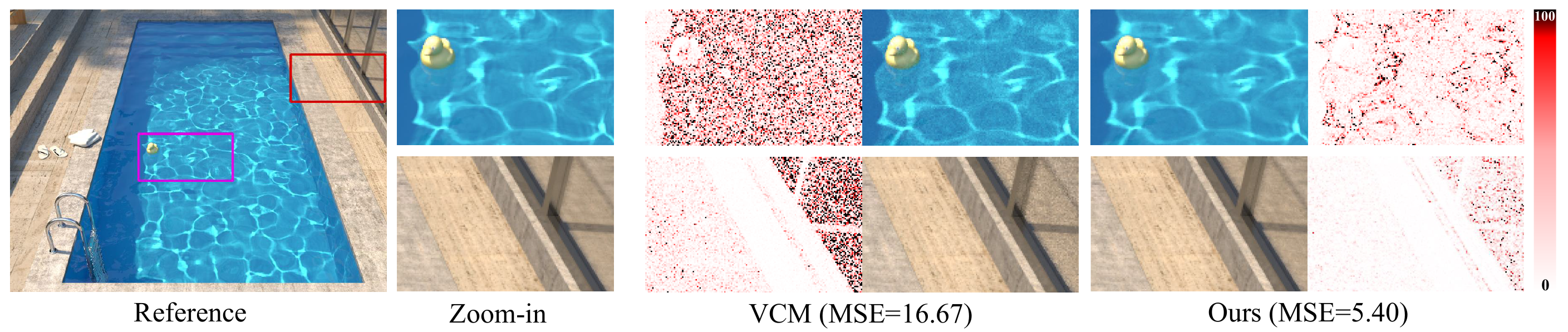}
    \caption{Equal-iteration (10K iterations) comparison on Pool scene rendered by VCM ($2.51h$) and our VCM+ ($2.57h$). Our algorithm is less noisy and has a lower MSE (mean squared error), as shown in the zoom-in regions with error visualization.}
    \label{teaserfigure}
\end{figure*}




\IEEEPARstart{L}{ight} transport simulation has been well-studied in photo-realistic rendering and applied to a wide range of digital entertainment industries.
The key challenge is to efficiently synthesize a photo-realistic image with various lighting conditions. Considerable advances have been made in the past decades to improve the efficiency of light transport algorithms.
Bidirectional path tracing (BDPT)~\cite{LW1993BPT, veach1995bidirectional} is one of the most successful algorithms.
However, it struggles with caustics and other lighting effects caused by specular-diffuse-specular (S-D-S) paths~\cite{hachisuka2008progressive}.
Photon mapping (PM)~\cite{jensen1996global} can handle S-D-S paths by emitting many photons from light sources and performing density estimation on diffuse surfaces but with bias induced.
A large kernel reduces variance by collecting more photons but potentially brings more bias. Progressive photon mapping (PPM)~\cite{hachisuka2008progressive} runs multiple iterations and gradually shrinks the support radius of the kernel to alleviate the bias progressively. 
However, the convergence rate of PPM (optimally $O(N^{-2/3})$) is slower than that of BDPT ($O(N^{-1})$).

Chi-squared progressive photon mapping (CPPM)~\cite{lin2020cppm} improves the theoretical convergence rate under ideal conditions by employing hypothesis testing to verify the unbiasedness of kernel estimation.
Based on a theory that uniformly distributed photons lead to unbiased estimates, CPPM uses the $\chi^2$-test to verify whether the photons within the kernel are subject to a uniform distribution.
CPPM shrinks the kernel radius only when it finds its test statistics are inconsistent with the hypothesized \emph{statistical model}, where the statistical model usually refers to a set of statistical assumptions concerning the generation of sample data.
However, CPPM's statistical assumption about this model is frequently violated in many cases, thus computing inaccurate radii and leading to biased or noisy results.

\begin{figure*}
   \begin{tabular}{c@{\hskip2pt} @{\hskip2pt}c@{\hskip2pt} @{\hskip2pt}c@{\hskip2pt} @{\hskip2pt}c@{\hskip2pt} @{\hskip2pt}c@{\hskip2pt} @{\hskip2pt}c}
        \includegraphics[height=0.9in]{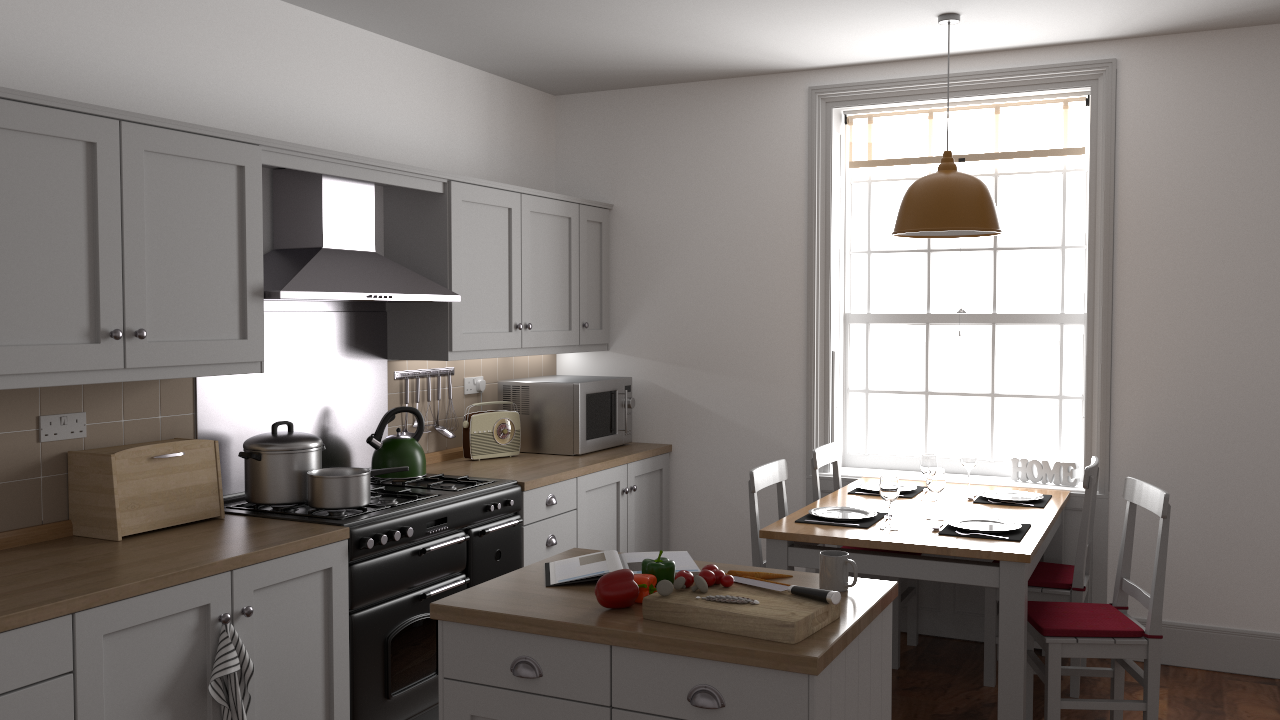} &
        \includegraphics[height=0.9in]{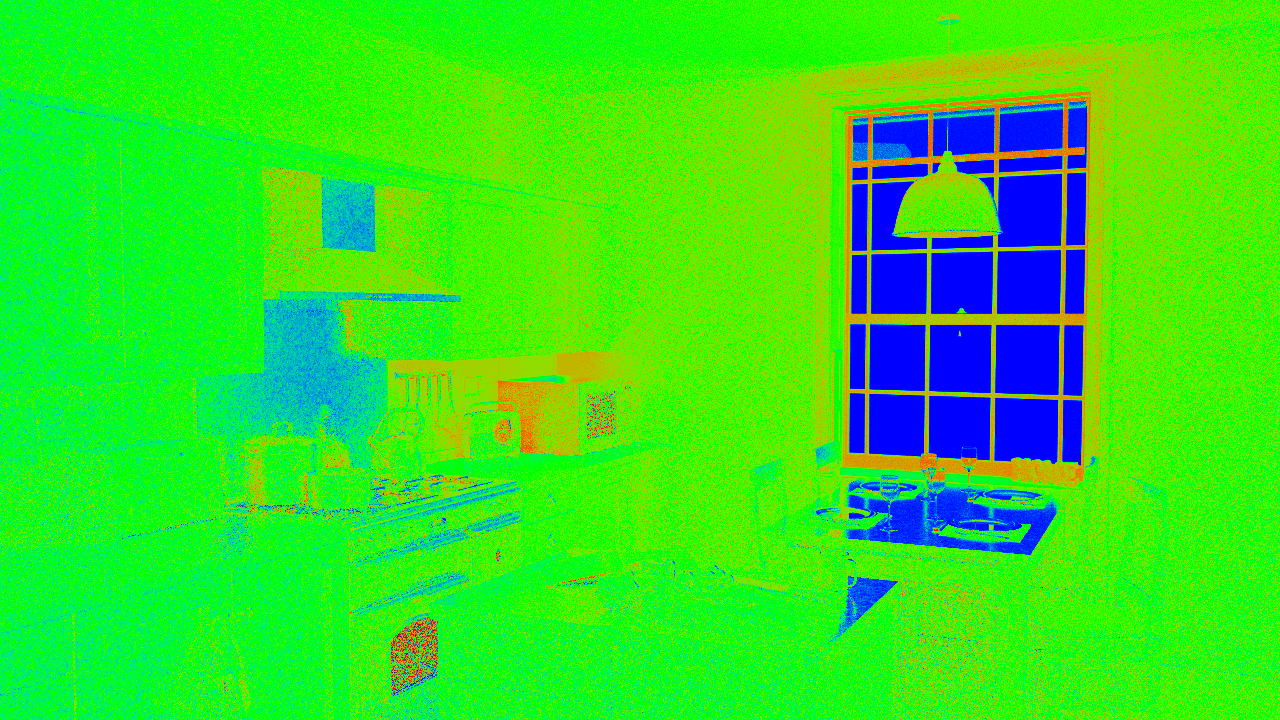} &
        \includegraphics[height=0.9in]{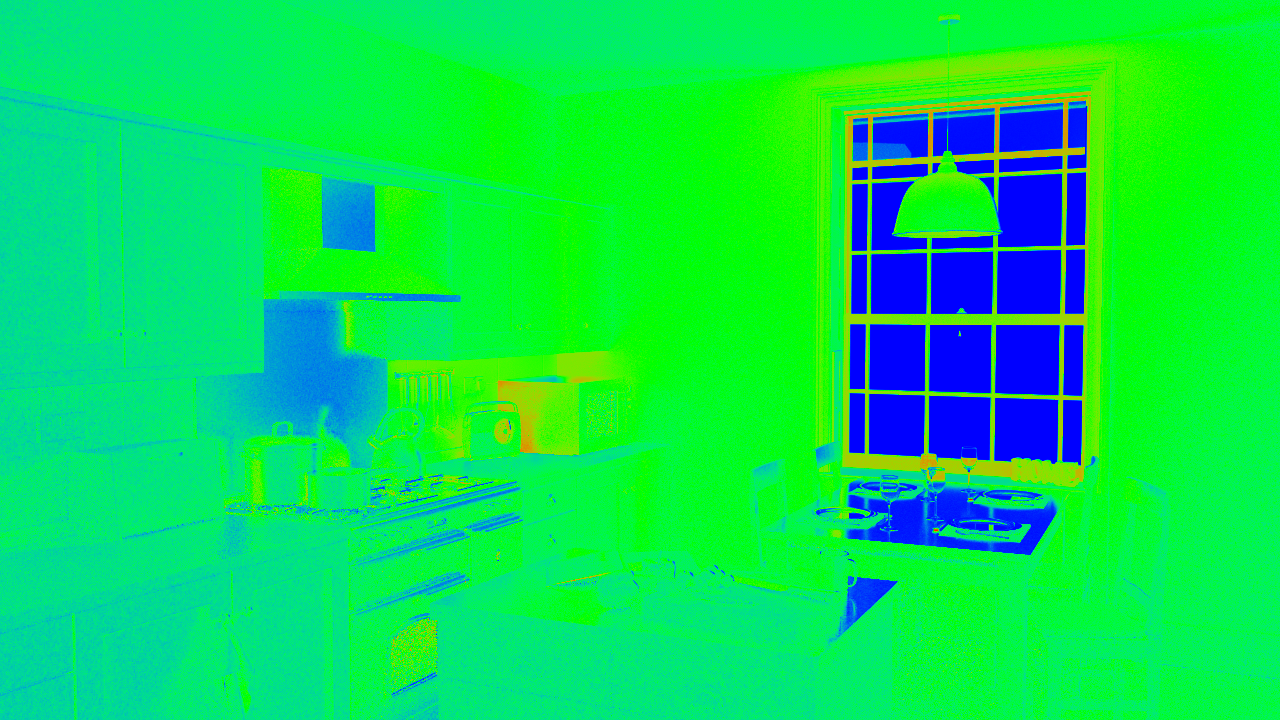} &
        \includegraphics[height=0.9in]{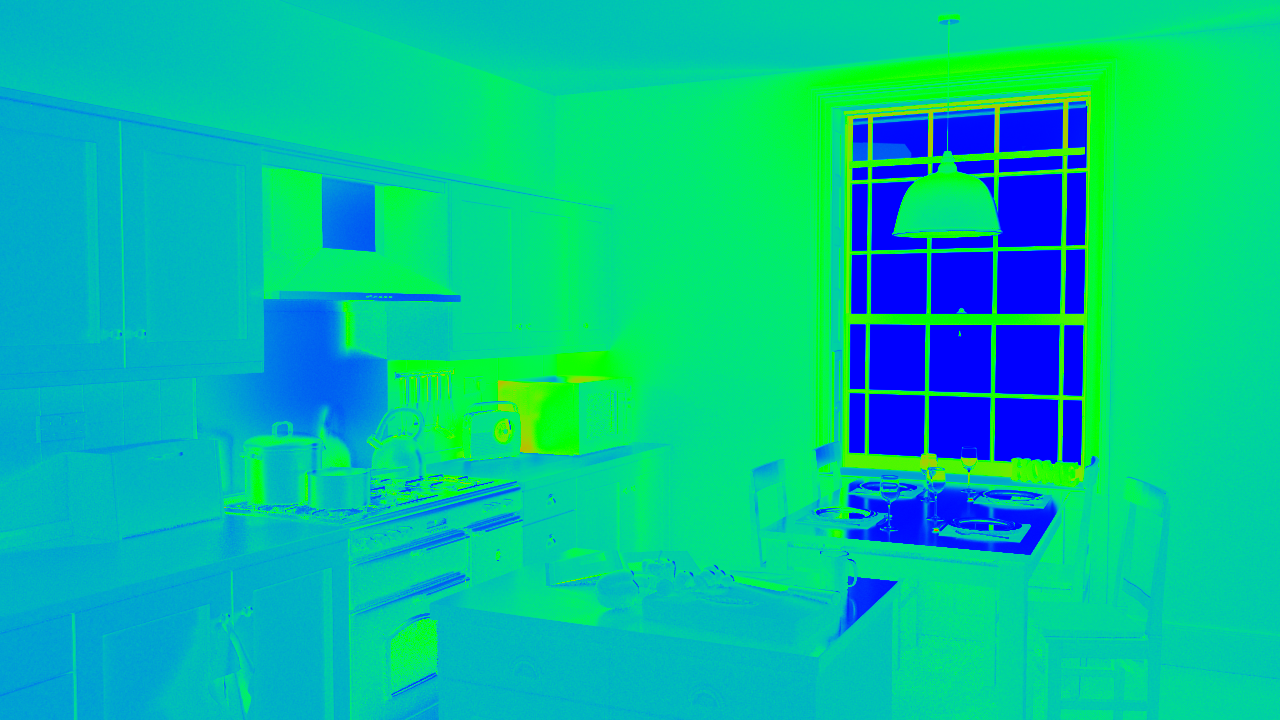} &
        \includegraphics[height=0.9in]{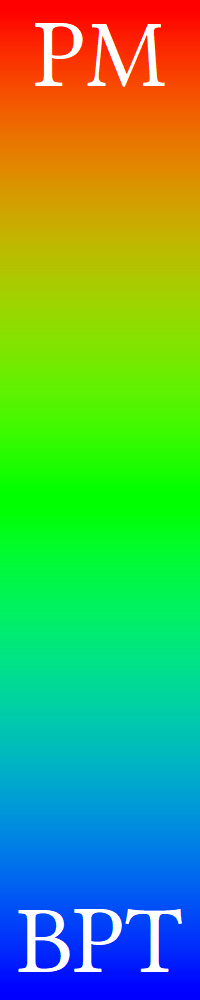}
        \\
        Reference &
        100 iterations &
        1,000 iterations &
        10,000 iterations
   \end{tabular}
   \vspace{-0.1in}
   \caption{Contribution visualization of BDPT and PM in VCM, respectively, taking Kitchen scene as an example. The relative contribution from PM gradually diminishes over iterations, and drags down VCM's overall performance consequently.}
   \label{fig:contrib-convergence}
  \vspace{-0.08in}
\end{figure*}

Combining BDPT with PM technique via multiple importance sampling (MIS) to obtain advantages from both sides, Vertex Connection and Merging (VCM)~\cite{Georgiev:2012:VCM} or Unified Path Sampling (UPS)~\cite{hachisuka2012ups} is widely accepted as a robust and efficient solution for light transport simulation with respect to various lighting conditions. 
VCM/UPS employs bidirectional path sampling strategy and expresses the probability density functions of BDPT and PPM w.r.t. the same measure.
However, PPM will gradually become a short slab of VCM/UPS over iterations and will consequently slow down the overall performance~\cite{Georgiev:2012:VCM}. 
This is because their MIS weight for PPM is subject to the radius for kernel estimation. Whereas, as the kernel radius of PPM converges to zero, the MIS weights will gradually shrink to zero due to PPM's higher variance than that of BDPT. 
Consequently, BDPT gradually takes a dominant role in rendering (see \autoref{fig:contrib-convergence}), and the computational overhead of PPM becomes increasingly worthless, i.e., the resulting algorithm is nearly equivalent to BDPT as the kernel shrinks. This is a key issue of VCM, as pointed out in Figure 7 of VCM~\cite{Georgiev:2012:VCM}.



Our method is mainly dedicated to resolving the efficiency problem in these kernel estimation related methods. Identifying an appropriate kernel radius is crucial for both the efficiency of kernel estimation of PPM and the performance of VCM/UPS. As mentioned above, a larger radius tends to induce more bias meanwhile a smaller one may lead to higher variance and performance downgrade. We aim to maximize the radius that can minimize variance while keeping the result unbiased. We begin with a large kernel radius and shrink it when the kernel estimation is detected as biased. To determine whether the estimation is biased by photon samples within the kernel, we employ a parametric test that focuses on analyzing and comparing the mean and variance of observations (photon samples), i.e., quantitative analyzing the feature of irradiance distributed on surfaces.
In this regard, we propose an efficient statistical testing-based approach to select the kernel radius for unbiased estimation, which tackles the limitation of CPPM. Furthermore, we extend our statistical model with a theoretical formulation applicable to the VCM framework. By hypothesis testing based on this model to help reduce the bias and variance and then optimize the contribution from vertex merging (PPM), our method eventually improves the efficiency of VCM in dealing with diverse scenarios with different lighting settings. We highlight our results and compare our method against VCM, as shown in \autoref{teaserfigure}.

To summarize, our main contributions are two-fold:
\begin{itemize}
%
%

    \item We propose a statistical model for hypothesis testing on photon samples along with a testing-based progressive kernel estimation algorithm, which employs the Analysis of Variance (ANOVA) F-test to decide whether the observations follow the null hypothesis of the statistical model or not, and therefore help find an . Our algorithm is more general in handling diverse photon distributions and can synthesize images with better quality.

    \item In addition, we propose a hypothesis testing-based formulation for the VCM/UPS framework, named VCM+, in which we derive a sufficient condition for a theoretically unbiased estimate. As a reinforcement of VCM, VCM+ can theoretically reach $O(N^{-1})$ convergence. Our algorithm improves the overall performance, wherein PPM can contribute more to the final pixel measurement in those light paths where PPM can handle better than BDPT.
\end{itemize}

\section{Related Work}

\subsection{Path Tracing}

As a classic solution to the rendering equation, the path tracing (PT) algorithm \cite{kajiya1986rendering} traces random paths from the eye to find light sources.
Bidirectional path tracing (BDPT) \cite{veach1995bidirectional} traces eye sub-paths from the eye and light sub-paths from light sources synchronously and connects these paths via multiple importance sampling (MIS).
Path tracing, as well as BDPT, provides an unbiased estimator of the pixel measurement.
BDPT can efficiently sample a wide range of paths, but its ability to handle S-D-S paths is inferior to photon mapping~\cite{hachisuka2008progressive}. BDPT also provides a bidirectional path sampling framework in which one eye sub-path and one light sub-path will construct $l-1$ samples for the path with length $l$.

\subsection{Photon Mapping}
Photon mapping (PM)~\cite{jensen1996global} provides an alternative solution but with bias introduced. It also traces eye sub-paths from the eye and light sub-paths from light sources.
Unlike BDPT, it first samples a large number of light sub-paths and stores the light vertices (a.k.a photons) on the paths into a range searching data structure (a.k.a photon map).
Then it estimates radiance at eye vertices with density estimation using nearby light vertices \cite{hachisuka2012state}.
Some other methods leave the kernel function simple and lightweight and choose its support radius wisely, such as adaptive progressive photon mapping~\cite{Kaplanyan2013appm} that leverages the bias and variance with theoretically optimal support radii. 
Recently, the neural network was employed to predict a kernel function for density estimation \cite{shilin2020deep}.

Progressive photon mapping (PPM)~\cite{hachisuka2008progressive} diminishes the bias by gradually reducing the support radius of the kernel.
Both PM and PPM work with the unidirectional path sampling scheme, i.e., one eye sub-path and one light sub-path construct one sample for each path length. 
From then onwards, many studies have followed this progressive line. 
Stochastic progressive photon mapping (SPPM)~\cite{toshiya2009stochastic} enables PPM to render more lighting effects like depth-of-field, motion blur, and glossy reflections. 
Vorba~\cite{vorba2011bidirectional} extended PM and presented bidirectional photon mapping that combines the radiance estimated from different vertices on an eye sub-path together via MIS.
Although PPM can converge to the correct result, it has a lower convergence rate than BDPT because shrinking the radius leads to fewer samples and resultant higher variance. 
Hachisuka et al.~\cite{hachisuka2010progressive} presented a progressive error estimation framework for PPM.
Knaus and Zwicker~\cite{knaus2011progressive} presented a probabilistic analysis of PPM.

Lin et al.~\cite{lin2020cppm} proposed chi\nbd-squared progressive photon mapping (CPPM). It gives a better estimation of support radius that minimizes bias and variance with chi\nbd-squared tests.
An assumption about the correlation between photon distribution and radiance distribution is required, and this assumption is further relaxed as the \textit{location-independent assumption (abbr. independence assumption)} that the contribution of a light vertex is independent of the relative position to the eye vertex, wherein a full path is constructed by this light vertex and eye vertex. $\chi^2$-test gives good estimations of support radius when this assumption holds.
However, this independence assumption may still be violated in practice. $\chi^2$-test malfunctions in those circumstances, resulting in such artifacts as noise, light leaks, and visual blur. 
result.
To improve the generality and thereby promote efficiency, we investigate a new model for hypothesis testing that removes this less commonly held assumption.

\subsection{Vertex Connection and Merging}

Vertex Connection and Merging (VCM)~\cite{Georgiev:2012:VCM} and Unified Path Sampling (UPS)~\cite{hachisuka2012ups} are bidirectional frameworks that both combine the BDPT and PM techniques by putting their probability density functions (pdf) under the same measure. MIS leverages the relative contribution between BDPT and PPM to minimize the variance of this combined estimator.
However, the variance of PPM gradually increases with the kernel shrinking. Whereas, the variance of BDPT will eventually be much lower than that of PPM for the paths that BDPT and PPM can both sample.
Consequently, the MIS weight for PPM gradually reduces to zero, making the rendering less efficient.
Because VCM and UPS share the same key technical features, we use VCM as a proxy of VCM/UPS for simplicity in the rest sections.

Thereafter, unbiased photon gathering~\cite{qin2015unbiased} based on the bidirectional framework was proposed to process each photon individually to create an unbiased path sample. Still, it cannot handle S-D-S paths well. As an orthogonal technique to the VCM, Markov chain Monte Carlo (MCMC) methods were combined with VCM to improve the rendering efficiency \cite{mcmcvcm}.

As a combination of BDPT and PPM, reducing the variance from PPM would reduce the overall variance of VCM.
Our algorithm expects to reduce the variance brought by the inappropriate radius of kernel estimation, elevate the contribution from PM with low variance and present an improved estimator under the VCM framework, thereby speeding up the rendering process.

\subsection{Statistical Hypothesis Testing}
\emph{Statistical hypothesis testing} is a method of statistical inference~\cite{calder1953statistical}.
It first sets up a \emph{null hypothesis} that is testable based on observations.
Hypothesis testing requires constructing a \emph{statistical model} of the shape the data would be in. The hypothesis that the statistical divergence of the observations is due to chance alone is called the \emph{null hypothesis}.
Next, it computes the test statistic from these observations. If the resultant statistic is unlikely to occur when the null hypothesis is true (according to a predetermined threshold probability, i.e., a significance level), it rejects the null hypothesis.
It's possible for a statistical hypothesis testing to incorrectly reject a true null hypothesis or incorrectly not reject a false null hypothesis.
The probability of the former event is controlled by the significance level and is usually equal to a scalar value of the significance level.
For the latter, increasing the sample size is always a feasible solution to decrease its probability with a fixed significance level~\cite{everitt1998cambridge}.

CPPM~\cite{lin2020cppm} assesses whether the photons (or light vertices) are uniformly distributed by Pearson's $\chi^2$-test, which can be used to evaluate the difference between sets of categorical data and a hypothetical frequency distribution~\cite{william1952goodness}.
ANOVA~\cite{miller1997beyond} developed by Ronald Fisher~\cite{fisher1921probable} is a collection of statistical tools that partitions the observed variance into different sources to analyze the differences of means.
The ANOVA F-test is used to assess whether the expected values of a quantitative variable within several groups differ from each other.
Therefore, it can be used to assess whether the radiance (or the contribution of full paths) is uniform within the kernel, so it can work for situations where the independence assumption is invalid.
Compared with non-parametric ANOVA alternatives such as Kruskal–Wallis test~\cite{kruskal1952use}, which requires ranking the samples, the F-test is more lightweight.
We propose a null hypothesis for unbiased estimation concerning the contribution of full paths, and then use the Analysis of variance (ANOVA) F-test for hypothesis testing.

\section{Hypothesis Testing for Kernel Estimation}
\label{sec:hypothesis-testing}

\begin{figure}
    \centering
    \includegraphics[width=0.8\linewidth]{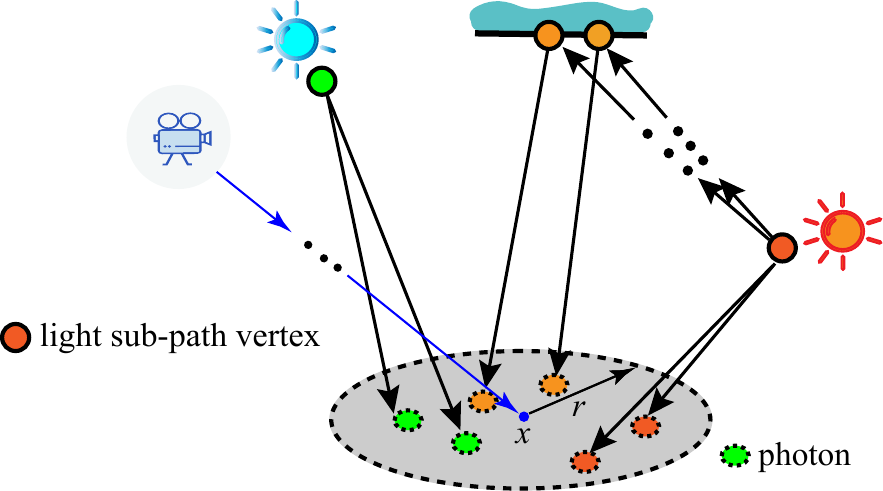} 
    \caption{Illustration of photon samples with different contributions within a searching area. In general, different light sources or light sub-paths during transport varies the associated contribution of each sample. These photons should not be treated equally. }
   \label{fig:differentcontribution}
   \vspace{-0.1in}
\end{figure}

\begin{figure}[t]
    {
    \centering
    \includegraphics[trim={2.2cm 1cm 0cm 1cm}, clip,width=0.9\linewidth]{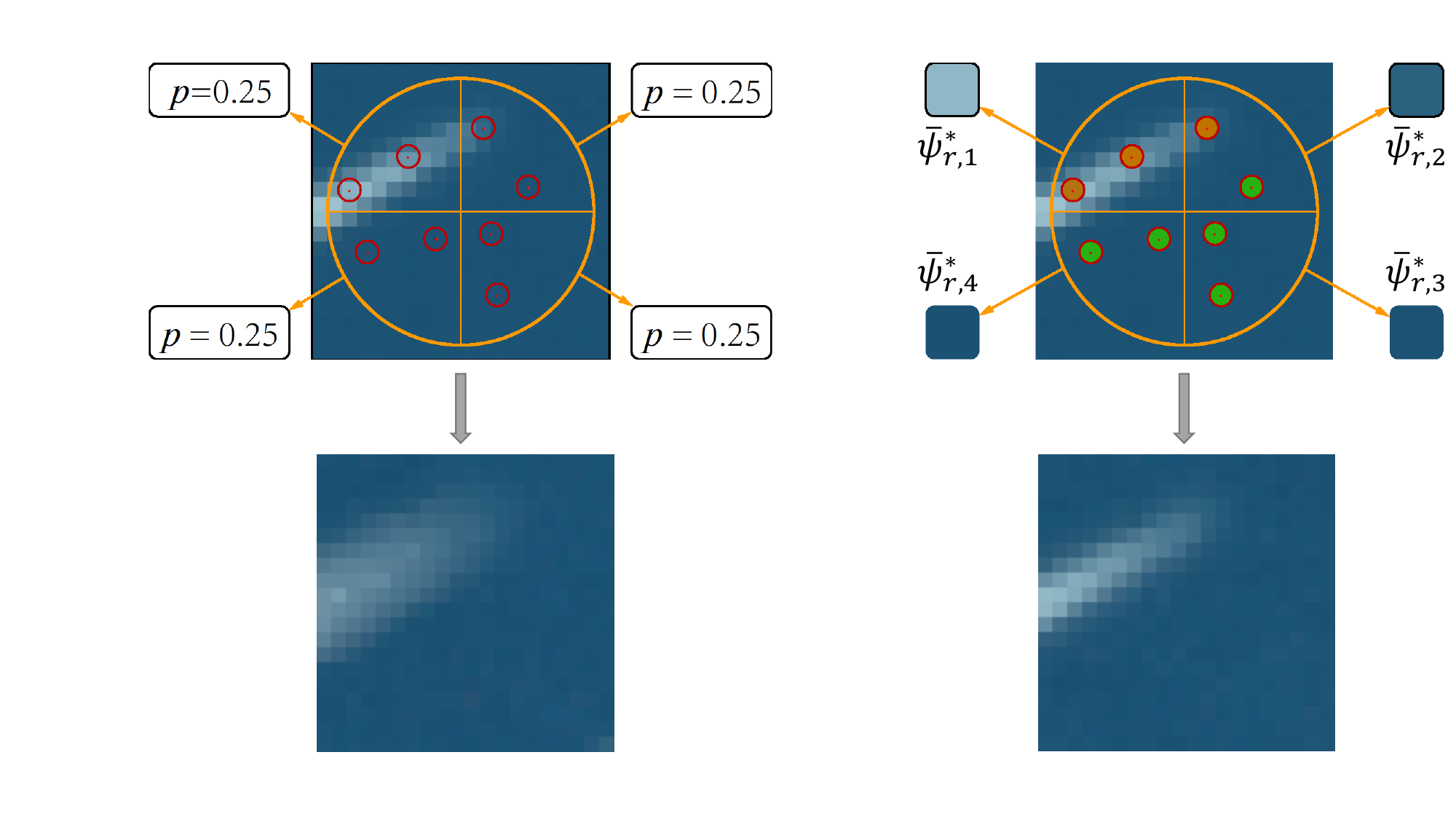}} 
    \caption{Illustration of a simple case that the independence assumption is violated. The photons have different contributions within a kernel (upper-right). Here, photons on the top are likely to have more contributions. However, CPPM detects no bias through $\chi^2$-test due to equal contribution assumption (left), whereas our method observes variational contributions across photons and detects bias through F-test for clearer result (right). }
    
   \label{fig:spotlight}
   \vspace{-0.1in}
\end{figure}

\changed{Quantitative analysis of the feature of irradiance distributed on surfaces can help improve the efficiency of radiance estimation. 
CPPM focuses on analyzing the distribution of photon samples, and its statistical model assumes independence of the populations (i.e., photon samples) and uses the $\chi^2$\nbd-test to verify the unbiasedness~\cite{lin2020cppm}.} However, in practical scenarios, various lighting conditions such as multiple area light, chromatic light/textured light, and glossy reflection may vary the contribution of each sample during light transport, introducing correlations between the position and the contribution of photons, as illustrated in \autoref{fig:differentcontribution}. This violates the independence assumption and leads to biases in some local regions, resulting in blurry images, as shown in \autoref{fig:spotlight} (left). Meanwhile, CPPM tends to underestimate the kernel radius for radiance estimation, leading to noisy results frequently.

\changed{
Our statistical model is designed to eliminate the independence assumption in kernel estimation; herein, our approach can guarantee unbiasedness when a sufficient condition is satisfied, regardless of the scene settings. The key to achieving this goal is to identify a statistical model with generality and its corresponding unbiased conditions for kernel estimation (as detailed in Sec.~\ref{sec:uni-hypo-condition}), as well as a method to verify the unbiasedness of radiance within a kernel radius (as described in Sec.~\ref{subsec:verifying}).
Based upon the proposed statistical model (Sec.~\ref{sec:uni-hypo-condition}) as a theoretical foundation, our hypothesis testing methodology works as follows: firstly, the irradiance mean values for various regions can be calculated using the samples within a given kernel; next, hypothesis testing is employed to analyze whether there is a statistically significant difference in irradiance values among these regions; lastly, whether the estimation is unbiased or not can be determined based on the statistical results of difference analysis. The ANOVA F-test is employed for hypothesis testing, and the principle behind verifying unbiasedness will be explained in detail in Sec.~\ref{subsec:verifying}.
}



\subsection{Statistical Model for Unbiased Estimation in PPM}
\label{sec:uni-hypo-condition}
Given that the accuracy of any conclusion drawn from a statistical inference depends on the validity of the \emph{statistical assumptions}, the assumptions are expected to be met practically.
We derive a formulation without extra assumptions that can support the hypothesis of our statistical model as follows.

We start with the integral expression of radiance $I$ in PM by Kaplanyan and Dachsbacher~\cite{Kaplanyan2013appm}:
\begin{equation}
    I \approx \int_{\mathcal{M}\times \mathcal{M}} K_r(\x-\x^*) \Psi(\x, \x^*) d\x d\x^*,
    \label{equ:kaplanyan-integral}
\end{equation}
where $K_r$ is a 2D normalized kernel function with support radius $r$; $\mathcal{M}$ is the scene surface; and $\Psi(\x, \x^*)$ is the contribution of all possible full paths constructed by the eye vertex $\x$ and the light vertex (i.e., photon) $\x^*$.

To align the light vertices from multiple passes associated with different eye vertices, we map these vertices to a unified 2D space similar to CPPM \cite{lin2020cppm}, and the map function and its inverse function are
\begin{equation}
    \begin{cases}
        f_{\mathrm{map}, \x}(\x^*)=\left(\langle \x-\x^*,{\bf{u}}(\x)\rangle,\langle \x-\x^*, {\bf{v}}(\x)\rangle\right)\\
        f_{\mathrm{map}, \x}^{-1}(\y) = \x + {\bf{u}}(\x)\y_u+{\bf{v}}(\x)\y_v
    \end{cases} ,
    \label{equ:map-function}
\end{equation}
where $\x$ is the location of an eye vertex; $\x^*$ is a light vertex in the scene space; $\bf{u}(\x)$ and $\bf{v}(\x)$ are two orthogonal directions on the tangent plane at $\x$; $\langle \cdot,  \cdot \rangle$ is the inner product; ${\y=(\y_u, \y_v)}$ is a point in the unified 2D space.
In fact, this mapping aligns multiple gather points onto the unified space, and thus it adapts to multiple eye paths.

Then, one of the domains of integration in \autoref{equ:kaplanyan-integral} can be substituted by the unified space as
\begin{align*}
    I &\approx \int_{\mathcal{M}\times \Omega_r} K_r(\y) \Psi(\x, f_{\mathrm{map}, \x}^{-1}(\y)) d\x d\y \\
    &\approx \int_{\Omega_r} K_r(\y) \int_{\mathcal{M}} \Psi(\x, f_{\mathrm{map}, \x}^{-1}(\y)) d\x \  d\y \ ,
    \numberthis
    \label{equ:unify-integral}
\end{align*}
where $\Omega_r$ is the region within a radius $r$ from the origin in this unified 2D space.
For simplicity, we define a contribution function $\Psi^*(\y)$ in the unified space, which is $$\Psi^*(\y)=\int_{\mathcal{M}} \Psi(\x, f_{\mathrm{map}, \x}^{-1}(\y)) d\x \ .$$
Then we can simplify the estimation as
$$I \approx \int_{\Omega_r} K_r(\y) \Psi^*(\y) d\y \ .$$

According to the principle of kernel estimation of PM, if the kernel function $K_r$ is substituted by a Dirac delta function, $I$ in \autoref{equ:unify-integral} will be unbiased and $$ I=\Psi^*({\bf 0}) \approx  \int_{\Omega_r} K_r(\y) \Psi^*(\y) d\y \ .$$

Thereby, we propose the unbiased condition we hope to hold, which is a simple and sufficient condition for unbiased kernel estimation:
\begin{equation}
    \forall \y \in \Omega_r, \Psi^*(\y) \equiv \Psi^*({\bf 0}) \ .
    \label{equ:sufficient-condition}
\end{equation}
That is, the contribution function $\Psi^*(\y)$ is a constant function within $\Omega_r$.
The intuition behind the condition is that the estimate is unbiased when the illumination is evenly distributed within the disk $\Omega_r$.
Therefore, the null hypothesis for our PPM algorithm is that the unbiased condition mentioned above holds. In other words, the observations of the contribution function are subject to a constant function, which can be tested using our hypothesis testing method in the following Sec.~\ref{subsec:verifying}.

Compared our unbiased condition (\autoref{equ:sufficient-condition}) with the one used in CPPM that the samples of $\y$ are assumed to be uniformly distributed in $\Omega_r$, we can infer that CPPM actually relies on the correlation of the contribution function $\Psi^*(\y)$ and the distribution of $\y$.
Therefore, its high efficiency benefits from the scene settings and the sampling strategy of $\y$, and it may fail if the distribution of $\y$ is inconsistent with the contribution function $\Psi^*(\y)$.
The independence assumption can be removed. In contrast, this condition used to verify unbiasedness will be more reliable than CPPM. The estimate will be theoretically unbiased when the observations follow our proposed statistical model. Note that the samples need not be uniformly distributed for unbiased estimation with regard to our statistical model.

\subsection{Hypothesis Testing to Verify Unbiasedness}
\label{subsec:verifying}

%

A lightweight hypothesis testing method is required to infer the properties of samples based on our statistical model for kernel radiance estimate.
In this subsection, we present a method to compute the test statistics within a kernel radius and obtain the inference through hypothesis testing. Therefore, an appropriate radius can be determined by shrinking the radius and testing over iterations until the observations within this radius are deemed unbiased.

\subsubsection{Domain Partition and Integration}
\label{subsection:domainpartition}
The hypothesis test requires grouping the observations into categories, i.e., the random samples should be classified into several mutually exclusive classes. Accordingly, we can specify the equal-areal sectors as categories, and their associated light vertices being collected are thereby grouped. 
By partitioning $\Omega_r$ into sectors and assuming that $\Psi^*(\y)$ is constant in any sector, a straightforward solution for hypothesis testing is to gather samples of $\Psi^*(\y)$ in each sector and compare each sector's sample mean.
However, a sector's sample mean will be unavailable when it receives no light vertex, which may frequently occur after sample collection during each pass of photon mapping.
We notice that the kernel estimation in photon mapping based on Monte Carlo integration can also be expressed as sample means. 
Therefore, we use the samples of Monte Carlo integration in estimating the integration of $\Psi^*(\y)$ over each sector instead of the grouped light vertices in the statistical tests to compare these sectors' integration.

We first partition $\Omega_r$ into $n$ sectors $\Omega_{r, 1}\dots \Omega_{r,n}$.
The integration of $\Psi^*(\y)$ over sector $\Omega_{r, i}$, i.e. 
$\int_{\Omega_{r,i}} \Psi^*(\y) d\y$, can be written as
\begin{align*}
   &\int_{\Omega_{r,i}} \Psi^*(\y) d\y= \\
    &\int_{\mathcal{M}^2} \Psi(\x, \x^*) \mathbf{1}_{\Omega_{r,i}}(f_{\mathrm{map},\x}(\x^*)) dA(\x)dA(\x^*) \ ,
    \numberthis
    \label{equ:one-sector-integral}
\end{align*}
where $\mathbf{1}_{\Omega_{r,i}}(\x)$ is the indicator function whose value is $1$ if $\x$ is in $\Omega_{r,i}$ and $0$ otherwise.
\autoref{equ:one-sector-integral} shows that the integral of $\Psi^*(\y)$ over $\Omega_{r,i}$ can be estimated like photon mapping as
\begin{align*}
    \int_{\Omega_{r,i}} \Psi^*(\y) d \y \approx \overline{\psi}^*_{r,i}
    \approx \frac{1}{m} \sum_{j=1}^{m} \psi^*_{r,i,j}  \ ,
    \numberthis
\end{align*}
where $\overline{\psi}^*_{r,i}$ is the average of totally $m$ samples, and $\psi^*_{r,i,j}$ is the $j$-th sample of the Monte Carlo integration.
With this interpretation, for one pair of light sub-path and eye sub-path, we can obtain one sample for each sector in which only the indicator function of the samples is different.
Therefore, for $m$ pairs of light sub-paths and eye sub-paths, we can obtain $m$ samples for any of the $n$ sectors to compare their sample means.



\subsubsection{ANOVA F-test}
\label{subsubsection:ANOVA-f-test}
Since ANOVA F\nbd-test can assess whether the expected values of a quantitative variable within $n$ groups differ from each other, it is applied to test our null hypothesis that the expected value of $\psi^*_{r,i,j}$ in each sector is the same.
The Monte Carlo samples used to estimate the integration of $\Psi^*(\mathrm{y})$ over the same sector are regarded as samples of the same group in the ANOVA F\nbd-test.
The F\nbd-test statistic is the ratio of between-group variability and within-group variability, and its equation for groups with equal sample size is
\begin{equation}
    F=\frac{m \sum_{i=1}^n (\overline{\psi}^*_{r,i}-\overline{\psi}^*_r)^2 / (n-1)}{\sum_{i=1}^n \sum_{j=1}^{m} (\psi^*_{r,i,j}-\overline{\psi}^*_{r,i})^2 / (n(m-1))}\ ,
    \label{equ:f-statistic}
\end{equation}
where $m$ is the number of samples of each sector, and $\overline{\psi}^*_r$ is the average of all the $nm$ samples.
Intuitively, if the mean values are close across these sectors, then the between-group variability should be close to the within-group variability, and the statistic should be greater than and approaching $1$.
Otherwise, the between-group variability should be greater than within-group variability, and the statistic should be significantly greater than $1$.

We use a significance level $\alpha_{F}$ (typically $0.01$, $0.05$ or $0.10$ in statistics) to obtain a critical value $F_c$, which is a quantile of $1-\alpha_{F}$ of the F-distribution with degrees of freedom $d_1=n-1$ and $d_2=n(m-1)$ in our case.
If $F$ is greater than $F_c$, our statistical model is considered violated by the real population, and the null hypothesis is rejected. We decrease the radius for kernel estimation in this circumstance to reduce potential bias.
The ANOVA F\nbd-test is conservative in rejecting the null hypothesis. It rejects the null hypothesis only when the between-group difference is statistically significant.

The F\nbd-test assumes that the samples within a group are independent and normally distributed, and all groups have the same standard deviation, which is rather strict.
The violation of these assumptions will decrease the accuracy of the F\nbd-test.
However, the F\nbd-test for ANOVA is generally robust and the accuracy increases as sample size grows~\cite{moore2021introduction}.
Also, F\nbd-test has shown its availability when the samples follow a wide range of distributions~\cite{blanca2017non,feir1974empirical,BATHKE2004413}.
F\nbd-test works well in our application because we keep accumulating samples when the F\nbd-test does not reject the null hypothesis, which will be described in our algorithm in Sec.~\ref{subsec:algorithm-description}.

\subsubsection{Analysis F-test vs.  \texorpdfstring{$\chi^2$}{chi-squared}-test }
\label{subsubsec:chi-vs-F}
$\chi^2$-test is used to assess whether samples are uniformly distributed in $n$ equi-areal sectors. 
Like F\nbd-test, if the statistic $\chi^2$ is greater than a critical value $\chi_c$ calculated using the significance level $\alpha_\chi$, the null hypothesis is rejected.
The $\chi^2$\nbd-test statistic can only evaluate the difference between a real population and the hypothesized distribution.
While according to \autoref{equ:f-statistic}, the ANOVA F\nbd-test statistic evaluates the difference between the population with its contribution and the hypothesized distribution function.
Therefore, the ANOVA F\nbd-test is more accurate in making statistical inferences for this type of sample, which can alleviate the potential bias, as illustrated in \autoref{fig:spotlight} (right).

\subsubsection{Test Samples}
\label{subsec:test-samples}

Since \autoref{equ:one-sector-integral} has a similar formulation to PM (as shown in \autoref{equ:kaplanyan-integral}), we can reuse the samples of PM to get $\psi^*_{r,i,j}$.
According to the definition of $\Psi(\x, \x^*)$, each sample $\psi^*_{r,i,j}$ is the summation of the contributions from all full paths constructed by one eye sub-path and one light sub-path.
Therefore, one eye sub-path and one light sub-path make up a sample for each sector, though most samples $\psi^*_{r,i,j}$ equal zero because their values of indicator function are zero.
By convention, we use a range searching data structure to obtain non-zero contribution eye and light vertices pairs, and therefore we can not directly get the value of $\psi^*_{r,i,j}$.
We have to identify the sub-paths that the vertices belong to and sum up the contributions from the same sub-paths, and this procedure is computationally expensive.

To reduce this heavy computational overhead, we assume that one eye sub-path and one light sub-path can only construct no more than one full path with a non-zero contribution.
This is based on an observation that a random eye vertex and a random light vertex generally have a very low probability of constructing a non-zero-contribution path since the kernel radius is relatively small compared with the scene size.
Consequently, each non-zero-contribution full path is treated as an individual sample to get $\psi^*_{r,i,j}$.


\section{Hypothesis Testing For VCM+}


PPM algorithm works in unidirectional sampling way, that is one eye sub-path and one light sub-path construct one sample for each path length.
While in the VCM framework, one eye sub-path and one light sub\nbd-path construct multiple samples for each path length using the MIS technique, and each sample is weighted to minimize the total variance~\cite{vorba2011bidirectional}.
Consequently, the statistical model used in PPM formulation is incompatible with VCM's bidirectional framework.
In this section, we first deduce the unbiased condition for bidirectional PPM of VCM without BDPT involved. Thereby, we present an unbiased VCM estimator provided that the unbiased condition is satisfied, and we obtain VCM+ by hypothesis testing for unbiasedness.

\subsection{Unbiased Condition for Bidirectional PM of VCM}
We start with the estimate of bidirectional PM as an ingredient of VCM~\cite{Georgiev:2012:VCM}, which can be written as
\begin{equation}
     I \approx \frac{1}{N} \sum_{i=1}^N \langle I \rangle_i \quad,
    \label{equ:pm-contrib}
\end{equation}
where $N$ is the number of eye sub-paths;
and $\langle I \rangle_i$ is the estimate using the $i$-th eye sub-path as:
\begin{equation}
    \langle I \rangle_i \approx \frac{1}{J}\sum_{j=1}^{J} \sum_{l=2}^\infty \sum_{s=2}^{l-1} w_{s,r}(\overline{\x}_{i,j,l,s}^*)
    \frac{K_{s,r}(\overline{\x}_{i,j,l,s}^*) \Psi_s(\overline{\x}_{i,j,l,s}^*)}{p_s(\overline{\x}_{i,j,l,s}^*)}.
    \label{equ:bidir-PPM-Ii}
\end{equation}
In this equation, $J$ is the number of light sub-paths for each eye sub-path;
$r$ is the support radius of the kernel;
$\overline{\x}_{i,j,l,s}^*$ is the full path of length $l$ constructed by the $s$-th vertex of the $i$-th eye sub-path and the ($l$-$s$+2)\nbd-th vertex of the $j$-th light sub-path;
$K_{s,r}$ is the kernel function over the two vertices;
$w_{s,r}$ is the MIS weight function concerning sampling probability;
$p_s$ is the probability density function;
$\Psi_s$ is the full path contribution function, which includes the emission, geometry, bidirectional scattering distribution function (BSDF), and eye importance terms.

To derive a theoretical formulation, we use path integral framework~\cite{veach1997robust}, and the expected value of $I$ in an integral form is:
\begin{equation}
    I = \sum_{l=2}^{\infty} \sum_{s=2}^{l-1} \int_{\mathcal{M}^{l+2}} w_{s,r}(\overline\x_{l}^*) K_{s,r}(\overline\x_{l}^*) \Psi_s(\overline\x_{l}^*) d\mu(\overline\x_{l}^*),
    \label{equ:e-c-ppm}
\end{equation}
where $\mathcal{M}$ is the scene surfaces in $\mathbb{R}^3$; ${\overline{x}_l^*=(\x_0 \dots \x_{l+1})}$ is a full path of length $l$ constructed by $l$+2 vertices; ${d\mu(\overline{x}_l^*)=dA(\x_0) \dots dA(\x_{l+1})}$ is the differential product area measure.

To focus on the kernel estimation, we apply Fubini's theorem to the above equation (\autoref{equ:e-c-ppm}) to change the order of integration and then integrate over $d\mu(\overline\x_{l}^*)$ except the ($s$-1)\nbd-th and the $s$\nbd-th differential into $F_{l,s,r}$, then we have:
\begin{equation}
    I = \int_{\mathcal{M}^{2}} \sum_{l=2}^{\infty} \sum_{s=2}^{l-1} K_r(\lVert \x-\x^* \rVert) F_{l,s,r}(\x, \x^*) dA(\x^*)dA(\x).
    \label{equ:PPM-reformulation}
\end{equation}
Let $\overline\x'$ be a path of length $l$ with $l+2$ vertices having $\x$ for the fixed $(s-1)$\nbd-th vertex and $\x^*$ for the fixed $s$\nbd-th vertex, then it lives within $\mathcal{M}^l$ and $d\mu(\overline{\x}')$ equals to $dA(\x_0)...dA(\x_{s-2})dA(\x_{s+1})...dA(\x_{l})$. $F_{l, s ,r}(\x, \x^*)$ can then be written as:  
\begin{equation}
    F_{l, s ,r}(\x, \x^*) = \int_{\mathcal{M}^l} w_{s, r}(\overline\x') \Psi_s(\overline\x') d\mu(\overline{\x}') .
\end{equation}

Same with the way discussed in \autoref{sec:uni-hypo-condition}, we change one of the domains of integration in \autoref{equ:PPM-reformulation} to the unified space as
\begin{equation}
    I \approx \int_{\Omega_r} K_r(\lVert \y \rVert) \int_{\mathcal{M}} \sum_{l=2}^{\infty} \sum_{s=1}^{l-1} F^*_{l,s,r}(\x, \y) dA(\x) \ d\y,
    \label{equ:VCM-approximate-estimation}
\end{equation}
where $F_{l,s,r}^*(\x, \y)=F_{l,s,r}(\x, f_{\mathrm{map}, \x}^{-1}(\y))$.

\begin{figure}
	\centering
	\includegraphics[width=0.75\linewidth]{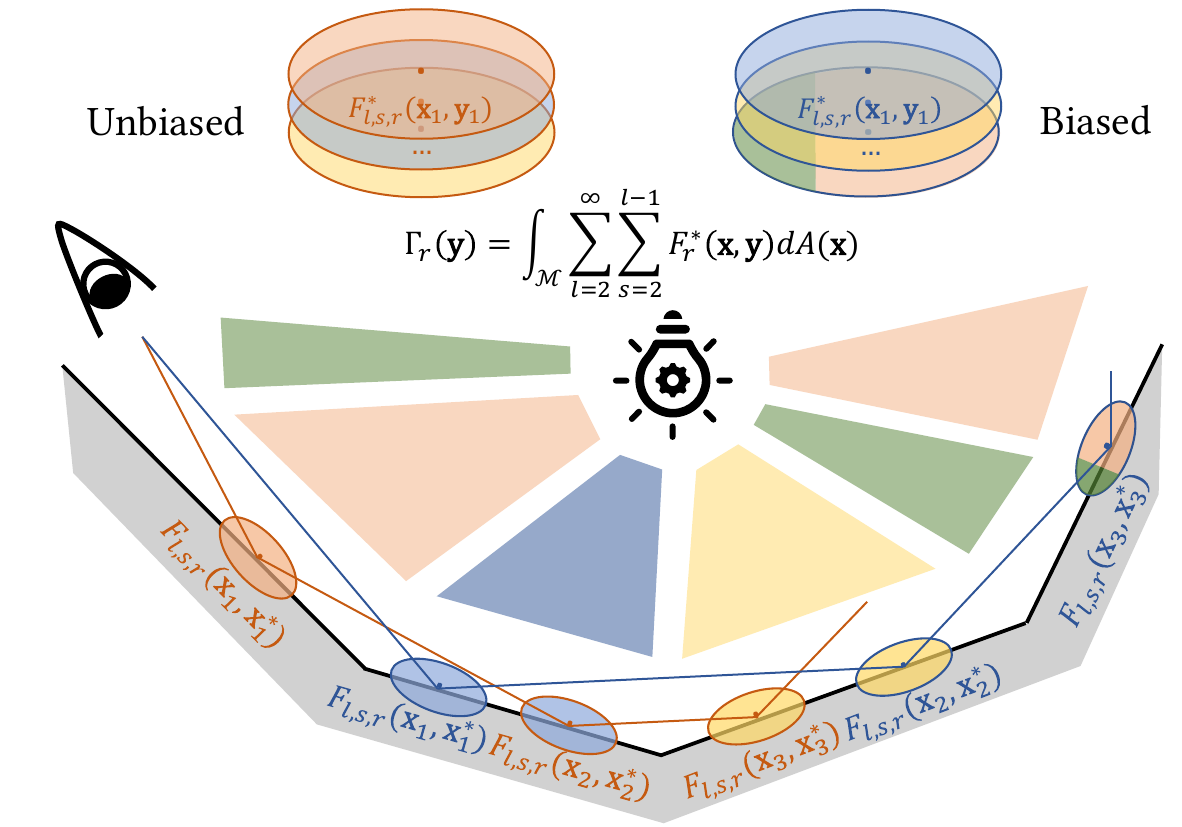}
	\caption{Illustration of unbiased condition for a bidirectional estimator. Two eye sub-paths are traced from the eye individually. The kernel estimation of the eye sub-path in red will be unbiased, while the blue ones will be biased. The discs at light vertices $\x$ indicate the contribution function $F_r(\x, \x^*)$ in path space. The contribution $F_r(\x, \x^*)$ around $\x$ are aligned as $F_r^*(\x, \y)$ and further integrated to obtian function $\Gamma_r(\y)$. The estimator is unbiased if $\Gamma _r ( \y )$ is a constant.
	}
	\label{fig:unbias-vs-bias}
  \vspace{-0.1in}
\end{figure}

In principle, the bias of PM algorithm is produced by the kernel function $K_r$, while the term $F^*_{l,s,r}$ will not introduce any bias.
Therefore, by substituting $K_r$ with Dirac delta function $\delta$, we can obtain an unbiased integral as
\begin{equation}
    \begin{aligned}
        I &= \int_{\Omega_r} \int_{\mathcal{M}} \sum_{l=2}^{\infty} \sum_{s=1}^{l-1} F^*_{l,s,r}(\x, \y) \delta(\lVert \y \rVert) dA(\x) d\y \\
        &= \int_{\mathcal{M}} \sum_{l=2}^{\infty} \sum_{s=2}^{l-1} F^*_{l,s,r}(\x, {\bf 0}) dA(\x). \\
    \end{aligned}
    \label{equ:VCM-dirac-estimation}
\end{equation}
For simplicity, we define function $\Gamma_r(\y)$ as
\begin{equation}
    \Gamma_r(\y) = \int_{\mathcal{M}} \sum_{l=2}^{\infty} \sum_{s=1}^{l-1} F^*_{l,s,r}(\x, \y) dA(\x),
\end{equation}
and we can obtain a simple and sufficient condition for an unbiased estimator as
\begin{equation}
    \forall \y\in \Omega_r, \Gamma_r(\y)\equiv\Gamma_r({\bf{0}}),
    \label{equ:bidir-sufficient-condition}
\end{equation}
i.e., function $\Gamma_r(\y)$ concerning contributions should be constant within a disk domain with a support radius $r$, as illustrated in the top of~\autoref{fig:unbias-vs-bias}.
\autoref{equ:bidir-sufficient-condition} is served as the unbiased condition for bidirectional PM (VCM without BDPT). 
Therefore, the null hypothesis for the bidirectional PM algorithm is that the unbiased condition mentioned above holds.

Compared to the unbiased condition for PM (\autoref{equ:sufficient-condition}), the intuition of \autoref{equ:bidir-sufficient-condition} is also evenly distributed illumination, but \autoref{equ:bidir-sufficient-condition} takes multiple sampling strategies together with MIS weights into consideration.
It implies the overall bias is also the MIS-weighted average of the bias from different sampling strategies.
Note that multiple estimates with bias can theoretically be unbiased when their bias happens to cancel out. 
Therefore, the MIS weights affect bias, and the unbiased condition (\autoref{equ:bidir-sufficient-condition}) should implicitly involve the MIS weights to safeguard the unbiasedness tightly.

\subsection{Unbiased VCM Estimator}

In the bidirectional framework, the VCM estimator considers one eye sub\nbd-path each iteration and combines multiple sampling techniques from VC (i.e., BDPT) and VM (i.e., PM) using the MIS technique.
The eye vertices on the eye sub\nbd-path are used to construct regular paths with $n_{\mathrm{VC}}$ independent light sub\nbd-paths and extended paths with $n_{\mathrm{VM}}$ independent light sub\nbd-paths using radius $r$.
VCM estimator can then be written as:
\begin{equation}
    \langle I \rangle =\langle C\rangle_{\mathrm{VC}}+\langle C \rangle_{\mathrm{VM}} \ ,
    \label{equ:VCM-estimator}
\end{equation}
where
\begin{align*} 
    \langle C\rangle_{\mathrm{VC}} &= \frac{1}{n_{\mathrm{VC}}}\sum_{l=1}^{n_{\mathrm{VC}}}\sum_{k=1}^{\infty}\sum_{s=0}^{k+1} w_{\mathrm{VC},s,r}(\overline\x_{k,s,l})\frac{f(\overline\x_{k,s,l})}{p_{s}(\overline\x_{k,s,l})} \\
    \langle C \rangle_{\mathrm{VM}} &= \frac{1}{n_{\mathrm{VM}}}\sum_{l=1}^{n_{\mathrm{VM}}}\sum_{k=2}^{\infty}\sum_{s=1}^{k-1} w_{\mathrm{VM},s,r}(\overline\x_{k,s,l}^*)\frac{f_{s,r}(\overline\x_{k,s,l}^*)}{p_{s}(\overline\x_{k,s,l}^*)}.
    \numberthis
    \label{equ:C-CV-VM}
\end{align*}
In this equation, $\overline\x_{k,s,l}$ is the regular path constructed by $k$-$s$+1 vertices from the eye sub\nbd-path and $s$ vertices from the $l$-th light sub\nbd-path among $n_{\mathrm{VC}}$ light sub\nbd-paths; $\overline\x_{k,s,l}^*$ is the extended path constructed by $k$-$s$+1 vertices from the eye sub\nbd-path and $s$+1 vertices from the $l$-th light sub\nbd-path among $n_{\mathrm{VM}}$ light sub\nbd-paths; $f$ is the measurement contribution function measuring the contribution along the path; $f_{s,r}$ is the measurement contribution function for VM; $p_s$ is the probability density function;
and $w_{v,s,r}$ is the power heuristic function of MIS for corresponding techniques where $v^*$ denotes $\mathrm{VC}$ or $\mathrm{VM}$:
\begin{equation}
    \begin{aligned}
     w_{v^*,s,r}(\overline\x_k) &= \frac{n_{v^*}^\beta p_{v^*,s,r}^\beta(\overline\x_k)}{\Upsilon} \ , \\
    \Upsilon &= {
    n_{\mathrm{VC}}^\beta \sum_{s=0}^{k+1}p_{\mathrm{VC},s,r}^\beta(\overline\x_k)
    +
    n_{\mathrm{VM}}^\beta \sum_{s=1}^{k-1}p_{\mathrm{VM},s,r}^\beta(\overline\x_k)
    } \ .
    \end{aligned}
    \label{equ:mis-weight}
\end{equation}
The balance heuristic corresponds to $\beta$=1.
In the above equation, $p_{\mathrm{VM},s,r}$ is proportional to $r^2$, while $p_{\mathrm{VC},s,r}=p_{\mathrm{VC},s}$ is independent of $r$. That is, the kernel radius is the key to determining the contribution from VM.

Then, the progressive estimator is the average of $N$ iterations:
\begin{equation}
    \langle I \rangle = \frac{1}{N} \sum_{i=1}^N \langle I \rangle_{i} \ ,
    \label{equ:progressive-VCM-estimator}
\end{equation}
where $\langle I \rangle_{i}$ is the result computed by \autoref{equ:VCM-estimator} using the $i$-th radius $r_i$, the $i$-th set of eye sub\nbd-paths and the $i$-th set of light sub\nbd-paths.

Besides the MIS weight, VM ingredient in the estimate shares the same components of \autoref{equ:bidir-sufficient-condition}.
Therefore, we can integrate the MIS weight of both VM and VC into the full paths' contribution. Then the unbiased condition for bidirectional PM is also available for the VCM framework.

In our VCM+ algorithm, we use ANOVA F-test to test whether the unbiased condition holds or not in the same way as discussed in Sec.~\ref{subsec:verifying}.
We see the null hypothesis as an equivalent of the unbiased condition for VM in VCM+ in practice. We will reject/not reject the null hypothesis given the testing of observations of the contribution. If not rejecting the null hypothesis, the radius $r$ will be identified for unbiased estimation. VCM+ will find the right radius as early as possible during the iterations, and hold this radius unchanged for kernel estimation.
That is, the kernel radius needs not shrink progressively to mitigate the bias of VM. Thus we reduce variance by collecting more photons with larger kernels.


\section{Algorithm}
\label{sec:algorithm}
In this section, we will outline our algorithm based on hypothesis testing. As no modifications have been made to the BDPT component, we will focus solely on the updates made to the PPM algorithm, which involve general enhancements to VCM under our theoretical model. The statistical models proposed for both PPM and VCM frameworks only affect the sample collection and radius reduction module solely, so that they can share the same working pipeline thereof.


\subsection{Kernel Radius and Radiance Estimation}
\label{subsec:algorithm-description}
The PPM part of our algorithm overall follows the working flow similar to CPPM: In each iteration, we trace $J$ light sub-paths, build a range search data structure of the light vertices, trace multiple eye sub-paths for each pixel, collect the light vertices around the eye vertices for kernel estimation, update the searching radius by hypothesis testing, update the power heuristic function of MIS for VCM+ according to \autoref{equ:mis-weight}, and update the estimate of the radiance finally (\autoref{equ:VCM-estimator}).
\changed{
The key steps with regard to hypothesis testing at each gather point are as the following:
\begin{enumerate}
    \item Disk $\Omega_r$ is partitioned into sectors (Sec.~\ref{subsection:domainpartition}), using the same strategy as CPPM.
    \item The statistics of the light vertices that fall in the sectors are accumulated. By applying light path tracing, each light vertex that falls inside the disk produces one sample for each sector. For each sector, we maintain the sample count ($m$), the sums of sample contributions ($\overline{\psi}_{r,i}^*$) and the sums of their squares ( $\sum_{j=1}^m(\psi_{r,i,j}^*)^2$) individually. They correspond to the terms presented in Sec.~\ref{subsubsection:ANOVA-f-test}. These statistics are collected to compute an F-statistic using \autoref{equ:f-statistic} for the ANOVA F-test. 
    \item ANOVA F-test (Sec.~\ref{subsubsection:ANOVA-f-test}) is used to check whether the null hypothesis can be rejected. The null hypothesis is that the unbiased condition holds itself (see \autoref{equ:bidir-sufficient-condition}) as mentioned in Sec.~\ref{sec:uni-hypo-condition}. We compare the computed F-statistic against a threshold based on a fixed confidence level of the statistical test (typically a global parameter) to reject or not reject the null hypothesis. In this way, we determined if the observations within the current kernel radius can lead to unbiased estimation.
    \item Update the kernel radius according to \autoref{equ:radiusupdate}.
    \item Update the power heuristic function of MIS for VCM+ according to \autoref{equ:mis-weight}.
\end{enumerate} }
The radius is updated according to the F\nbd-test as
\begin{equation}
\label{equ:radiusupdate}
    r_{i+1}=
    \begin{cases}
        r_i, & \text{if } F \le F_c \\
        \mathrm{max}\left(k r_i, r_{\mathrm{min}}\sqrt{(i+1)^{\alpha-1}}\right), & \text{if } F > F_c
    \end{cases}.
\end{equation}
Here, the parameter $\alpha \in (0,1)$ bounds the rate of radius reduction, $k \in (0,1)$ is a user-defined shrinking ratio, and $r_{\mathrm{min}}\sqrt{(i+1)^{\alpha-1}}$ serves as a lower bound to ensure convergence.
This is effective because the F-test is conservative in rejecting the null hypothesis, and a small sample size does not significantly increase the probability of wrongly rejecting a true null hypothesis~\cite{william1952goodness,jan2014sample}.
We remove the unnecessary parameter (e.g., $\beta$) designed for tuning in CPPM pipeline, which can improve the flexibility of our algorithm based on the convergence analysis in the later Sec.~\ref{subsec:convergence-analysis}.


When performing the ANOVA F\nbd-test, $J$ new samples are collected in one iteration for each sector.
As known to all, a large sample size can decrease the probability of wrongly not rejecting a false null hypothesis.
To increase the sample size, we retain sample statistics of former iterations where the radius does not shrink.
Therefore, denoting $t$ as the number of iterations for which the current radius remains fixed, the F\nbd-test has a total of $tJ$ samples for each sector.
According to \autoref{equ:f-statistic}, with an infinite number of samples, the F\nbd-test statistic goes to infinity if bias still exists. Our proposed F-test will finally reject the null hypothesis with sufficient samples. The only exception is, by chance, that differences in contribution occur not between sectors but within. It is a rather rare circumstance and can hardly occur in practice.
Generally, our algorithm performs F\nbd-test only on the luminance channel. 
In this way, the probability of wrongly rejecting an unbiased estimator equals $\alpha_F=0.01$.
When handling the chromatic light sources, we perform F\nbd-test on RGB channels individually.

\subsection{Convergence Analysis}
\label{subsec:convergence-analysis}
In this subsection, we analyze the convergence rate of our algorithm.
In the worst case where the hypothesis testing always rejects the null hypothesis, i.e., the condition for unbiased estimation is never satisfied, the radius will always shrink once the algorithm gets a non-zero sample.
Without the lower bound of the radius, we can deduce that $k$ does not affect the convergence of the kernel radius, and the radius convergence rate is $O(N^{-1/2})$ in this case~\cite{knaus2011progressive}.
This allows our algorithm to shrink the kernel radius faster before reaching an appropriate radius compared to CPPM and PPM.
However, applying Knaus and Zwicker's probabilistic framework~\cite{knaus2011progressive}, we learn that the variance of PPM does not converge in this case.
Therefore, with a lower bound, the convergence rate of PPM in our algorithm is asymptotically identical to PPM under the worst situation. The optimal mean squared error (MSE) convergence rate $O(N^{-2/3})$ is achieved by setting $\alpha=2/3$ in this situation.
In practice, the lower bound can hardly be reached since most pixels can pass the hypothesis testing with a relatively large kernel and stop reducing the kernel radius.
Assuming the hypothesis testing produces the correct result, the convergence rate becomes $O(N^{-1})$.
With this convergence rate, we can deduce that our algorithm also obtains the optimal convergence rate of $O(N^{-1})$ for VCM~\cite{Georgiev:2012:VCM}. That is, the VCM+ algorithm can reach a convergence rate of $O(N^{-1})$ under ideal conditions.

The parameter $k$ only affects initial bias and variance.
A larger $k$ generally introduces more bias but less variance.
Therefore, the optimal $k$ may vary from scene to scene.
We will provide some empirical values from our experiments in Sec.~\ref{subsubsec:radius-parameter-k}.

\begin{figure}
    \centering
    \includegraphics[trim={2.8cm 1cm 1.2cm 1.5cm}, clip, width=0.7\linewidth]{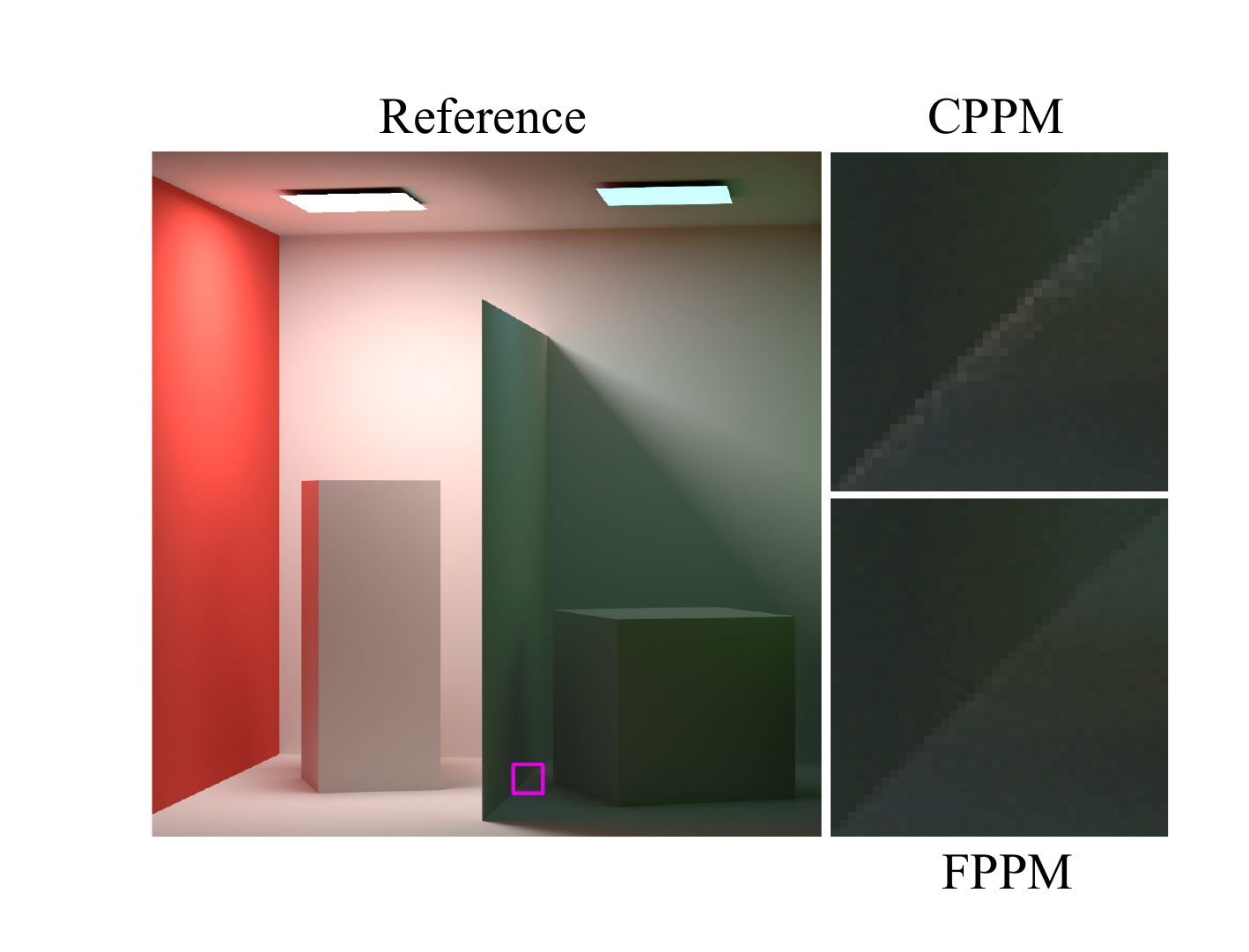}
    \caption{FPPM (ours) vs. CPPM after 1,000 iterations, using Split Cornell Box as an example. Our algorithm helps alleviate the light leakage artifacts.}
    \label{fig:light-leak-1}
\end{figure}

\section{Experiments and Discussions}
\label{sec:result}

We implement our algorithm under both the PPM framework and the VCM framework.
In this section, we provide experiments and asymptotic performance on diverse scenarios, and we also discuss the effect of different parameter values related to our algorithm in the experiments. 
The results of all comparative experiments are obtained under the same environment on the same machine. All the algorithms are developed based on the Mitsuba Renderer~\cite{Mitsuba}.

%


\subsection{Results and Comparisons}
\label{subsec:results}

Firstly, we compare two progressive photon mapping algorithms: FPPM (our F\nbd-test-based PPM) and CPPM (state-of-the-art method).
We then conduct various experiments and compare VCM+ with VCM (BDPT+SPPM).

\subsubsection{FPPM vs. CPPM}

\textbf{Settings:}
Both CPPM and FPPM (our algorithm) are based on hypothesis testing. 
The common parameters used by the CPPM and ours are the same unless otherwise stated. To perform the statistics test, we partition $\Omega_r$ into $n_a$ equi-areal annuli and each annulus into $n_s$ equi-areal sectors.
Specifically, we generally set $n_a=2$, $n_s=6$, $k=0.7$ and significance level $\alpha_\chi = \alpha_F=0.01$.
The critical value for the $\chi^2$\nbd-test ($\chi_c$) for CPPM is $24.724$, and the critical value for the F\nbd-test ($F_c$) update in run time according to the sample size, and is approximately $2.25$ when the total sample count approaches infinity.
\begin{figure*}
    \centering
  \begin{tabular}{c@{\hskip1pt} @{\hskip1pt}c@{\hskip1pt} @{\hskip1pt}c @{\hskip1pt} @{\hskip1pt}c}
        & \multicolumn{2}{c}{CPPM (MSE=5.308)} &\\
       \includegraphics[height=1.5in]{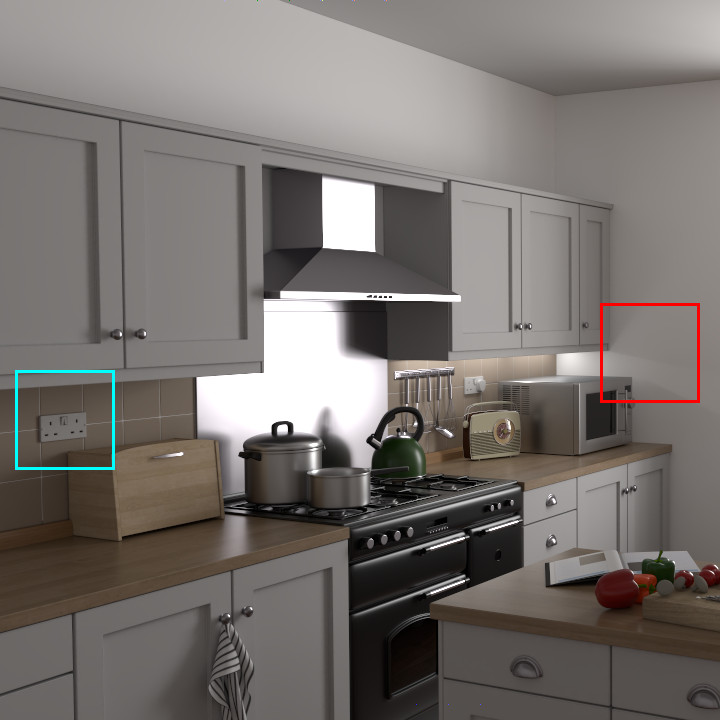}
       \includegraphics[height=1.5in]{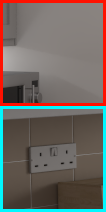}
       & \includegraphics[height=1.5in]{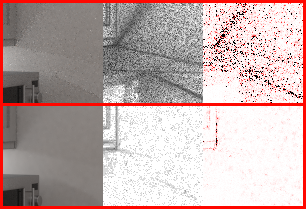}
        & \includegraphics[height=1.5in]{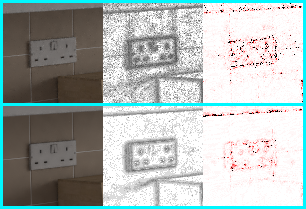}
        & \includegraphics[height=1.5in]{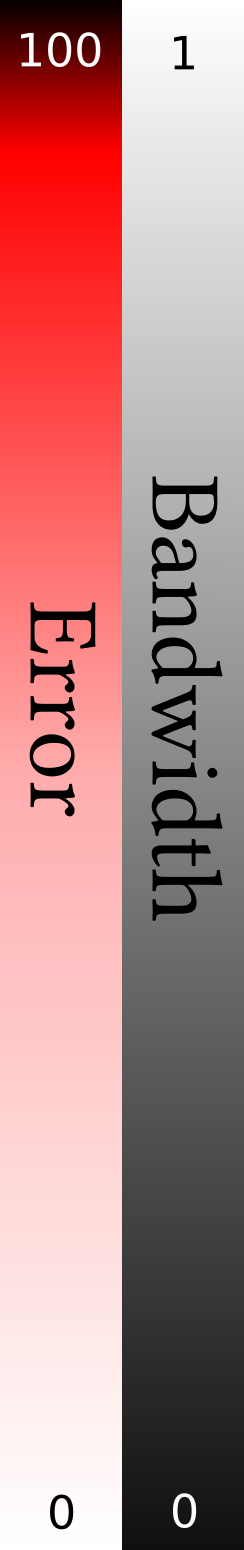}  \\
       Kitchen (Reference) & \multicolumn{2}{c}{FPPM (MSE=\textbf{3.457})} &\\
       [5.5pt]
         & \multicolumn{2}{c}{CPPM (MSE=3.830)} &\\
       \includegraphics[height=1.5in]{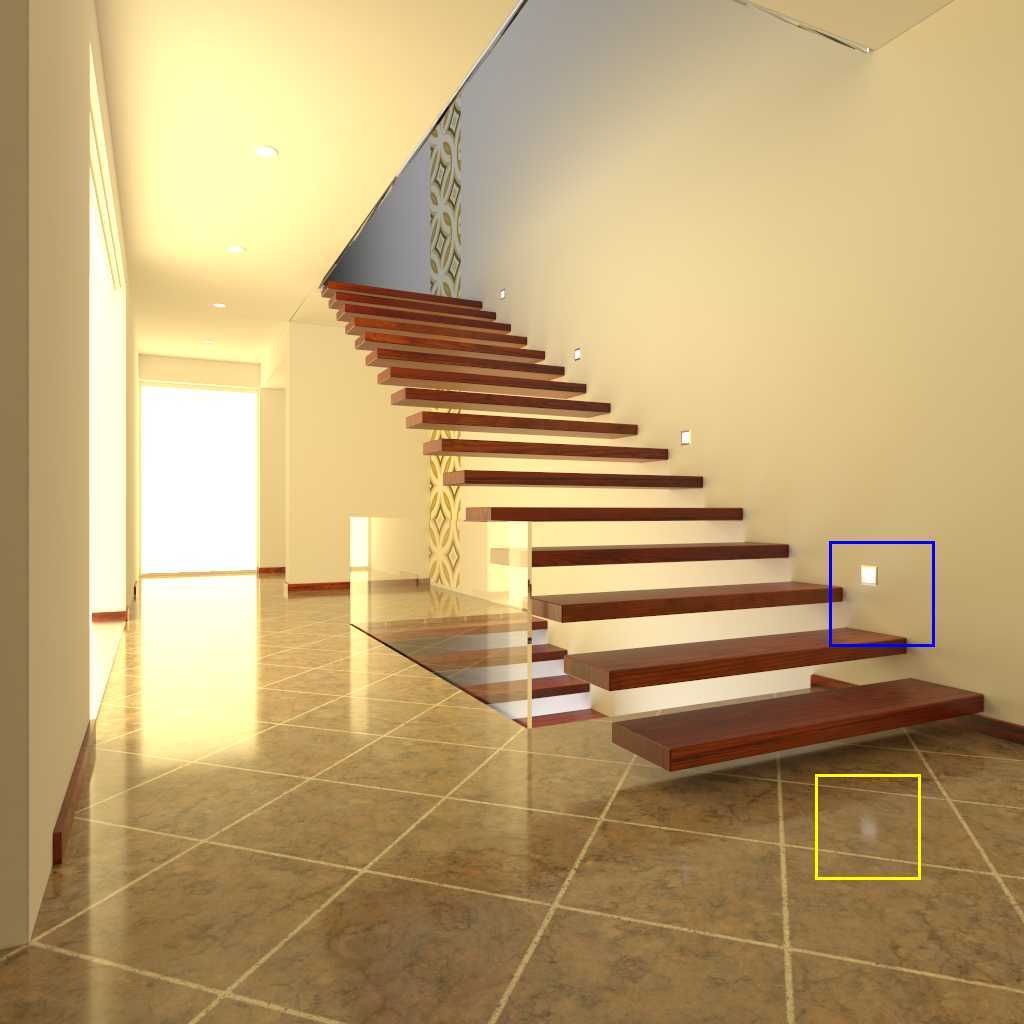}
       \includegraphics[height=1.5in]{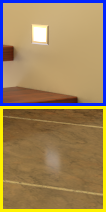}
       & \includegraphics[height=1.5in]{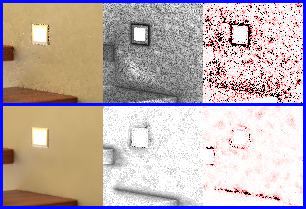}
        & \includegraphics[height=1.5in]{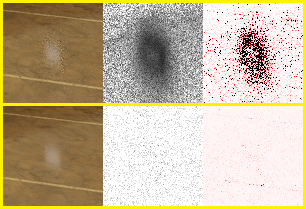}
        & \includegraphics[height=1.5in]{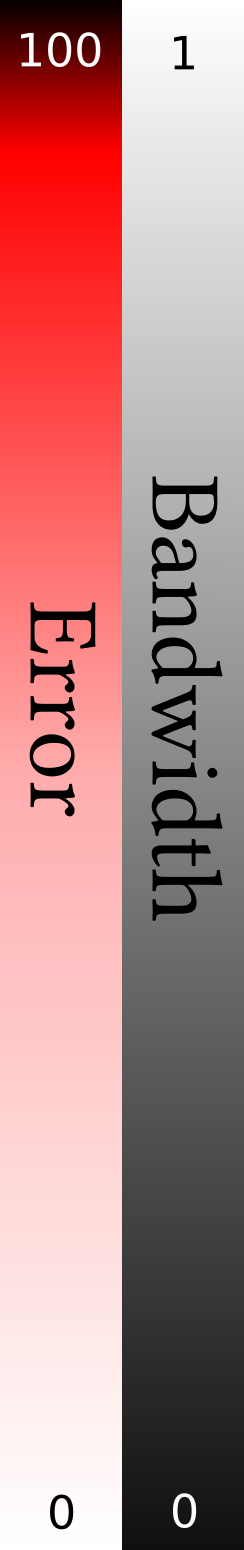} \\
    Hall (Reference) & \multicolumn{2}{c}{FPPM (MSE=\textbf{1.373})}  &\\
    [5.5pt]
     & \multicolumn{2}{c}{CPPM (MSE=4.445)} &\\
     \includegraphics[height=1.5in]{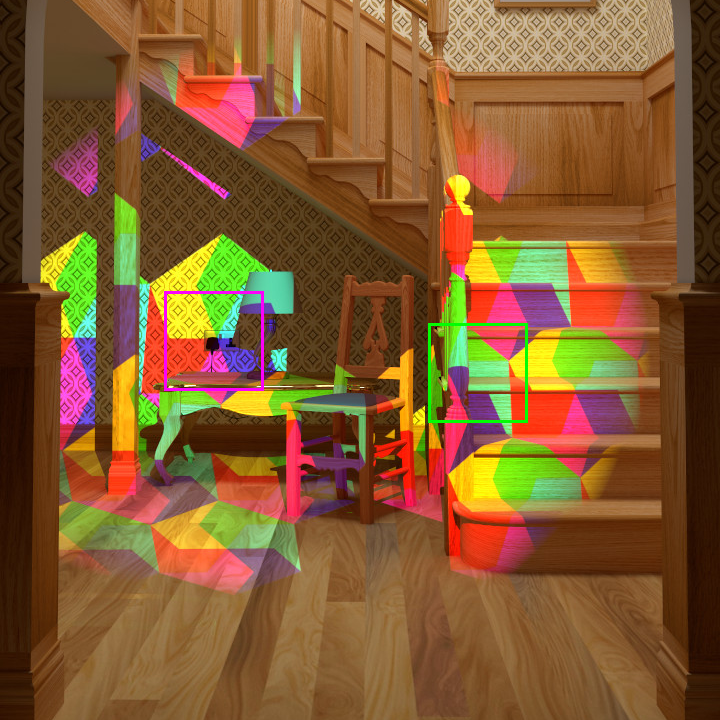}
       \includegraphics[height=1.5in]{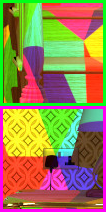}
       & \includegraphics[height=1.5in]{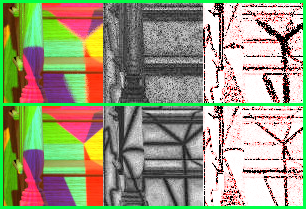}
        & \includegraphics[height=1.5in]{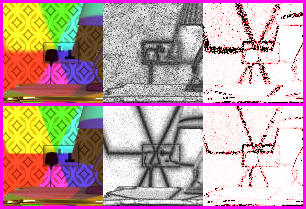}
        & \includegraphics[height=1.5in]{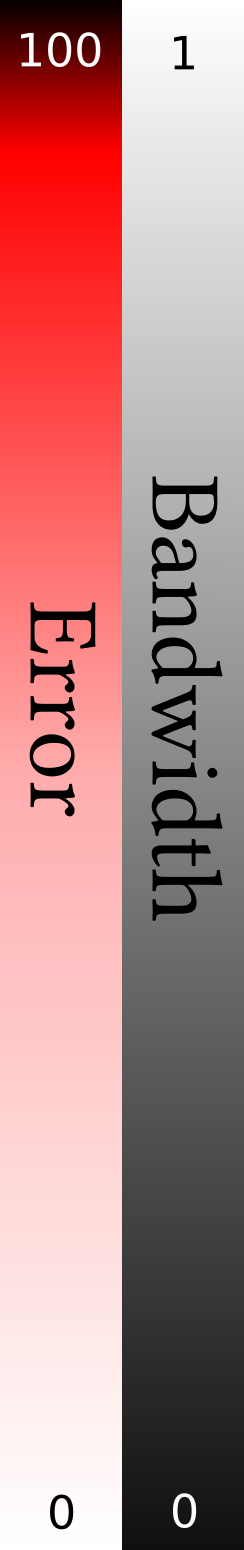} \\
    Staircase (Reference) & \multicolumn{2}{c}{FPPM (MSE=\textbf{3.659})} &\\[4pt]
  \end{tabular}
    \caption{Our algorithm (FPPM) outperforms CPPM in equal-pass comparisons on different test scenes (10K iterations for Kitchen and Hall, and 5K iterations for Staircase). We visualize the kernel radius in greyscale (middle of each zoom-in block) and the squared error (right of each zoom-in block). Our algorithm yields less noisy results and correctly handles false rejections produced by {$\chi^2$}\nbd-test where emitted photons failed to capture the actual irradiance distribution. Moreover, F-tests on different spectral channels help improve the quality of textured lights. } 
  \label{fig:pm-comparison-aggregate}
\end{figure*}

\begin{figure*}
  \centering
  \begin{tabular}{c@{\hskip1pt} @{\hskip1pt}c@{\hskip1pt} @{\hskip1pt}c}
      \includegraphics[height=1.5in]{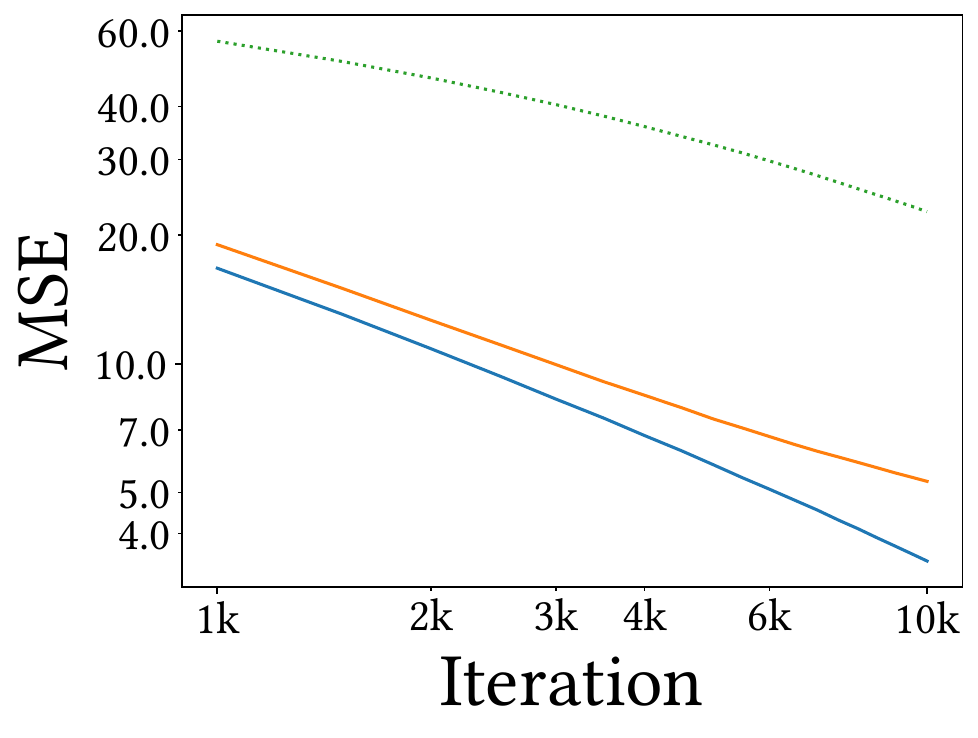}
      &\includegraphics[height=1.53in]{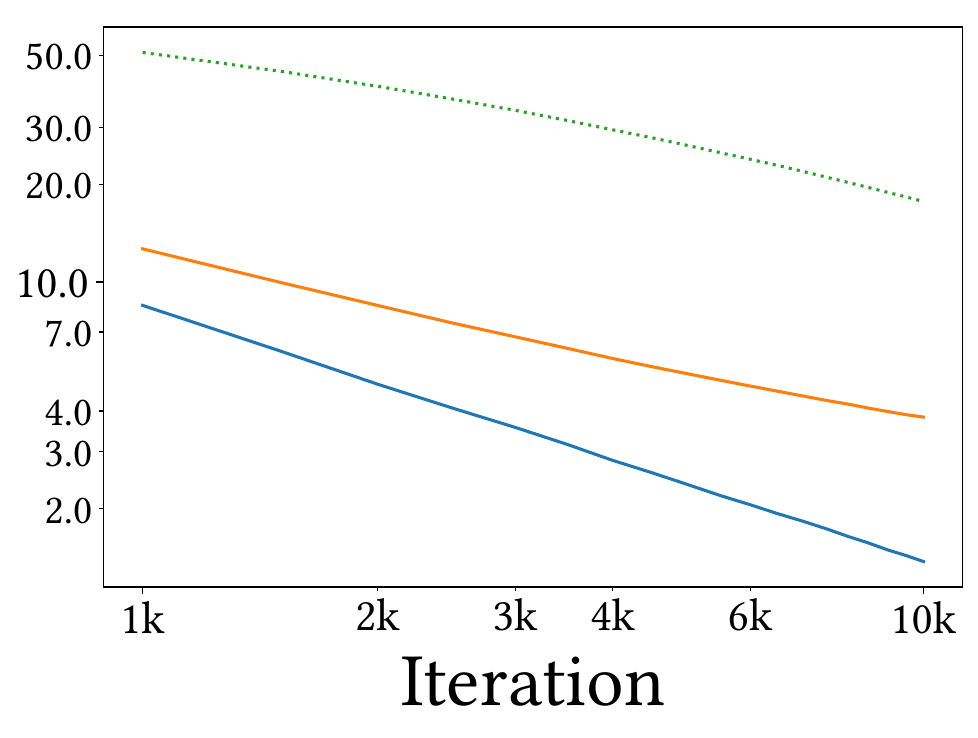} 
      & \includegraphics[height=1.5in]{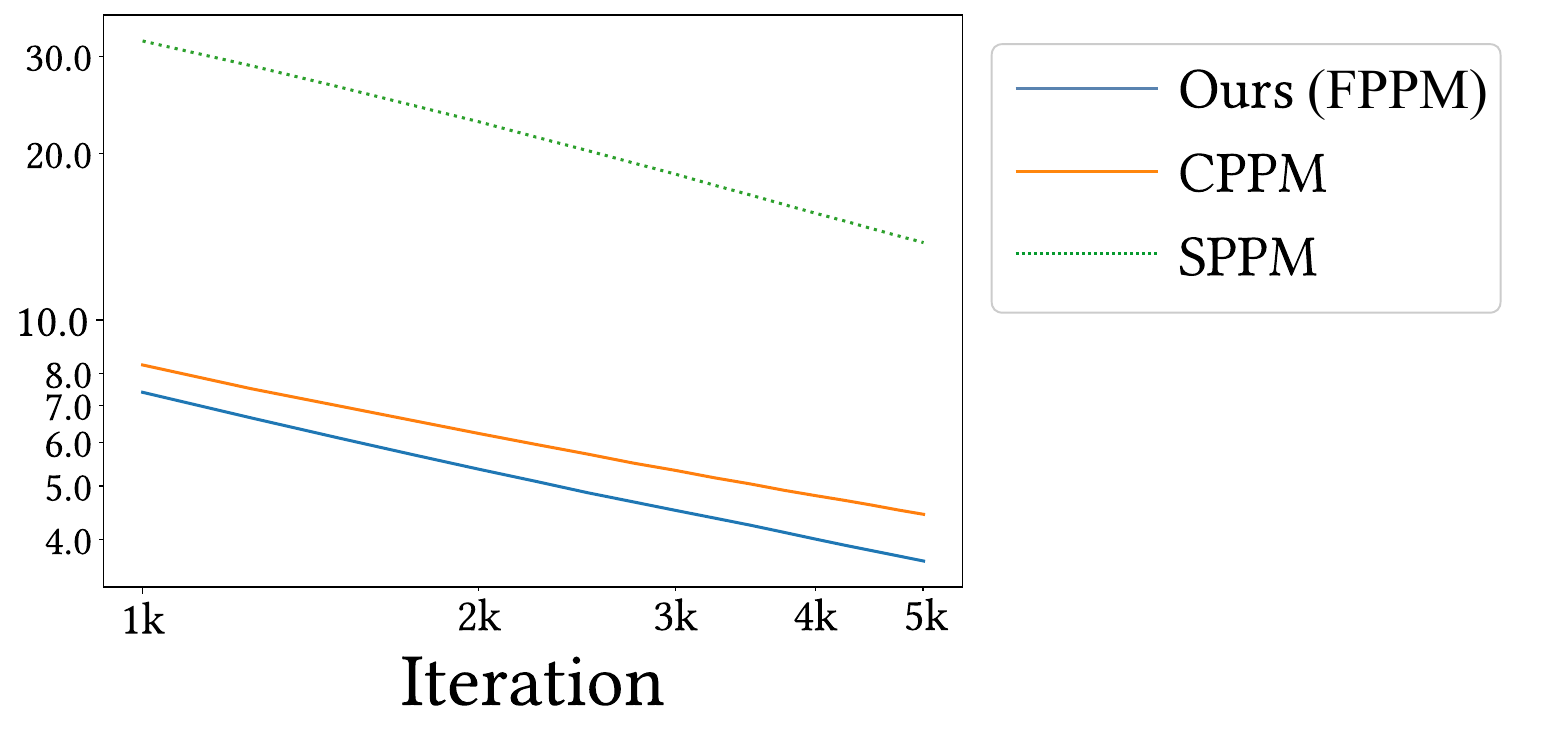}
       \\
    (a) Kitchen & (b) Hall & (c) Staircase \quad \quad \quad \quad \quad \quad \quad \quad
  \end{tabular}
    \caption{MSE over iterations of SPPM, CPPM, and Ours (FPPM). FPPM shows superior performance, converges faster than CPPM, and significantly outperforms the baseline.}
\label{fig:ppm-convergence}    
\end{figure*}





We first validate that our algorithm can minimize the common artifacts compared to CPPM. We test a special case, Split Cornell Box. As shown in \autoref{fig:light-leak-1}, our algorithm alleviates the light leaks and obtains more accurate results in equal-iteration comparison. 
Moreover, we conducted comparisons on three scenes, as shown in \autoref{fig:pm-comparison-aggregate} with highlighted details. These three scenes showcase diverse lighting conditions. The Kitchen contains numerous objects, small area lights above the oven, and a large area light from the window; the Hall scene is a two-storey building with glossy floors, and has many area lights as well as a large light source outside; the Staircase scene features a textured spotlight, which is a special case.

\begin{table*}[thb]
    \centering
    \begin{tabular}{c|c|c|c|c|c|c|c|c|c}
        \bottomrule
       \multicolumn{2}{c|}{} & Pool & Hall & Bathrooom & Dragon & Glass & Ball & Kitchen & Staircase   \\
        \multicolumn{2}{c|}{\raisebox{0.18in}[0ex][0ex]{Scene}}
         & \includegraphics[height=0.45in]{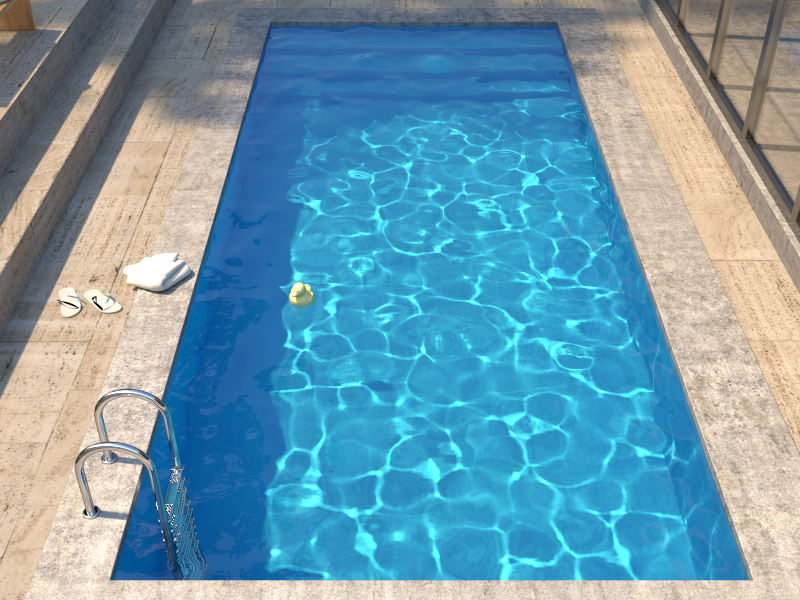}
         & \includegraphics[height=0.45in]{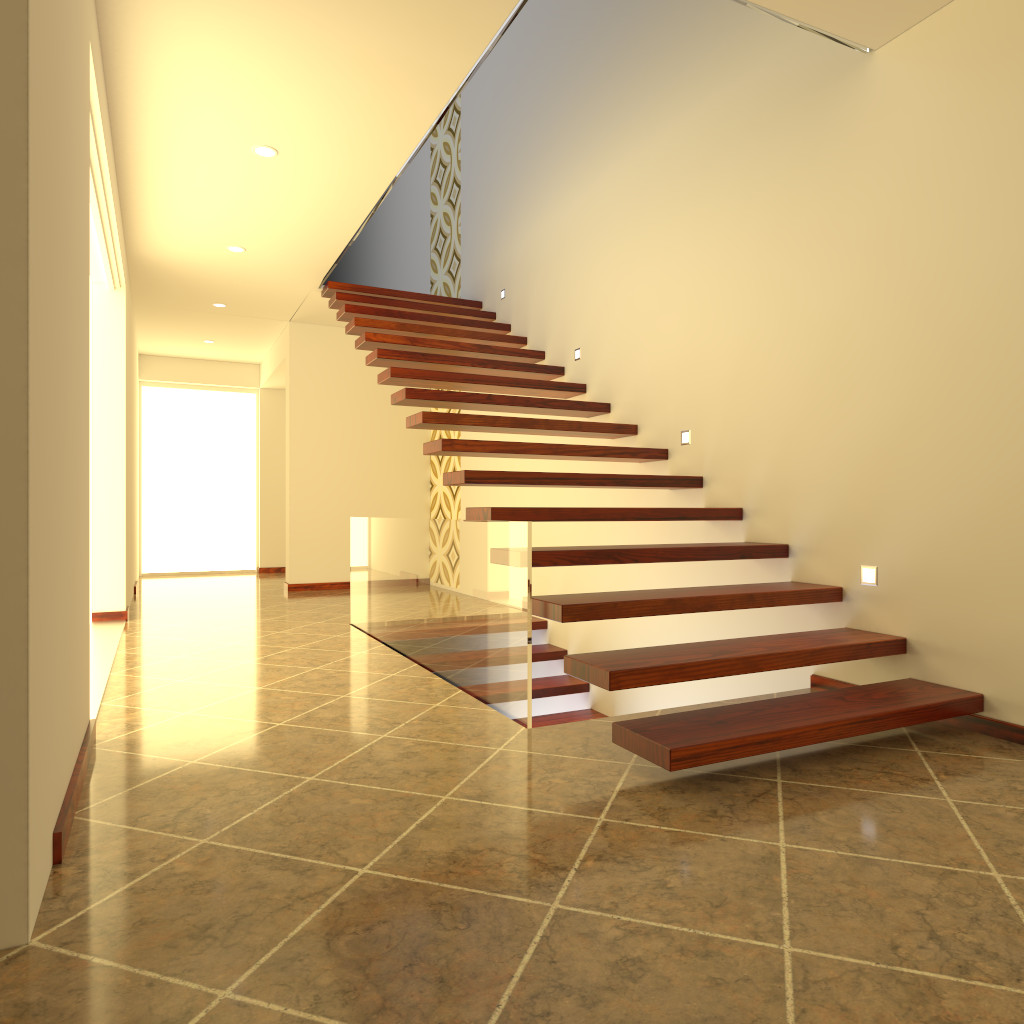}
          & \includegraphics[height=0.45in]{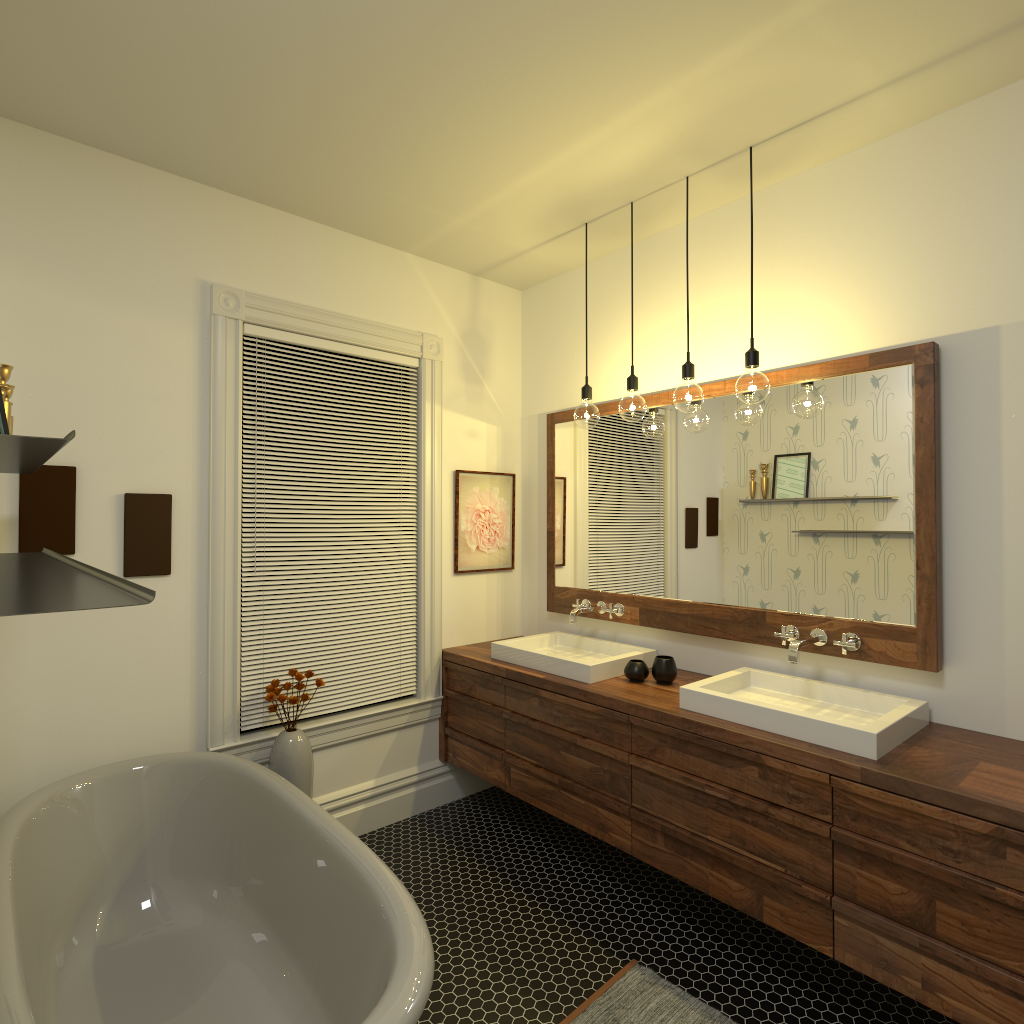}
         & \includegraphics[height=0.45in]{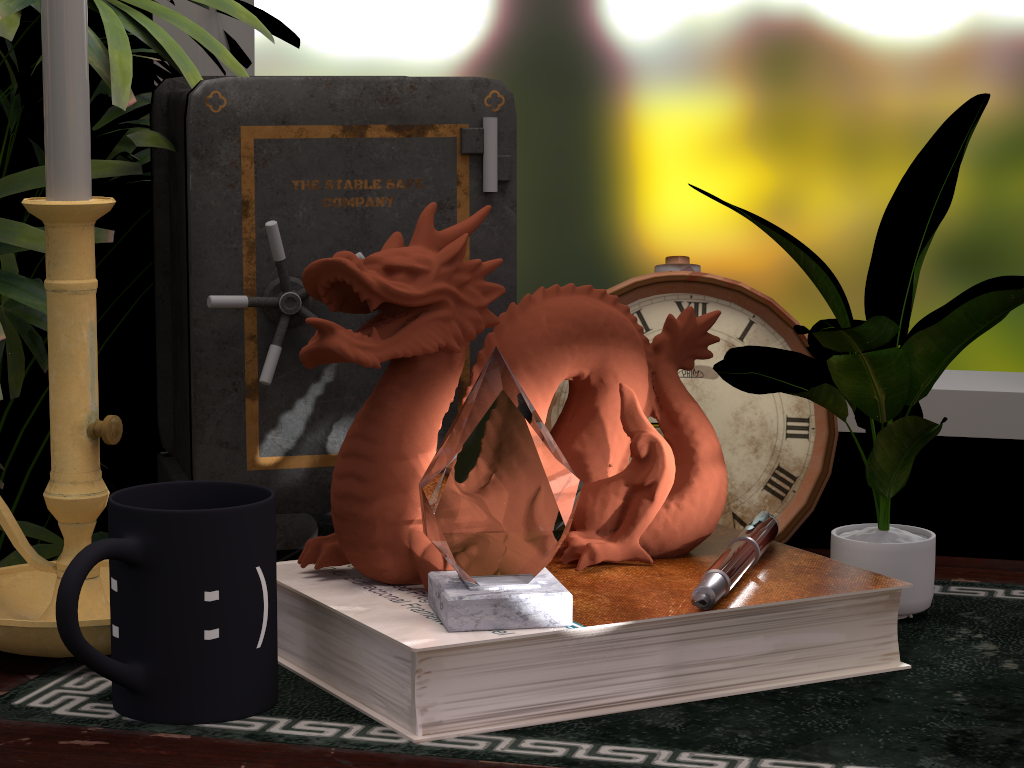} 
         & \includegraphics[height=0.45in]{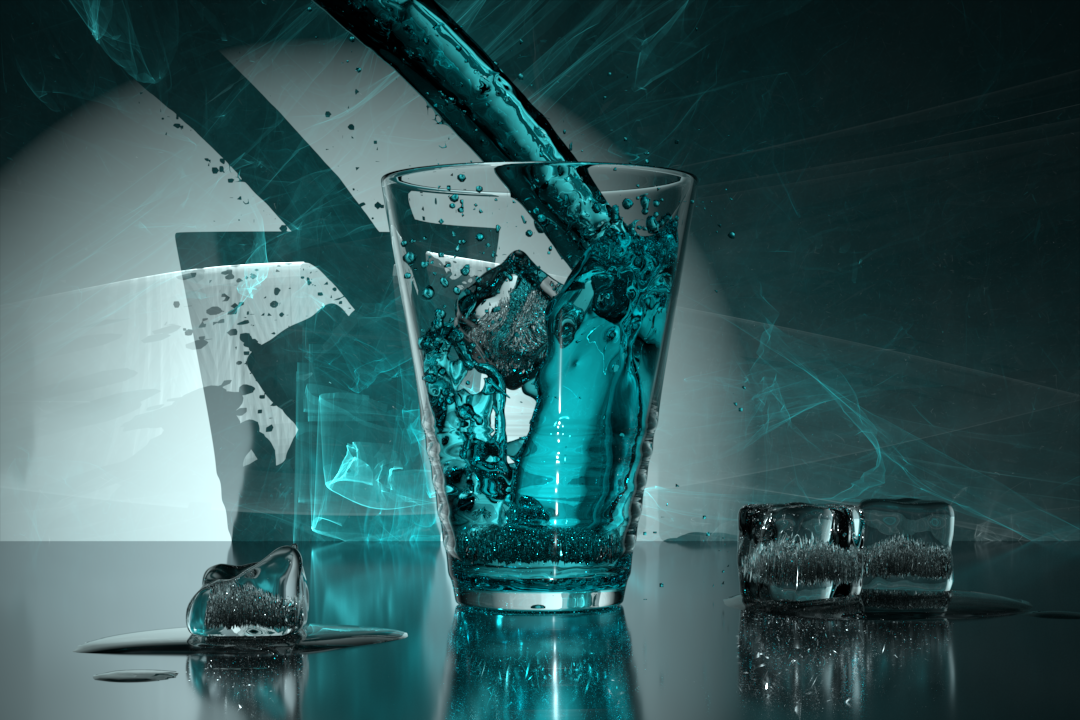}
         & \includegraphics[height=0.45in]{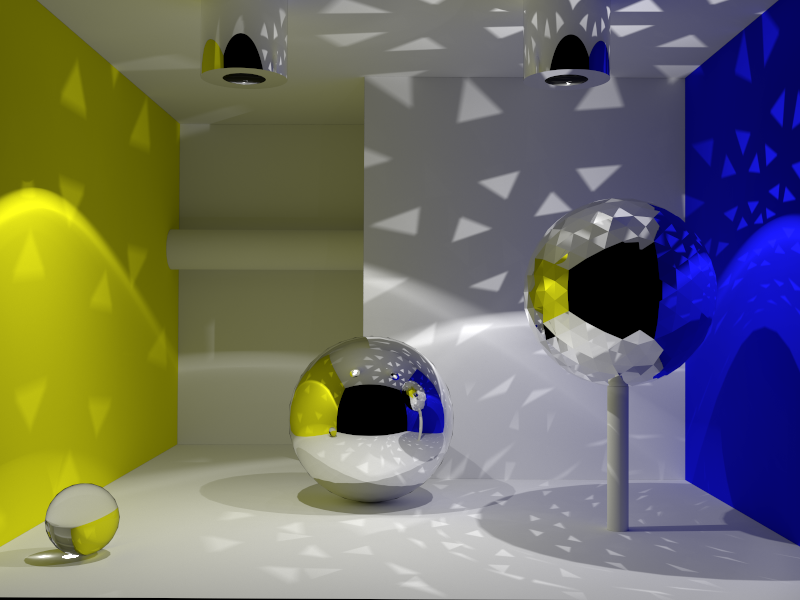}
         & \includegraphics[height=0.45in]{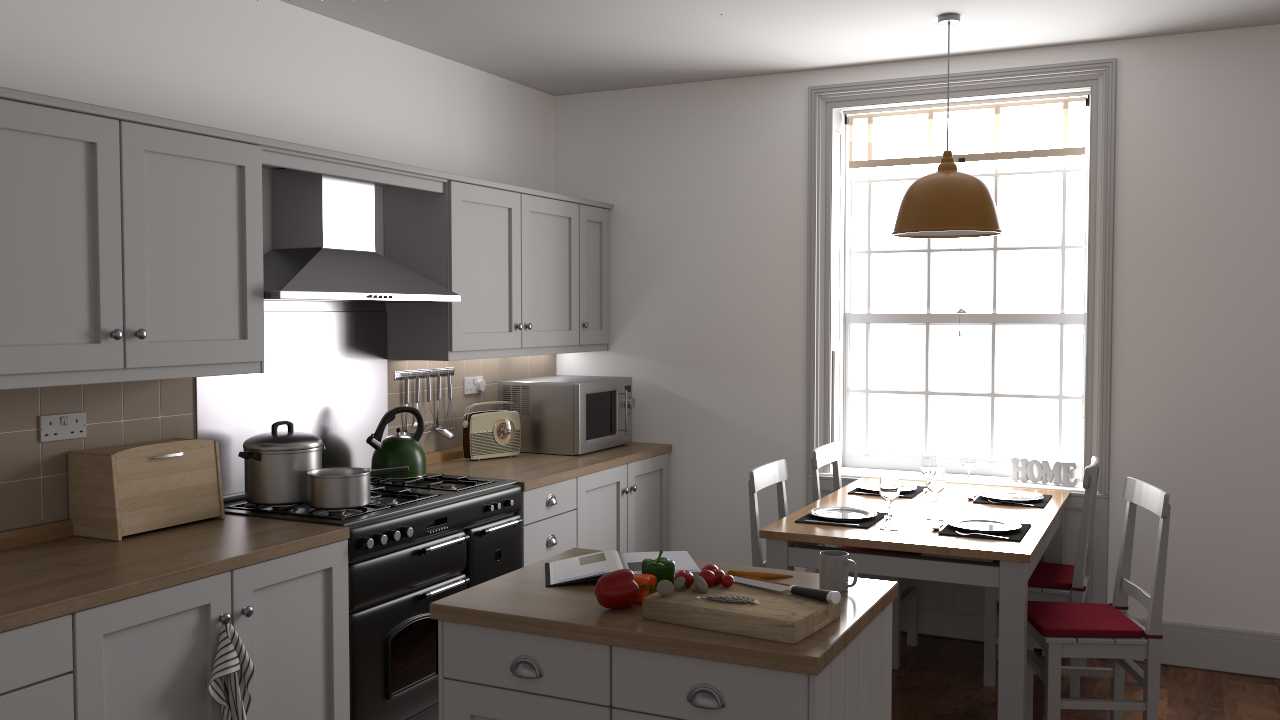}
         & \includegraphics[height=0.45in]{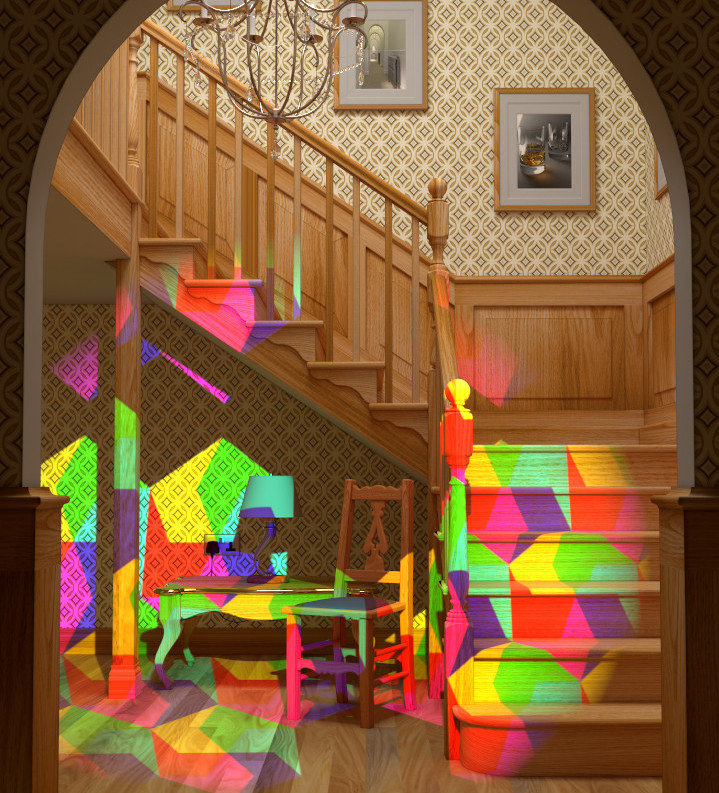}
\\ \hline 
        \multicolumn{2}{c|}{MSE $\le$}  & 50.0 & 20.0 & 10.0 & 10.0 & 10.0 & 10.0 & 10.0 & 10.0 \\ \hline
        \multirow{2}{*}{VCM} & Iter. &  $\mathrm{501}$ & $\mathrm{650}$ & $\mathrm{184}$
        & $\mathrm{226}$  & $ \mathrm{1361}$ & $\mathrm{149}$ & $\mathrm{808}$ & $\mathrm{573}$ \\
          & Time & $\mathrm{435.48s}$ & $\mathrm{2358.52s}$ & $\mathrm{807.03s}$ & $\mathrm{287.39s}$ & $\mathrm{2244.49s}$ & $\mathrm{115.85s}$ &  $\mathrm{3284.59s}$ & $\mathrm{2541.59s}$ \\ \hline
        \multirow{2}{*}{VCM+} & Iter. &  $\mathrm{222}$  
        & $\mathrm{397}$ & $\mathrm{153}$ & $\mathrm{183}$ & $\mathrm{807}$ & $\mathrm{98}$ & $\mathrm{599}$ & $\mathrm{101}$  \\
          & Time &  $\mathrm{214.16s}$ & $\mathrm{1496.16s}$ & $\mathrm{732.68s}$ & $\mathrm{241.83s}$ & $\mathrm{1445.52s}$ & $\mathrm{81.31s}$ & $\mathrm{2627.52s}$ & $\mathrm{586.70s}$ 
         \\
        \toprule
    \end{tabular}
\caption{Iterations and time budgets required to reach the target MSE for VCM+ and VCM on eight benchmarks. Our VCM+ shows superior performance in terms of iterations and time consumption.}
    
    \label{tab:mse}
\end{table*}

\begin{figure*}[thb]
    \vspace{-0.1in}
    {
  \centering
  \begin{tabular}{c c c c}
       \quad\quad (a) Pool & \quad (b) Hall & (c) Bathroom & (d) Dragon \quad \quad \quad \quad \quad \quad \quad \\ [-0.02in]
        \includegraphics[height=1.1in]{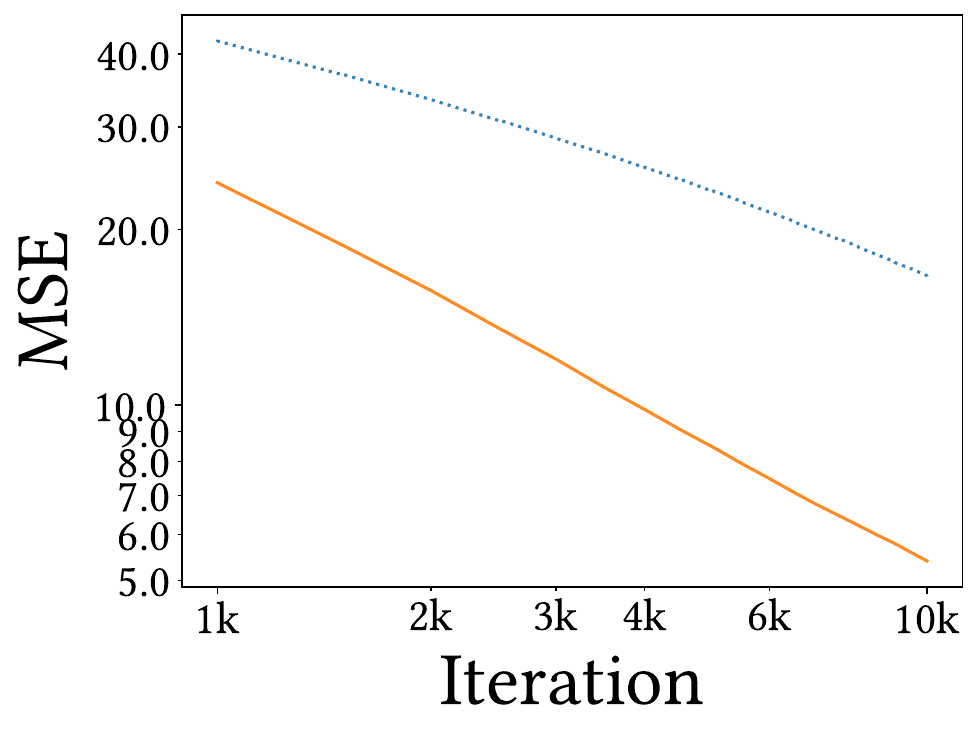} &
        \includegraphics[height=1.1in]{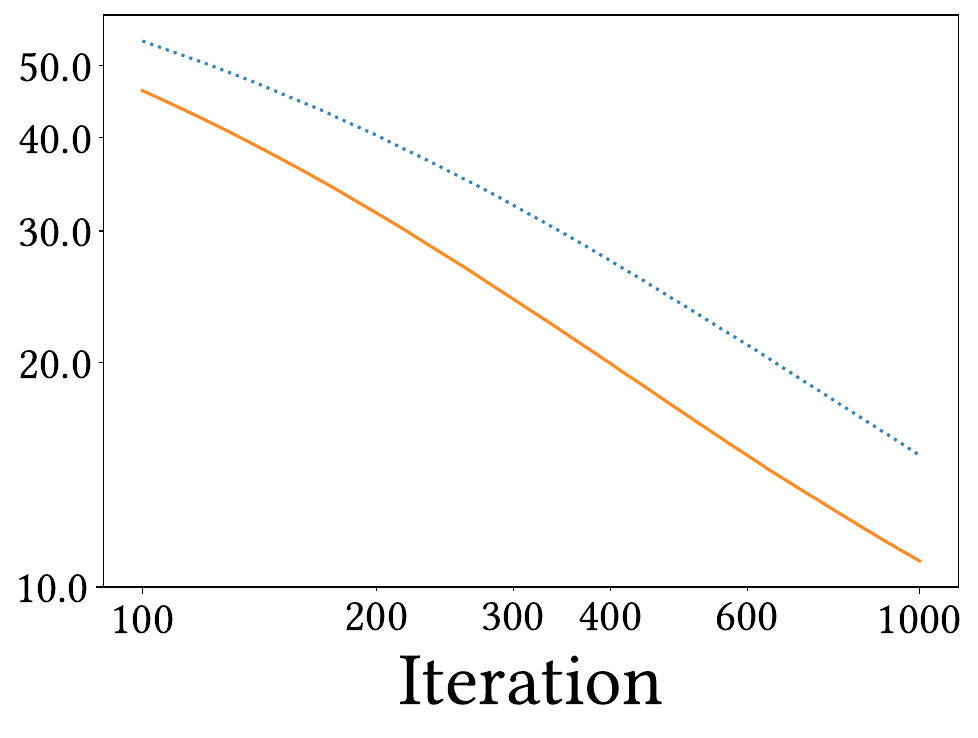} &
        \includegraphics[height=1.1in]{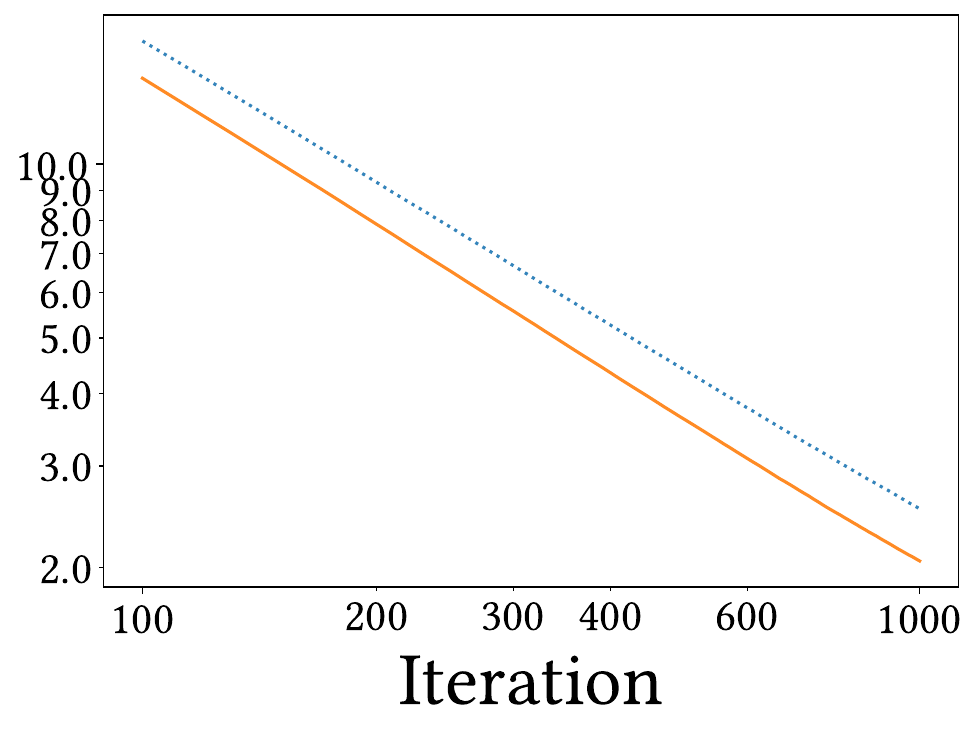} &
        \includegraphics[height=1.1in]{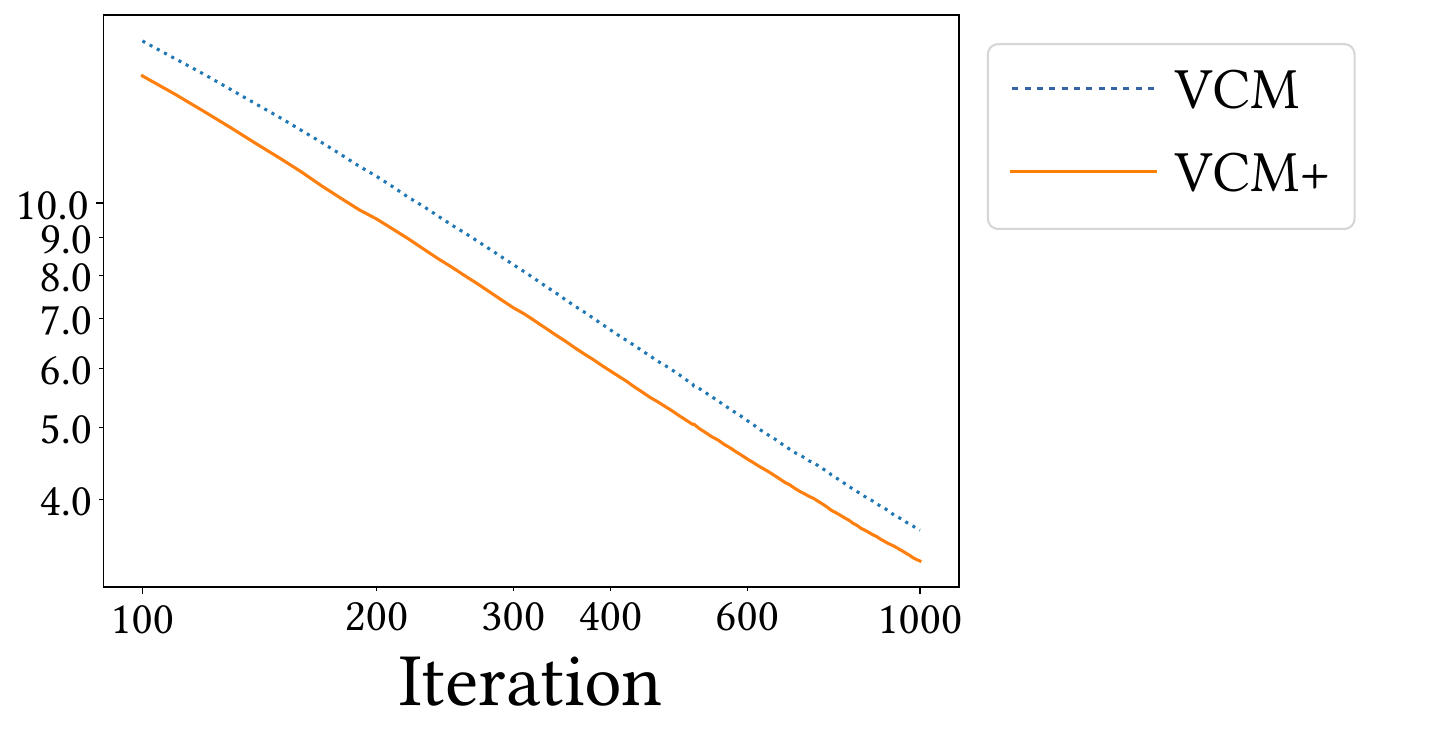} \quad \quad\\
        \quad\quad (e) Glass & \quad (f) Ball & (g) Kitchen & (h) Staircase \quad \quad \quad \quad \quad \quad \quad \\ [-0.02in]
        \includegraphics[height=1.1in]{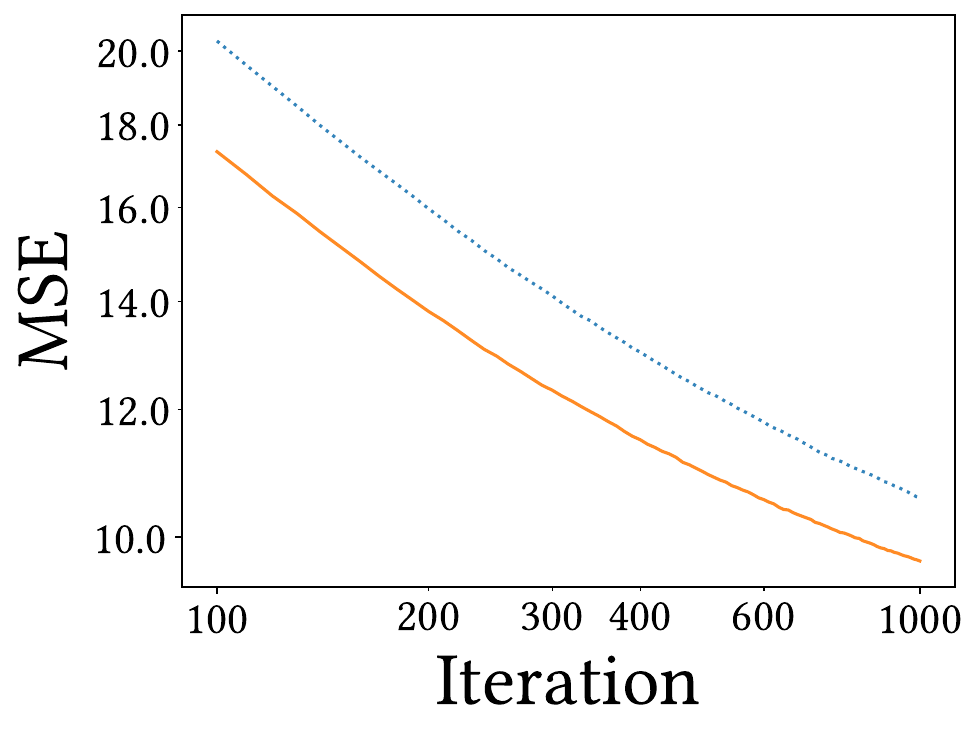} &
        \includegraphics[height=1.1in]{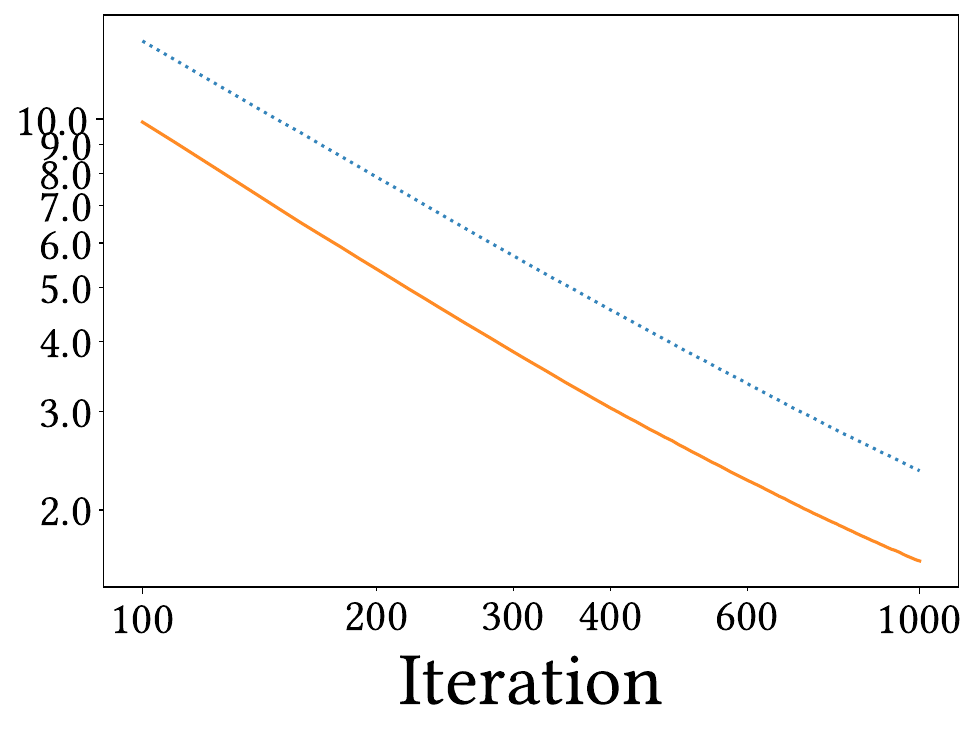} &
        \includegraphics[height=1.1in]{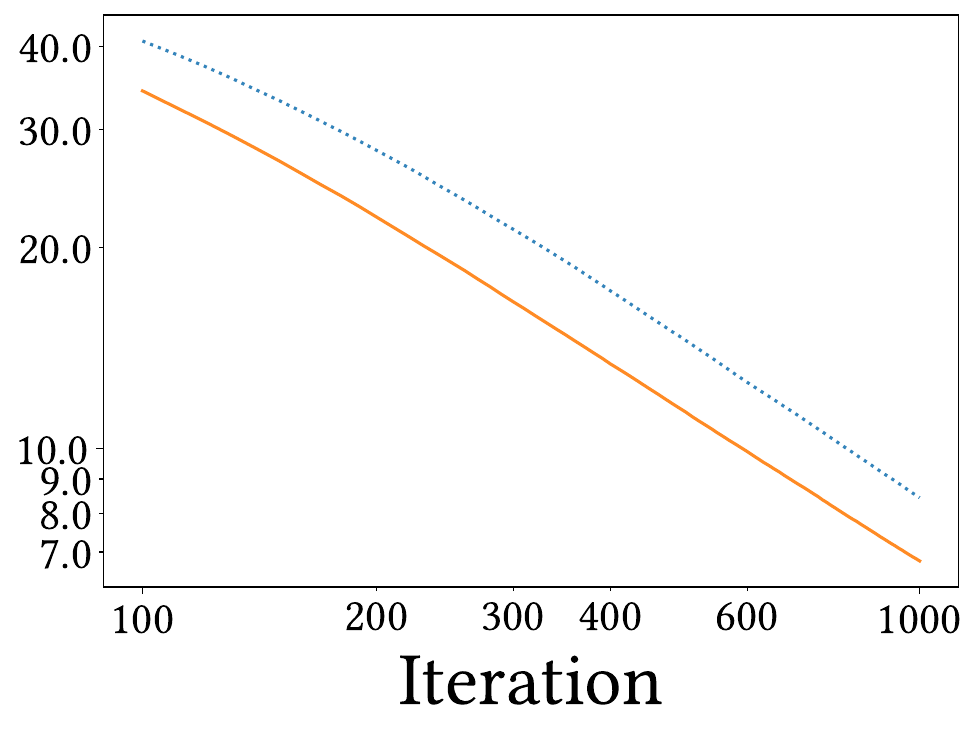} &\includegraphics[height=1.1in]{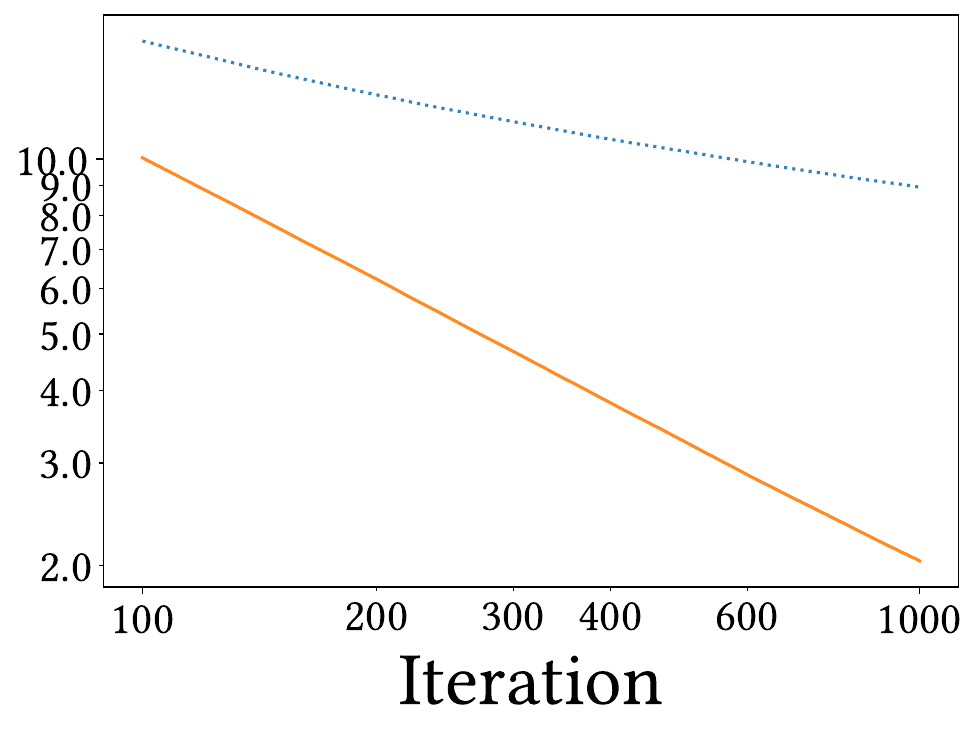} \quad \quad \quad \quad \quad \quad \quad \quad 
  \end{tabular}
  }
   \caption{MSE over iterations of eight test scenes by different methods. Our VCM+ algorithm using hypothesis testing has lower MSE and shows superior performance over VCM.}
  \vspace{-0.1in}
    \label{fig:mse}
\end{figure*}

In all tested scenes, our algorithm (FPPM) shows superior performance over CPPM in terms of MSE. Moreover, our algorithm can alleviate the noise, as highlighted in the zoom-in region of Kitchen and Hall; and it can alleviate the visual blur, as highlighted in the zoom-in region of Staircase. 
In CPPM, $\chi^2$-test works well relying on the independence assumption, a special case where unweighted photon statistics can model the local radiance. It is not always the case under complex indoor lighting conditions. $\chi^2$-test will malfunction with either glossy materials, indirect illumination with multiple area lights, or textured lighting. Where it incorrectly underestimated the kernel radius, noise may occur due to the lack of photons within a small kernel. Where the kernel size is overestimated, noticeable bias will arise. In contrast, our algorithm correctly determines the kernel radius to balance noise and bias by removing the independence assumption of CPPM, leading to better results.
We use greyscale to visualize the kernel radius of each zoom-in region in the middle of each block. White indicates a relatively large radius while black indicates a relatively small one. Our algorithm can clearly indicate the boundary at illumination discontinuities, implying that small radii should be used as expected to reduce bias. 

We also highlight the asymptotic performance of FPPM compared with CPPM, using SPPM as a baseline, as the convergence plots shown in \autoref{fig:ppm-convergence}. Both our algorithm and CPPM significantly outperform the baseline. Since our hypothesis-testing model is more general for samples with contributions attached under various lighting conditions, our algorithm shows superior performance and converges fast.

\subsubsection{VCM+ vs. VCM}

\textbf{Settings:}
As a reinforcement of VCM, VCM+ shares the same setting with VCM on common parameters unless otherwise stated. 
Specifically, initial radius $r_1$ is set to $0.1\% R_{\mathrm{bb}} \sim 0.3\% R_{\mathrm{bb}}$ for all algorithms, where $R_{\mathrm{bb}}$ is the radius of the bounding box of visible objects.
Lower bound ratio of the radius $r_\mathrm{min}$ is set to $0.1\%r_1$; and
$\alpha$ is set to $0.75$ for all algorithms, as recommended in VCM~\cite{Georgiev:2012:VCM}. Balance heuristic for MIS is used.
Eight benchmarks (including the three scenes previously used to test PPM algorithms) are used for testing. We collect the number of iterations and time consumption required to reach a specified MSE, and show the convergence plot over iterations. 
As seen from \autoref{tab:mse} and ~\autoref{fig:mse}, our VCM+ achieves the best performance with the lowest MSE over VCM in all testing scenarios.

The Pool is a typical scene that includes many complex caustics, and it is very hard to converge to a correct result. Our VCM+ shows more than 2$\times$ speedup over VCM with far fewer iterations, as shown in \autoref{tab:mse}; the speedup is going larger asymptotically, as shown in \autoref{fig:mse}. We also highlight the significant improvement of our algorithm in \autoref{teaserfigure}. VCM produces a much noisier result (MSE=16.67) than ours (MSE=5.54) after 10K iterations.
When compared with VCM, VCM+ has fewer iterations and less time consumption (\autoref{tab:mse}). Noticed that VCM+ spends more time on each iteration than VCM because it needs around 10\% extra computational overhead for hypothesis tests. However, this cost is worthwhile for prominent performance gain.

\begin{figure*}[tbh]
  \centering
  \begin{tabular}{c@{\hskip1pt} @{\hskip1pt}c@{\hskip1pt} @{\hskip1pt}c}
       \includegraphics[width=1.39in]{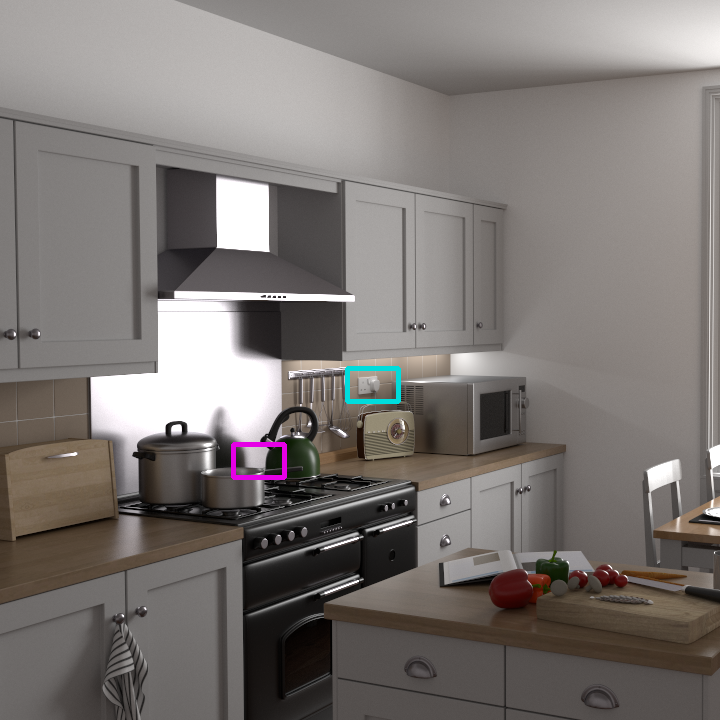}
        \includegraphics[width=1.5in]{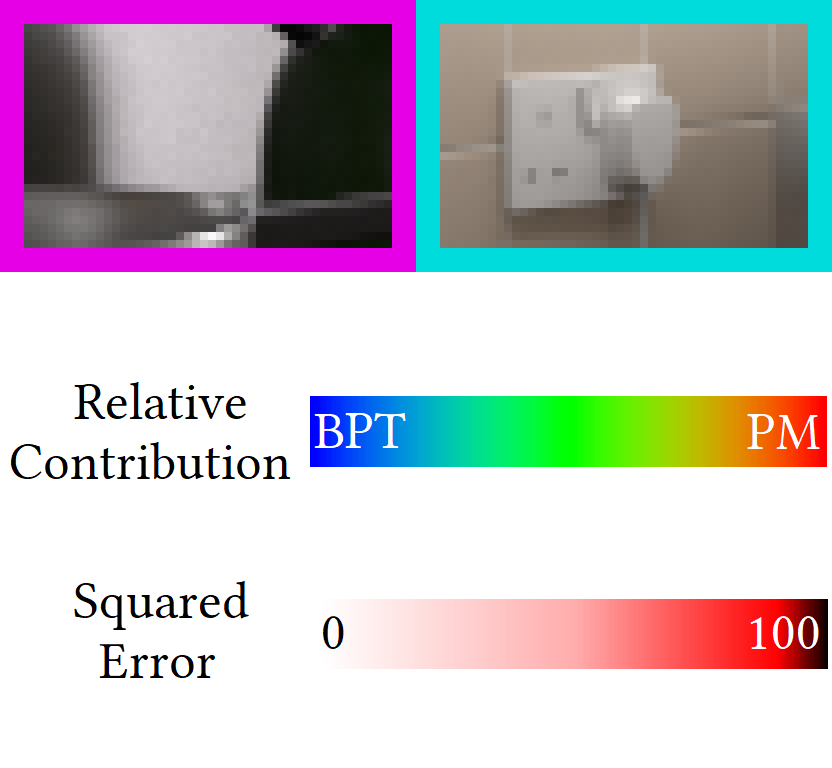} &
       \includegraphics[width=1.5in]{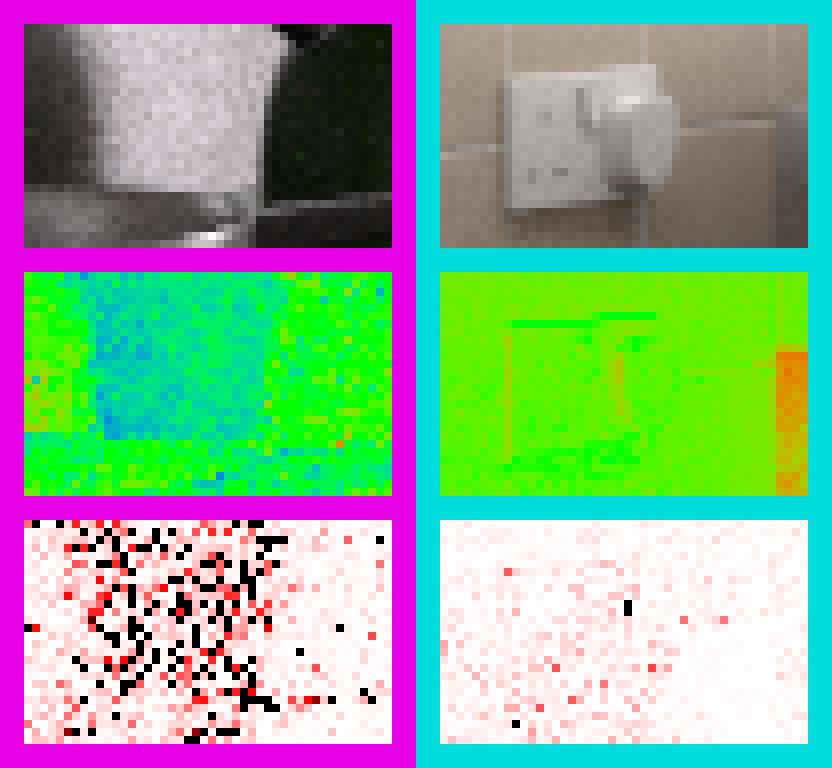} &
       \includegraphics[width=1.5in]{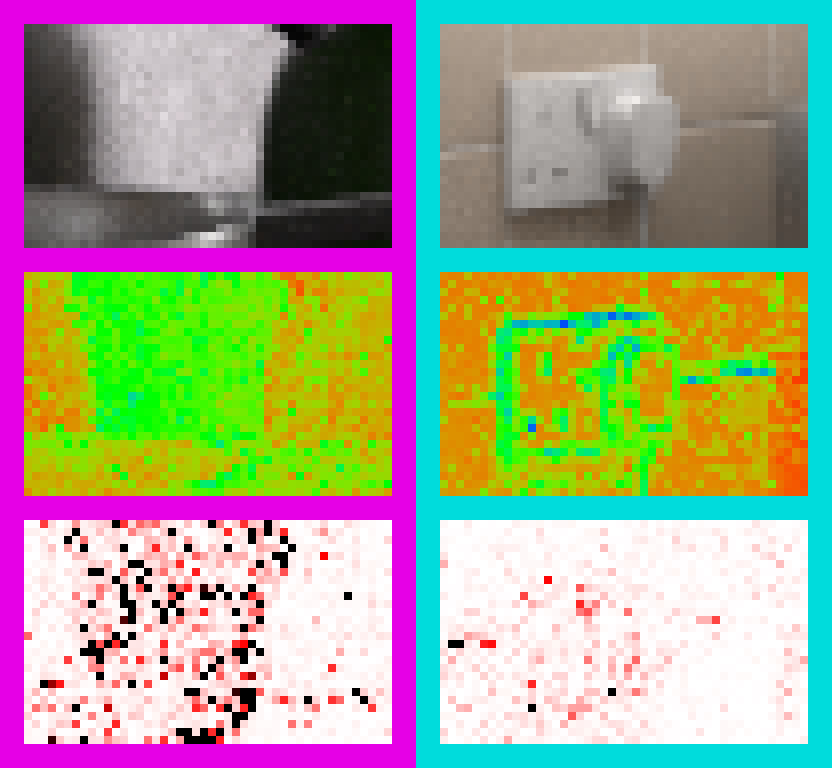} \\
       Reference & VCM & Ours (VCM+) \\ [-2pt]
       & (405 iter.; MSE=17.02) & (389 iter.; MSE=\textbf{13.66}) 
  \end{tabular}
  \caption{Equal-time (35 minutes) comparison on Kitchen scene. We visualize the relative contribution and the squared error. PM plays a more important role in VCM+ and compensates for the high variance brought by BDPT. VCM+ can compute appropriate contributions from BDPT and PM individually and obtains better results.}
  \label{fig:kitchen}
\end{figure*}

\begin{figure*}[tbh]
  \centering
  \begin{tabular}{c@{\hskip1pt} @{\hskip1pt}c@{\hskip1pt} @{\hskip1pt}c}
       \includegraphics[width=1.39in]{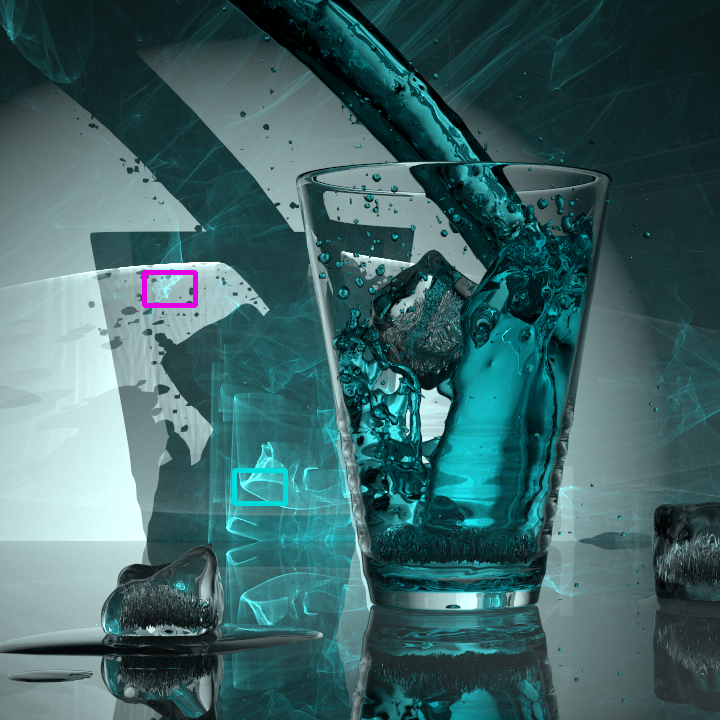}
        \includegraphics[width=1.5in]{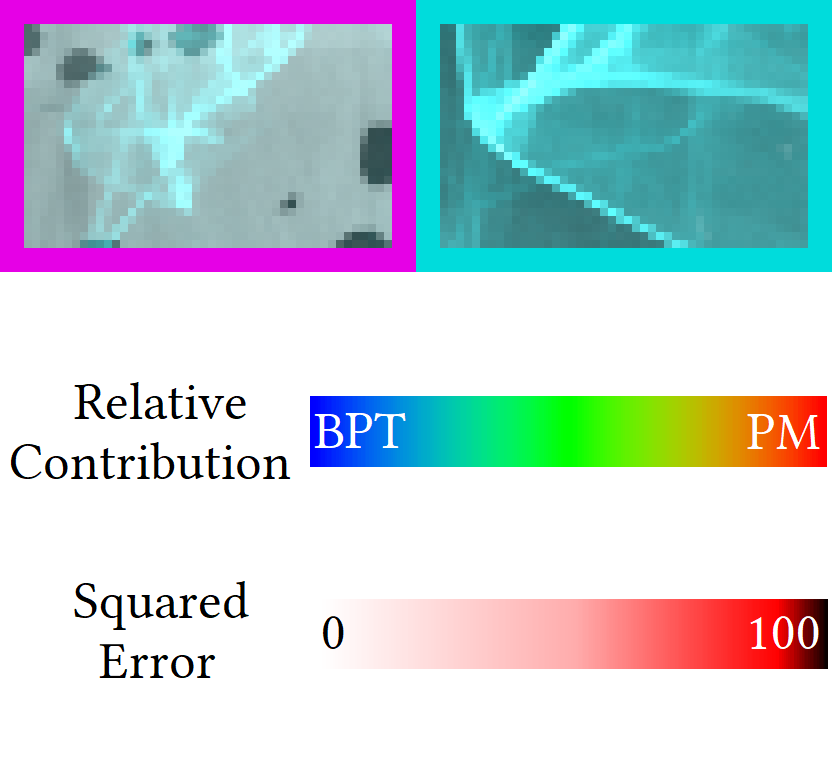} &
       \includegraphics[width=1.5in]{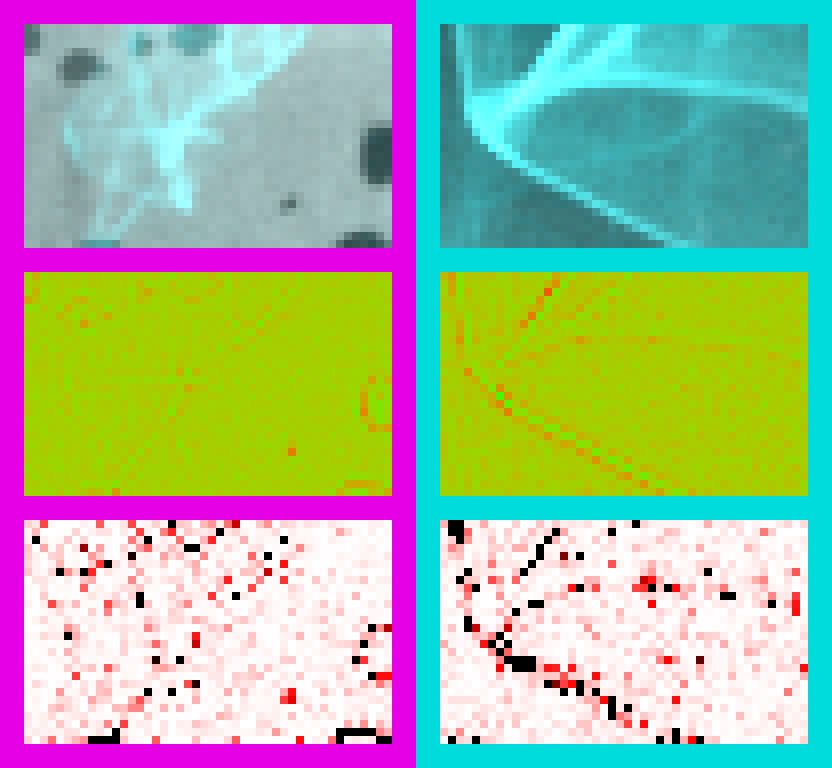} &
       \includegraphics[width=1.5in]{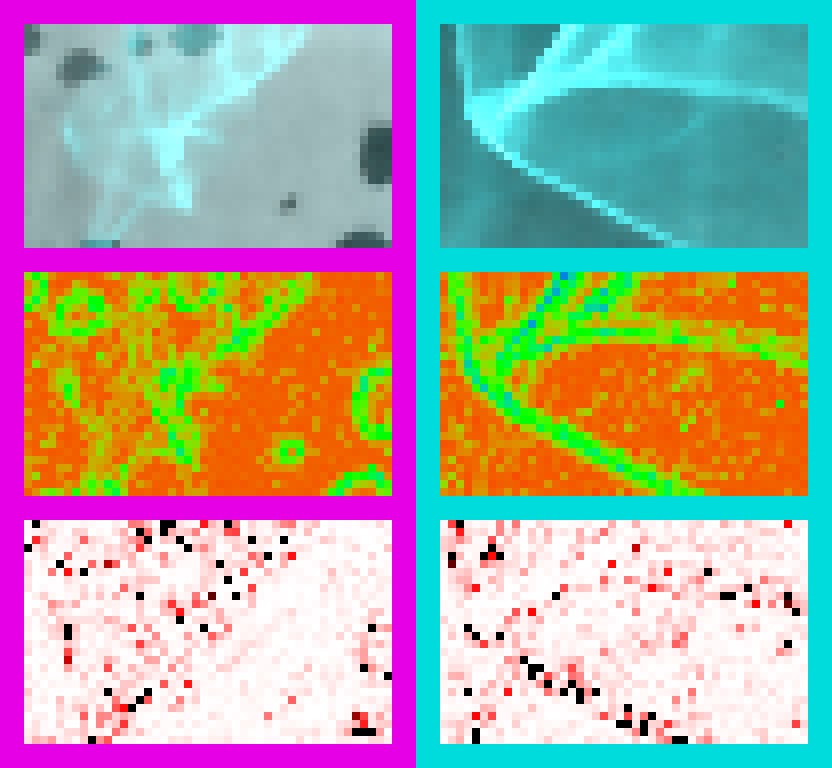} \\
       Reference & VCM & Ours (VCM+)\\ [-2pt]
       & (361 iter.; MSE=13.39) & (333 iter.; MSE=\textbf{12.05})
  \end{tabular}
  \caption{Equal-time (10 minutes) comparison on Glass scene. We visualize the relative contribution and the squared error. We highlight the contribution from PM that plays an important role in the region of caustics in our VCM+. Whereas, VCM is less efficient in handling these regions and leads to blurry or noisy results. }
    \label{fig:glass}
\end{figure*}


To figure out the rationale behind our algorithm's better performance, we conduct more comparisons between our VCM+ and the baseline VCM. 
We demonstrate the results ~\autoref{fig:kitchen} in and~\autoref{fig:glass}, respectively. 
The Glass scene (\autoref{fig:glass}) has many caustics, and the Kitchen scene (\autoref{fig:kitchen}) has few S-D-S paths but many glossy materials. Specifically, we illustrate the relative contribution between BDPT and PM in false color.
In our VCM+, PPM has a large relative contribution than BDPT in the regions that include S-D-S paths as expected, whereas PPM contributes less as it supposed to in other regions. VCM+ can compute appropriate contributions from photon mapping for most pixels.
Consequently, VCM+ achieves steady gains at the expense of computing photon mapping (i.e., vertex merging, VM).
And for the pixels initially estimated to be biased, PM in VCM+ has less contribution than that of VCM, thereby alleviating the PM-induced bias.
As a result, our algorithm has lower variance and bias overall and produces less noisy images with sharper edges.
Our results also indicate that the initial bias is typically lower in the regions where illumination changes smoothly, and higher at the boundaries of objects and caustics.

\begin{figure*}
  \centering
  \begin{tabular}{c@{\hskip1pt} @{\hskip1pt}c@{\hskip1pt} @{\hskip1pt}c@{\hskip1pt} @{\hskip1pt}c}
       \includegraphics[width=1.42in]{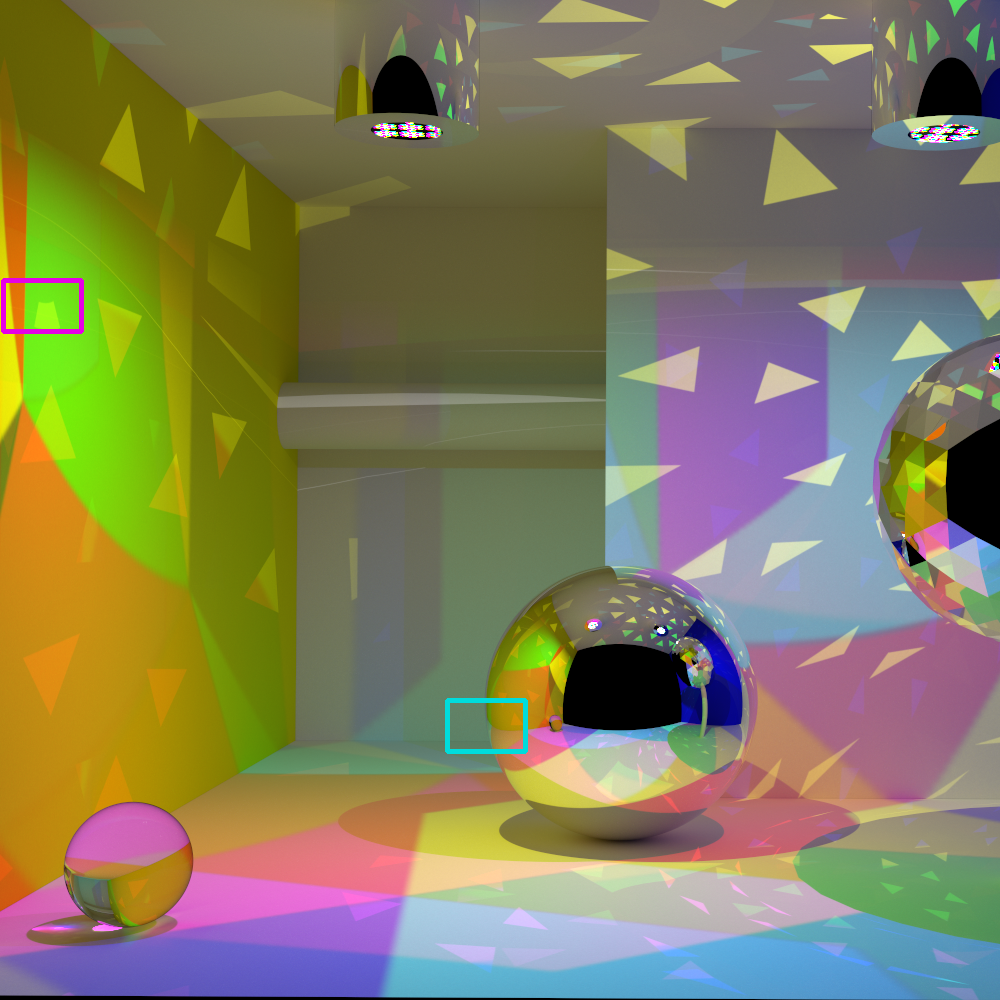}
       \includegraphics[width=1.48in]{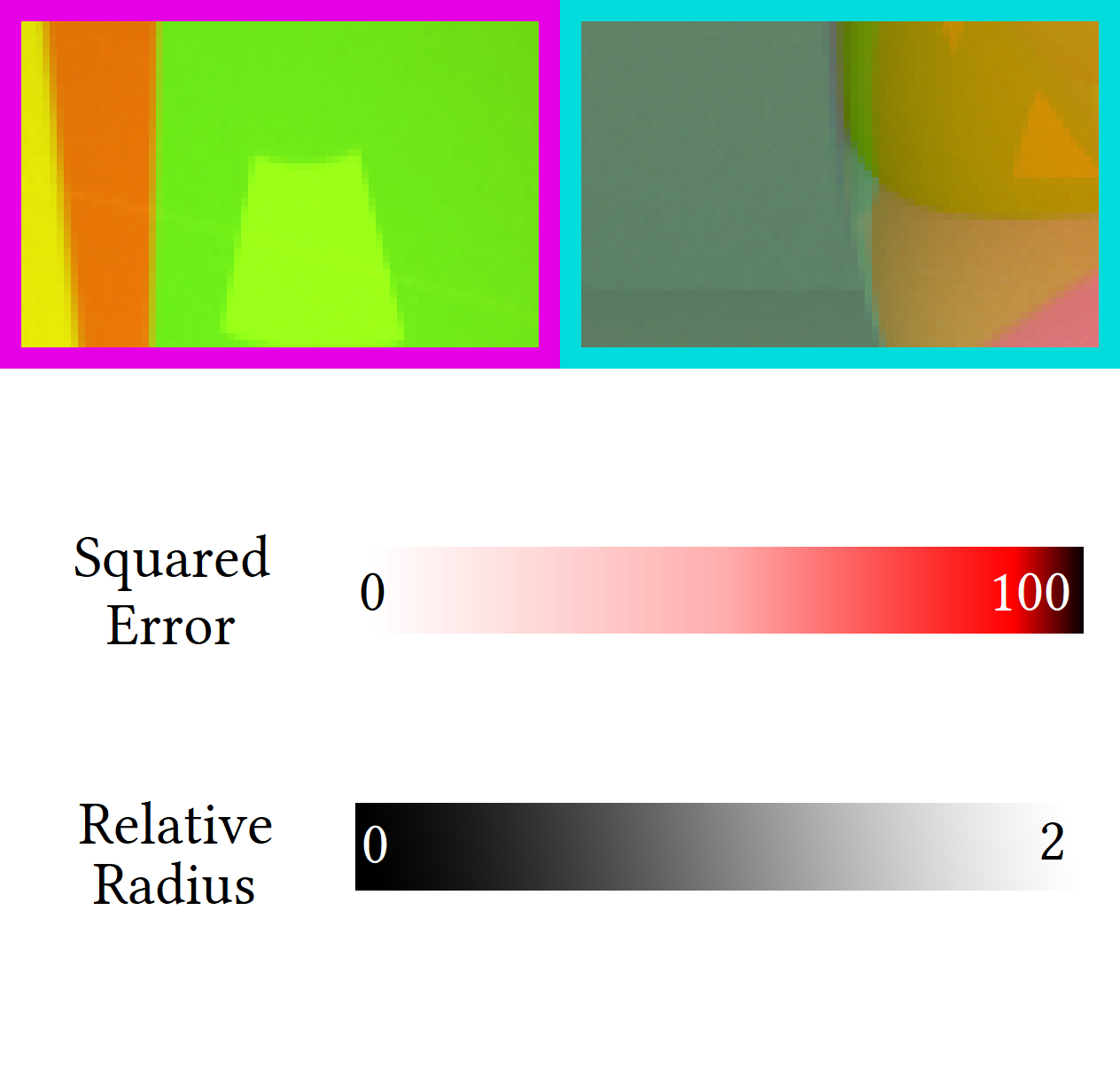} &
       \includegraphics[width=1.5in]{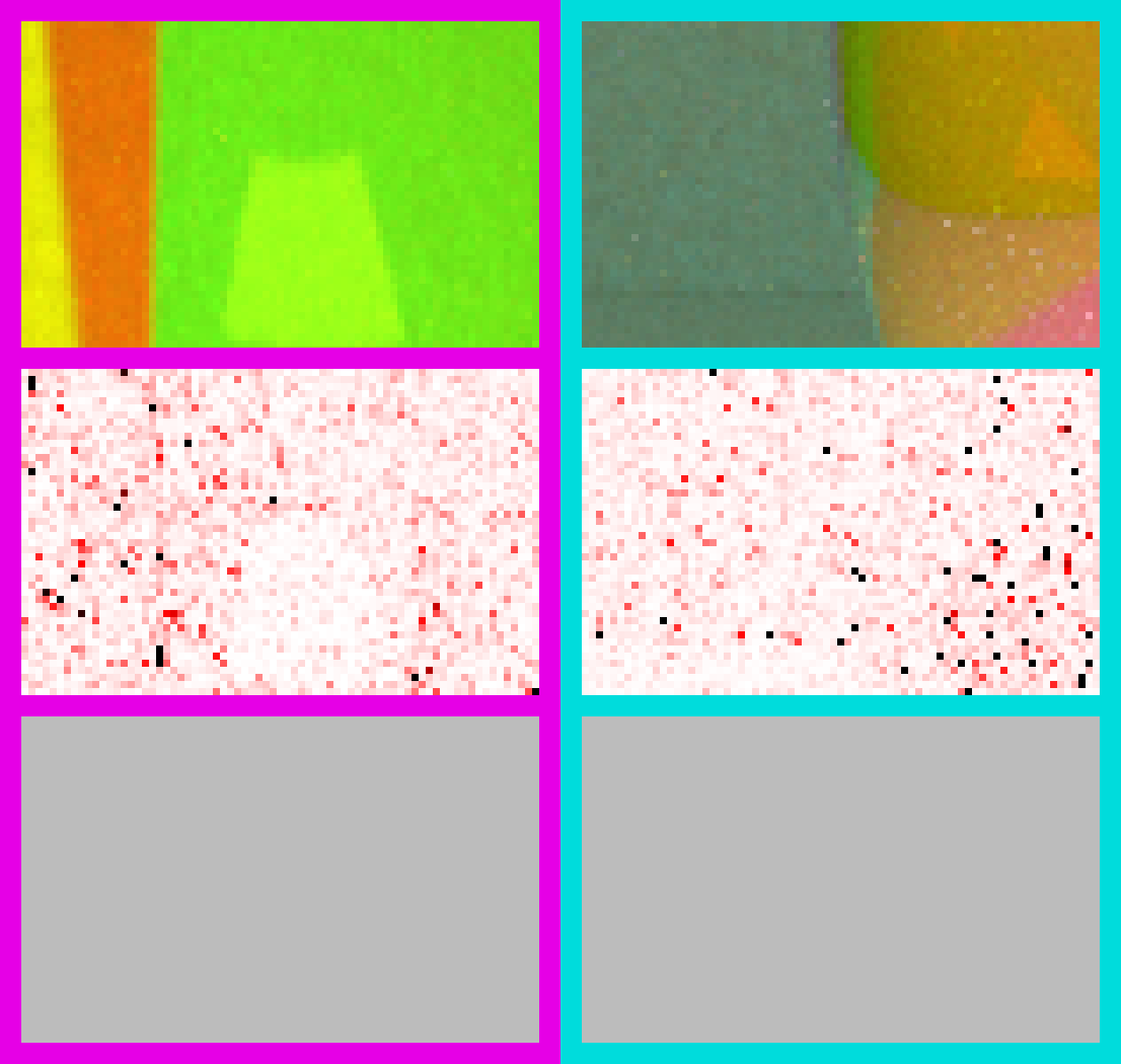} &
       \includegraphics[width=1.5in]{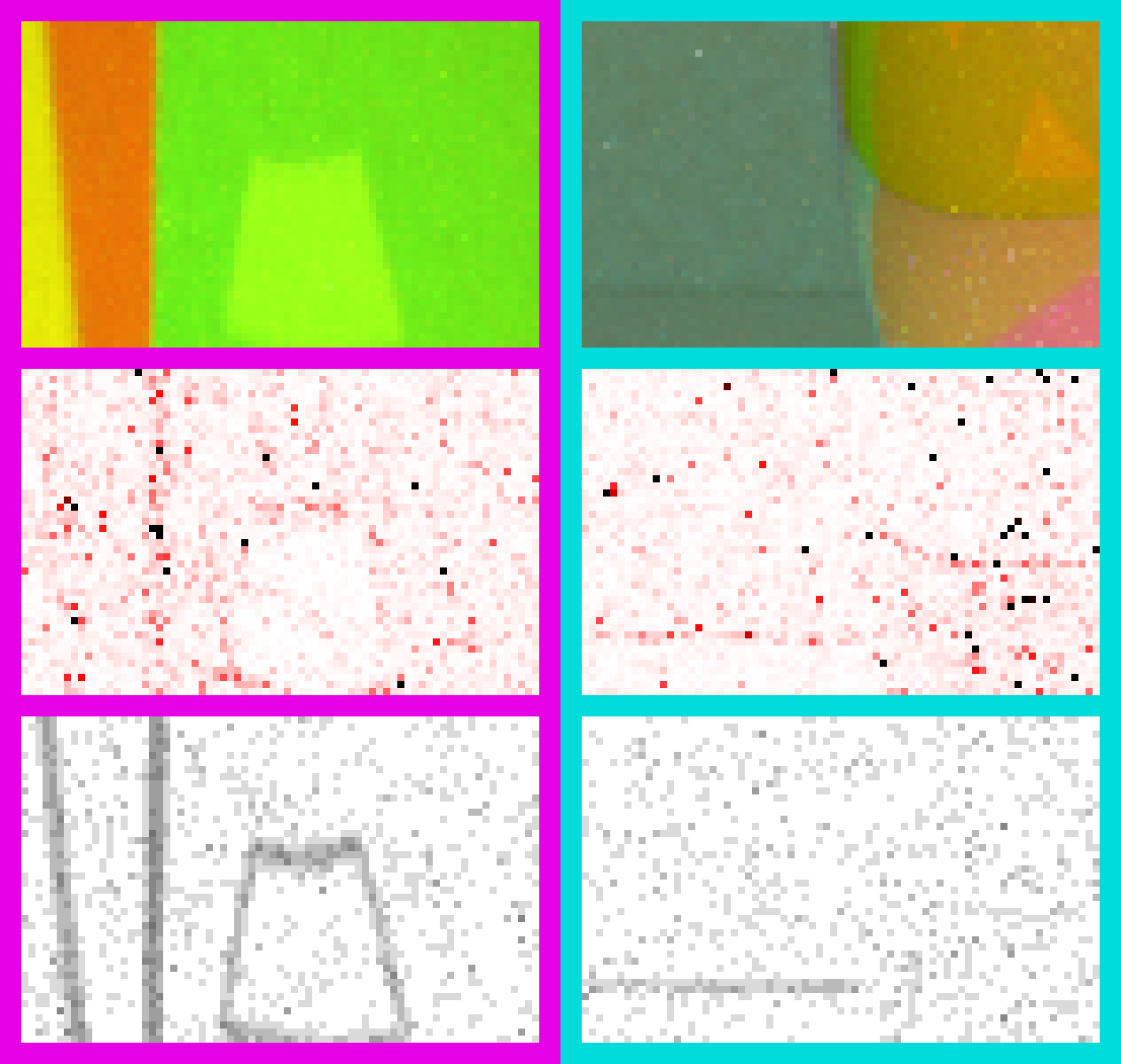}
       \\
       Reference & VCM & VCM+ 
       \\
        & (243 iter.; MSE=13.04) & (232 iter.; MSE=\textbf{9.50}) 
  \end{tabular}
    \vspace{-0.1in}
  \caption{Equal-time (10 minutes) comparison on the modified Ball scene with lights behind textured glasses. We visualize the squared error and relative radius (actual radius / VCM's radius). We highlight that our VCM+ can also handle special lighting setting well.
  }
  \vspace{-0.1in}
    \label{fig:box}
\end{figure*}

\begin{figure}
    \centering
    \begin{tabular}{ c c}
         \includegraphics[height=1.3in]{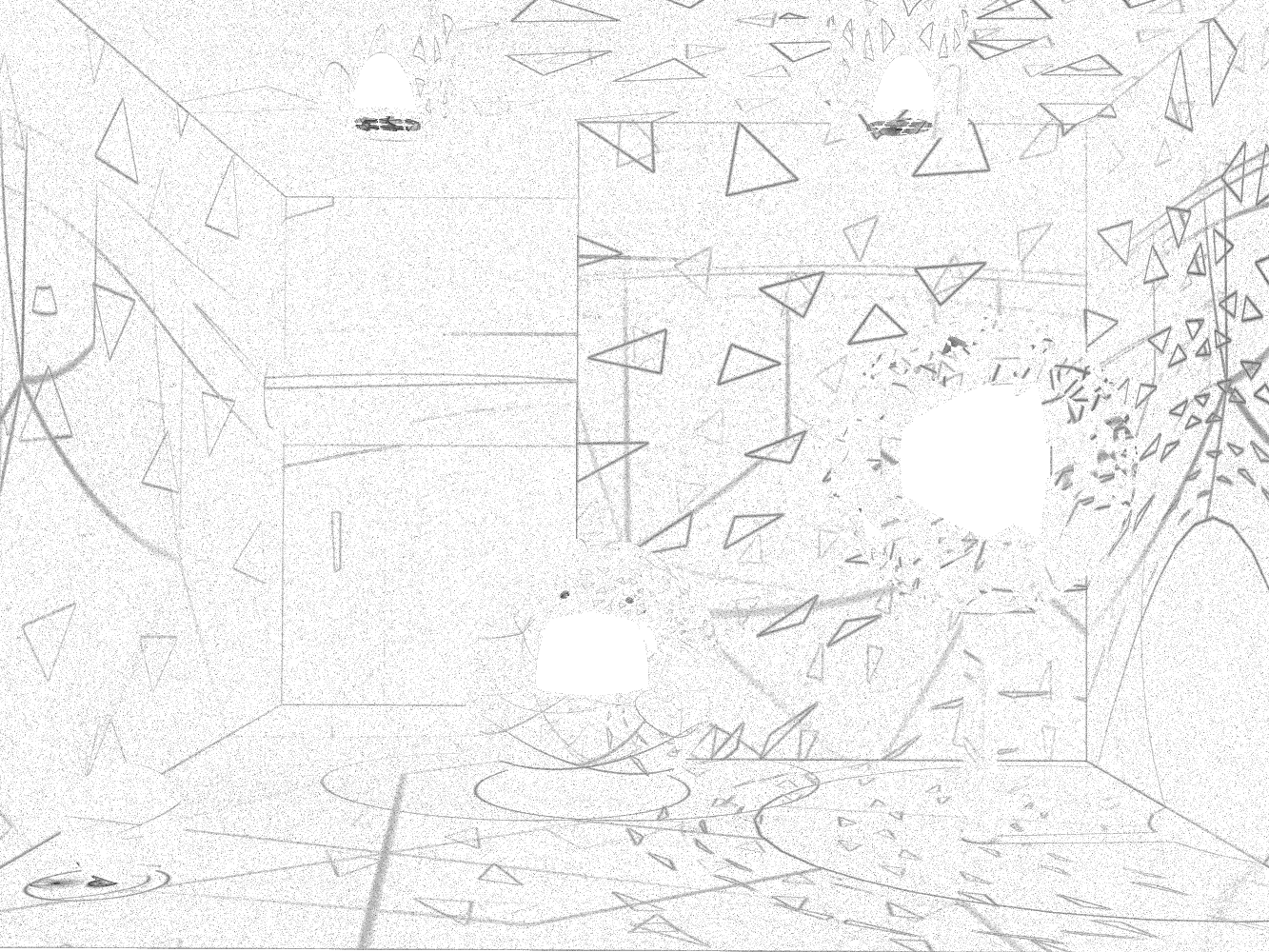}
         & \includegraphics[height=1.3in]{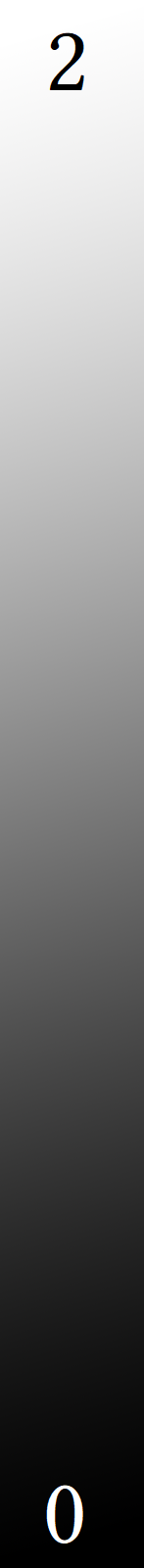}
    \end{tabular}
    \vspace{-0.1in}
    \caption{Relative radius (VCM+/VCM) visualization. We highlight that VCM+ can find the appropriate radius that leverages the contributions from VC (BDPT) and VM (PM).}
    \label{fig:radius-changing}
\end{figure}

In addition, we analyze the radius of each iteration in PPM, which determines the contributions between PPM and BDPT.
We show the results of a Ball scene with lights behind textured glasses (\autoref{fig:box}), and all illuminations come from this type of light source. The results show that our VCM+ can find appropriate radii for unbiased kernel estimation and synthesize superior images over VCM.
We visualize the relative radius (actual radius / VCM's radius) in this scene after 400 iterations in ~\autoref{fig:radius-changing}, which can explain the rationale for better performance by our statistical model with hypothesis testing using F\nbd-test. Our method can find an appropriate radius suitable for the minimization of variance and reduction of bias at different locations, especially a relatively small radius at the boundary or discontinuity of illumination while a relatively large radius at the region with a smooth lighting transition.
Overall, our algorithm leverages the contributions from VC (BDPT) and VM (PM) by finding the appropriate radius and hereby achieves better performance.


\subsection{Parameters}

\label{subsec:parameters}
We investigate the robustness of our VCM+ by tuning the values of different parameters used in the experiments that may affect the performance.
The values of parameters vary, but we still observe improvements over VCM.



\subsubsection{Initial Radius}
\label{subsubsec:initial-bandwidth}

\begin{figure}
    \centering
    \begin{tabular}{c c}
       \includegraphics[height=0.93in]{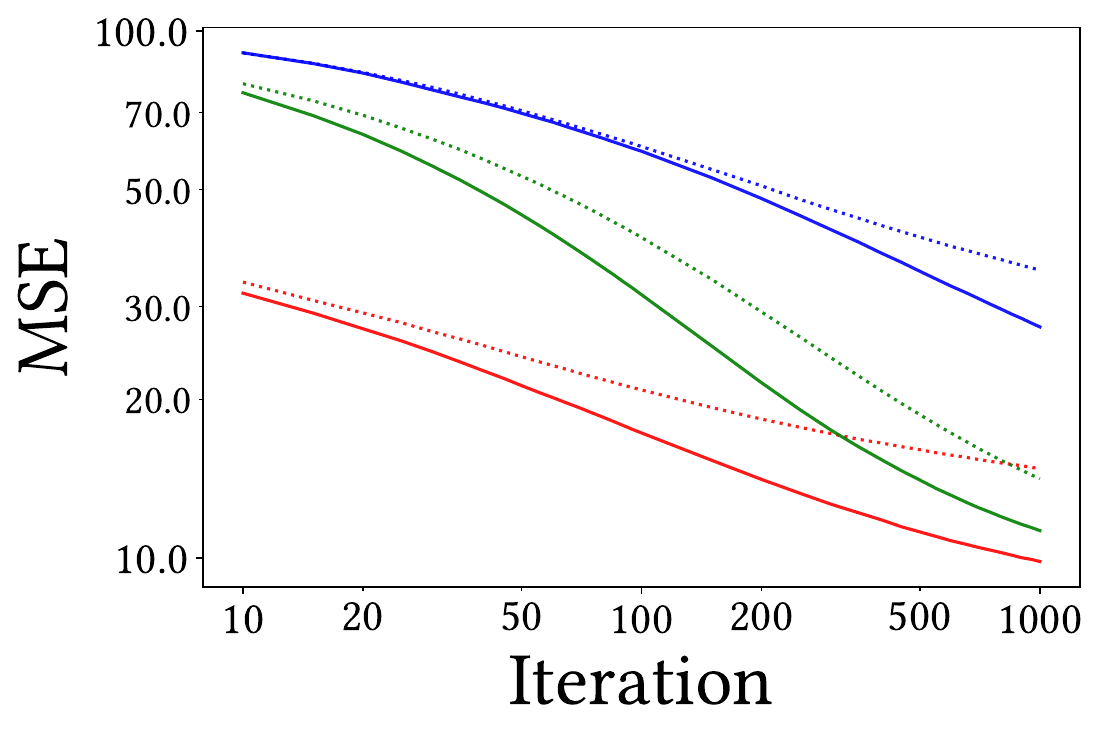} &
        \includegraphics[height=0.91in]{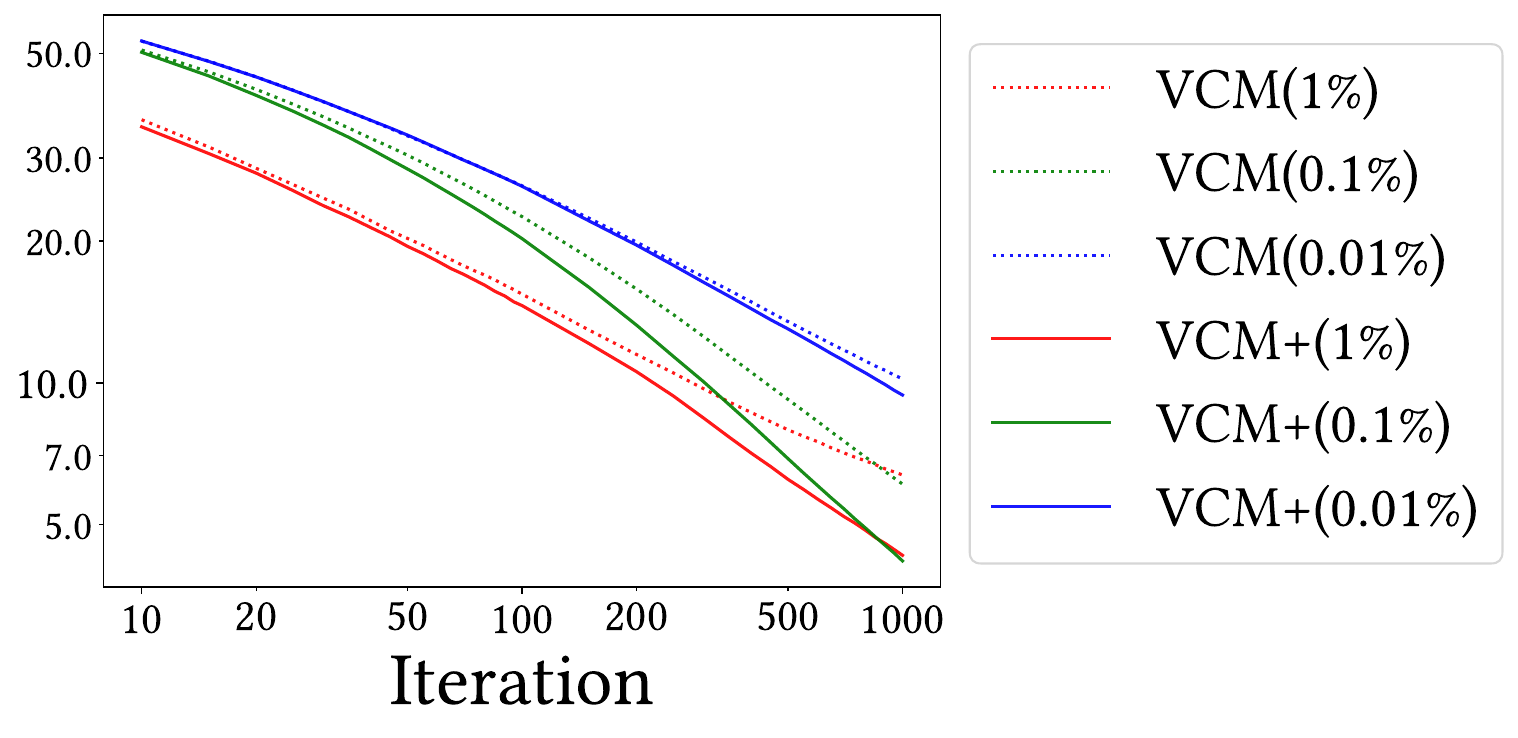} \\
        \quad  (a) Glass & (b) Dragon \quad \quad \quad \quad \quad
    \end{tabular}
    \caption{Asymptotic performance of VCM+ vs. VCM over iterations using different initial radius. VCM+ always exhibits better performance than VCM.}
    \vspace{-0.1in}
    \label{fig:initial-bandwidth}
\end{figure}

The initial radius may affect the performance of a progressive algorithm. Therefore, we test two algorithms (VCM and VCM+) with different initial radii.
Three values including $1\%, 0.1\%$, and $0.01\% R_{\mathrm{bb}}$ are used for the test of two scenes, Glass as the caustics representative and Dragon as the diffuse representative. These values range from coarse to fine-grained, and $0.01\%$ may be even pixel-level smaller. 
The asymptotic performance of these two algorithms using different initial radii are shown in \autoref{fig:initial-bandwidth}.

From the results, the initial radii have impact on both VCM and VCM+ performance. However, VCM+ outperforms VCM in all settings and exhibits good convergence behavior. A large initial radius (i.e. $1\%$) favors lower MSE in the early stage when variance dominates the error. As bias gradually accounts for a higher proportion of MSE, a relatively larger radius will slow down convergence. This can be manifested in that VCM with $0.1\%$ radius outperforms VCM with $1\%$ radius at around 1000 iterations in both scenes, with a steeper downward trend. our VCM+ benefits from the adaptive kernel radius selection scheme and hence receives a smaller impact from an overly large initial radius when compared to VCM. 
However, an aggressively small initial radius (i.e. $0.01\%$) makes the contribution from PM always trivial to VCM via MIS weights, resulting in a shift of VCM towards BDPT in the entire process. Obviously, the smallest initial radius leads to the worst performance for both VCM and VCM+, and their performances are very similar. 
Therefore, we suggest using a larger initial radius for VCM+ compared to the values suggested by \cite{Georgiev:2012:VCM}, because our test-driven unbiased radiance estimate can handle the bias well in the late rendering stages.


\subsubsection{Radius Shrinkage $k$}
\label{subsubsec:radius-parameter-k}
As discussed in \autoref{sec:algorithm}, we should select a moderate $k$. We compare different values of $k$ on a variety of scenes, as listed in~\autoref{tab:k}. For the Bathroom and Ball scenes, 100 iterations are tested.
For the Hall and Pool scenes that are difficult to converge, 1,000 iterations are tested.
Our approach consistently outperforms the baseline obtained by VCM, regardless of the value of $k$ used. From this table, $k=0.9$ obtains the best performance in most scenes, albeit different values do not introduce significant differences between each other. This may imply that a conservative shrinkage strategy may be appropriate for VCM+.


\begin{table}
    \caption{MSE obtained under different radius shrinkage values of $k$. Different scenarios are tested by our approach and VCM (baseline). Our approach using different $k$ values always outperforms the baseline; meanwhile, different $k$ values do not show significant discrepancies in the results.  }
    \centering
    \begin{tabular}{c|c@{\hskip2pt} @{\hskip2pt}c@{\hskip2pt} @{\hskip2pt}c@{\hskip2pt} @{\hskip2pt}c@{\hskip2pt} @{\hskip2pt}c@{\hskip2pt} @{\hskip2pt}c@{\hskip2pt}|c}
        \bottomrule
         \multirow{2}{*}{Bathroom (100 iter.)} & k & 0.5 & 0.6 & 0.7 & 0.8 & 0.9 & VCM  \\ \cline{2-8}
         & MSE & 14.09 & 14.08 & \textbf{14.06} & 14.20 & 14.42 & 16.36\\\hline
        \multirow{2}{*}{Ball (100 iter.)} & k & 0.5 & 0.6 & 0.7 & 0.8 & 0.9 & VCM  \\ \cline{2-8}
         & MSE & 10.25 & 10.03 & 9.81 & 9.66 & \textbf{9.56} & 13.79 \\ \hline
         
        \multirow{2}{*}{Hall (1,000 iter.)} & k & 0.5 & 0.6 & 0.7 & 0.8 & 0.9 & VCM  \\ \cline{2-8}
         & MSE & 11.02 & 10.92 & 10.83 & 10.70 & \textbf{10.60} & 15.01\\\hline

            \multirow{2}{*}{Pool (1,000 iter.)} & k & 0.5 & 0.6 & 0.7 & 0.8 & 0.9 & VCM  \\ \cline{2-8}
         & MSE & 24.40 & 24.30 & 24.05 & 23.87 & \textbf{23.71} & 41.99 \\
        \toprule
    \end{tabular}
    \vspace{-0.1in}
    \label{tab:k}
\end{table}


\subsubsection{F-test Significance Level $\alpha_F$}
If a smaller $\alpha_F$ is used, the probability of wrongly rejecting an unbiased estimator will be lower.
If a relatively larger $\alpha_F$ is used, the number of samples required to reject a biased estimator will be fewer.
We test three typical values $0.01$, $0.05$, and $0.10$ in statistics, and find that $\alpha_F=0.01$ obtains better performance than the other two, as shown in~\autoref{tab:alpha}.
This implies that a lower rejection rate is a better setting. Our algorithm is conservative about kernel radius reduction in this case. However, most pixels can still converge to unbiased radii quickly.
Overall, our VCM+ shows better performance than VCM regardless of the parameters used.

\begin{table}[t]
    \centering
    \caption{MSE value obtained under different $\alpha_F$. Kitchen and Glass are rendered with 100 iterations. The MSE obtained by VCM is the baseline. VCM+ with $\alpha_F=0.01$ obtains the best performance.}
    \begin{tabular}{c|c|ccc|c}
        \bottomrule
         \multirow{2}{*}{Kitchen} & $\alpha_F$ & 0.01 & 0.05 & 0.10 & VCM \\ \cline{2-6}
         & MSE & \textbf{34.30} & 35.58 & 37.05 & 40.90 \\ \hline
         \multirow{2}{*}{Glass} & $\alpha_F$ & 0.01 & 0.05 & 0.10 & VCM\\ \cline{2-6}
         & MSE & \textbf{17.44} & 18.29 & 19.70 & 22.93 \\
        \toprule
    \end{tabular}
    \vspace{-0.1in}
    \label{tab:alpha}
\end{table}


\section{Conclusion, Limitation, and Future Work}

In this paper, our improvement improves the efficiency of the PPM and VCM estimators.
Our key is a novel statistical model along with a hypothesis-testing method for an unbiased condition that is feasible for both PPM and VCM frameworks. From a statistical view, F-test, as one of the parametric tests, is useful when comparing statistical models to observations and can compute the reliable test statistic.
Our algorithm uses the F-test to detect potential bias and can find the appropriate radius for kernel estimation for most pixels. 
Consequently, our algorithm has lower variance and bias overall in practice and generates less noisy images with sharper boundaries. Our algorithm can help alleviate the light leakage artifacts as well.

\begin{figure}[t]
    \centering
  \begin{tabular}{c c c }
       \includegraphics[width=1.2in]{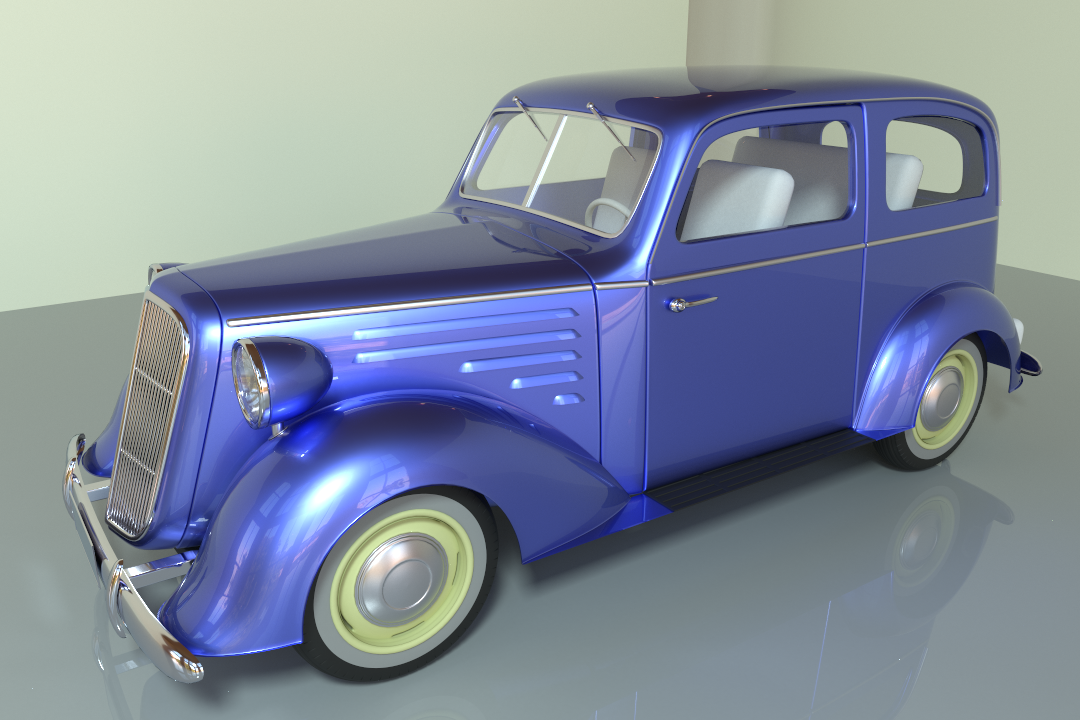} &  
       \includegraphics[width=0.75in]{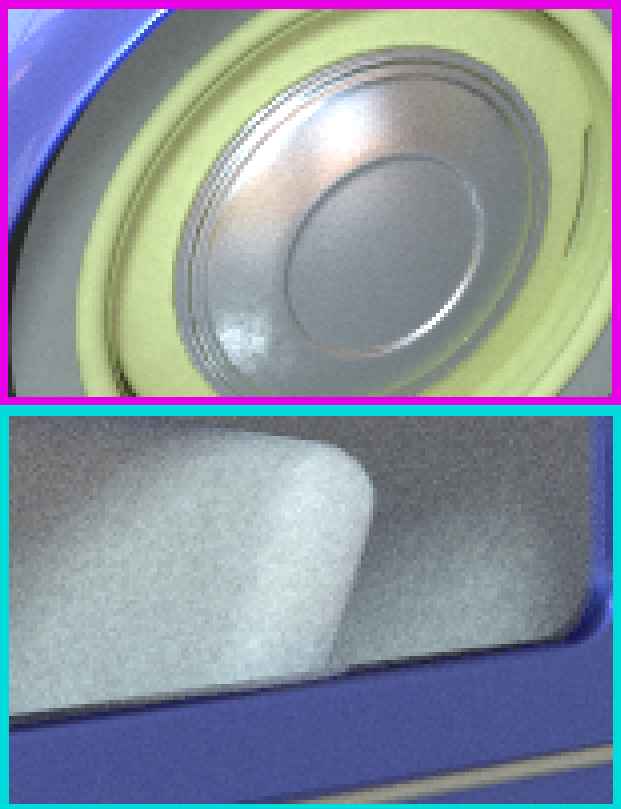} &
       \includegraphics[width=0.75in]{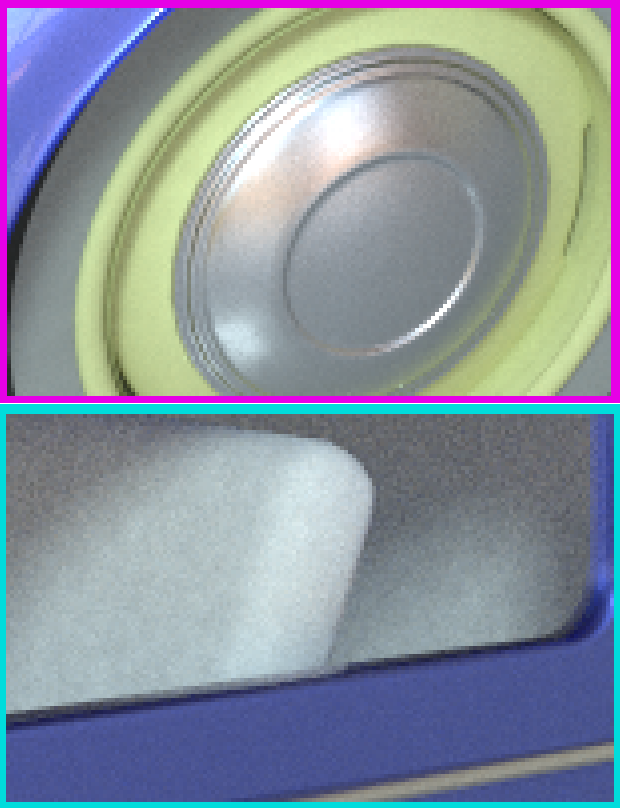}  \\
       &  VCM & VCM+ \\
       &  MSE=7.55 & MSE=7.29 \\
       \includegraphics[width=0.6in]{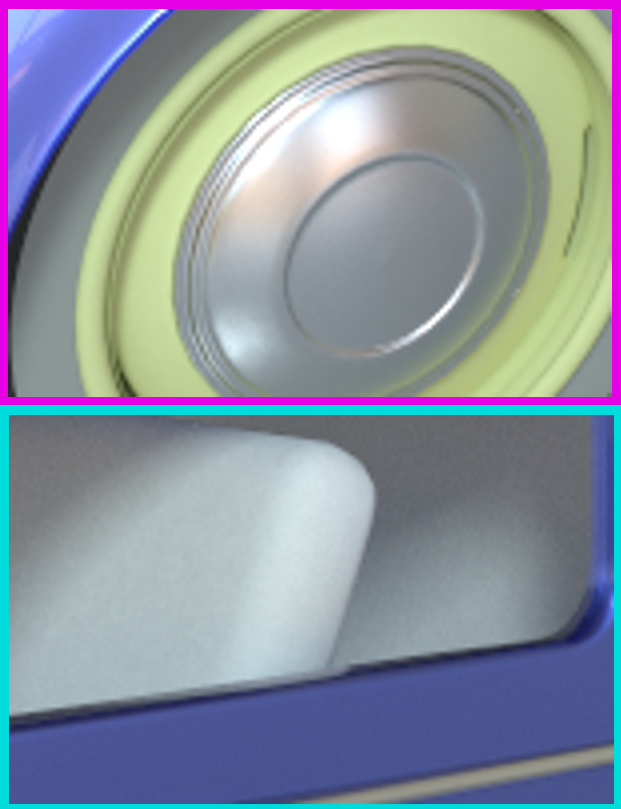}
       \includegraphics[width=0.6in]{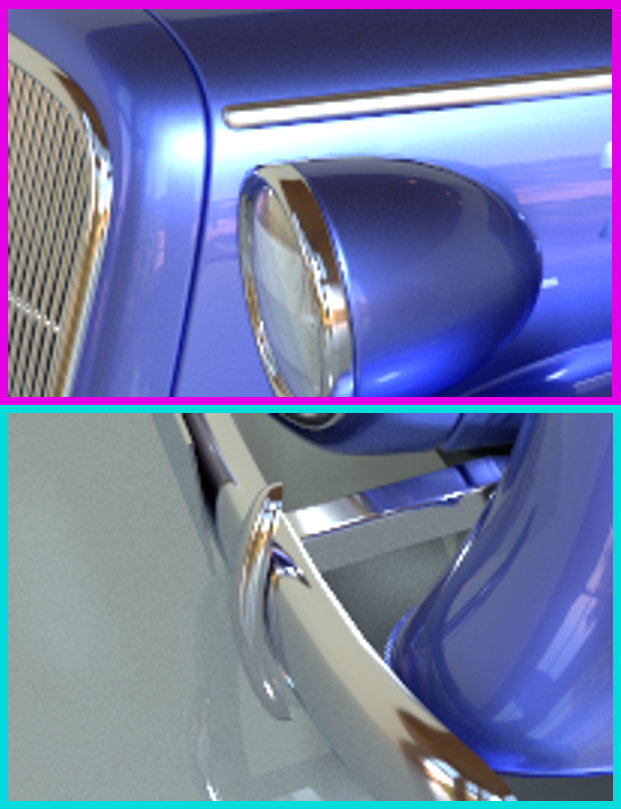}
       
       & \includegraphics[width=0.75in]{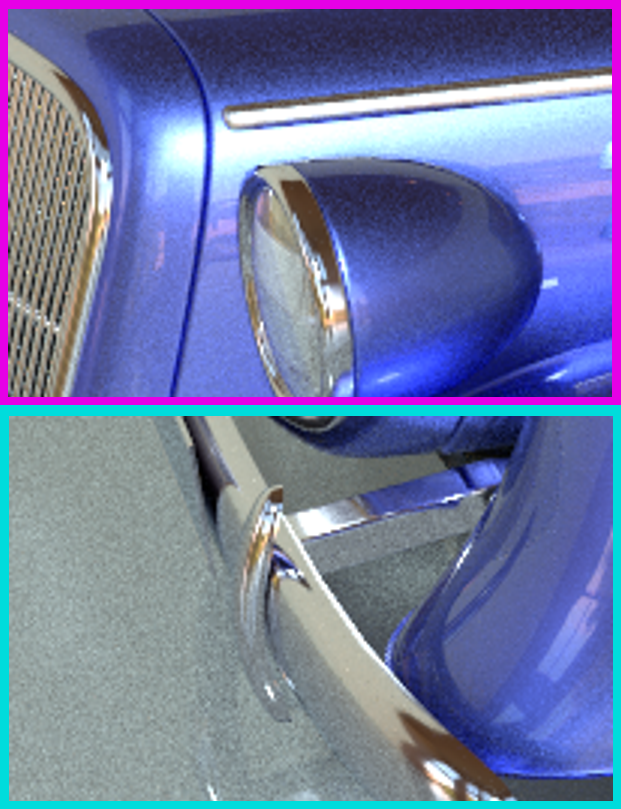}
       & \includegraphics[width=0.75in]{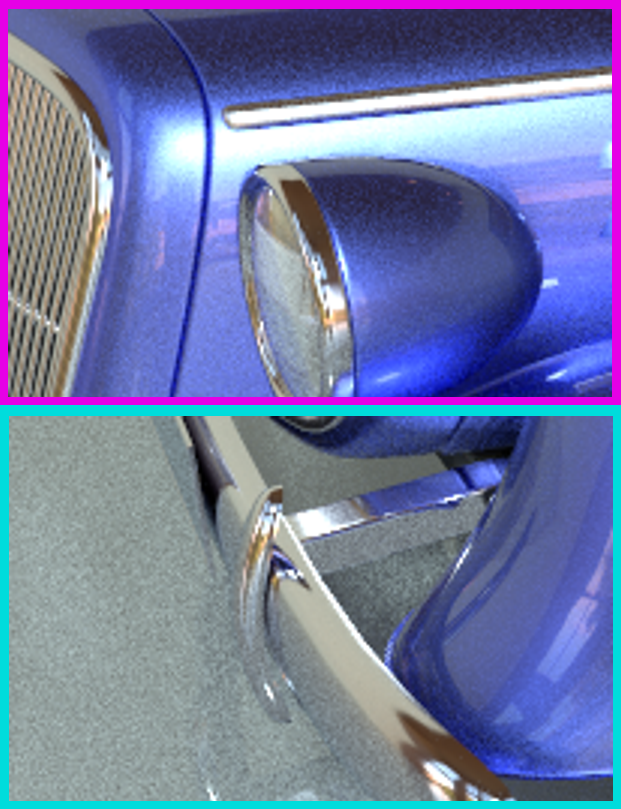} \\
       Reference & CPPM & FPPM\\
       & MSE=26.54 & MSE=25.04 
  \end{tabular}
  \caption{\changed{Equal-time (20 minutes) rendering of Car scene, VCM (829 iterations) vs. VCM+ (811 iterations), and CPPM (5619 iterations) vs. FPPM (4742 iterations). VCM+ and FPPM exhibit only marginal improvements over VCM and CPPM, respectively.}}
  \vspace{-0.1in}
  \label{fig:limitation}
\end{figure}

Our algorithm has some limitations.
First, the F\nbd-test requires sufficient samples to reject a false null hypothesis.
In those regions where samples are inadequate, the F\nbd-test may take a few iterations to reject a biased estimation, resulting in some initial bias. 
Next, our algorithm is more suitable for those complex scenes with multiple bounces during the light transport because this is apt to produce light vertices with different contributions within a kernel radius (as $\Psi_s$ of a path contribution function discussed in \autoref{equ:bidir-PPM-Ii}).
For those scenes where the light setup is simple and centered on one single object, which may imply the discrepancy of photon contribution is tiny, our algorithm no longer has an obvious advantage.
Our algorithms only slightly improve over the baseline, as shown in the Car scene in \autoref{fig:limitation}. 

There are many venues for future work to extend our work.
Combination with Markov chain Monte Carlo methods (MCMC)~\cite{Hachisuka2011Robust,mcmcvcm} or online learning methods~\cite{Vorba2014On-Line,muller2017practical,muller2019Neural,muller2020neural} to improve the sampling probability of high contribution paths may be able to alleviate the problem of initial bias.
The MCMC methods introduce severe violations of the assumption of independence (discussed in Sec.~\ref{subsubsection:ANOVA-f-test}), which needs further evaluation.
Our unbiased condition used for VCM+ involves MIS weights that contain pdf values (\autoref{equ:bidir-sufficient-condition}), which implicitly assumes pdfs do not change during the iterations, but a path in the online learning methods may have a different pdf at each iteration.
Therefore, seamless combination of online learning methods also needs further investigation.
In addition, the way of finding a better MIS strategy such as optimal MIS~\cite{optimalImportanceSampling}, or taking correlation of samples into consideration ~\cite{grittmann2021correlated}, or subspace MIS-aware sampling \cite{su2022SPCBPT} that can combine highly efficient BDPT with PM technique is also an interesting topic.

The key of our radiance estimate to dealing with the samples of light vertices is orthogonal to those signal sampling and adaptive methods \cite{stochasticsampling1986, jaume2003criteria}. Nevertheless, stochastic sampling of each pixel is employed in our method during the distributed ray tracing stage to generate multiple eye paths, same with SPPM \cite{toshiya2009stochastic}.
On the one hand, it may be possible to replace the unbiased condition for kernel estimation derived in \autoref{sec:hypothesis-testing} with a simpler criterion used in adaptive sampling. It may save computational overhead and make a possible improvement in the accuracy of bias detection. However, empirical criteria will not safeguard the unbiasedness of radiance estimation \cite{jaume2003criteria}.
On the other hand, an extension of our hypothesis-testing-based methodology to guide the sampling of a pixel for distributed ray tracing process may be an alternative solution to improve the efficiency further. However, a reasonable statistical model and assumption based on the prior about the morphology of each pixel should be established in advance for a parametric test. We can turn to nonparametric tests, but a larger sample size is required to conclude with the same degree of confidence as parametric tests, i.e., it may have less power.
An interesting idea will be to use unbiased samples to backward guide the adaptive sampling process in distributed ray tracing. 
These still need further investigation in the future.

\bibliographystyle{IEEEtran}
\bibliography{reference}

\begin{thebibliography}{10}
\providecommand{\url}[1]{#1}
\csname url@samestyle\endcsname
\providecommand{\newblock}{\relax}
\providecommand{\bibinfo}[2]{#2}
\providecommand{\BIBentrySTDinterwordspacing}{\spaceskip=0pt\relax}
\providecommand{\BIBentryALTinterwordstretchfactor}{4}
\providecommand{\BIBentryALTinterwordspacing}{\spaceskip=\fontdimen2\font plus
\BIBentryALTinterwordstretchfactor\fontdimen3\font minus \fontdimen4\font\relax}
\providecommand{\BIBforeignlanguage}[2]{{%
\expandafter\ifx\csname l@#1\endcsname\relax
\typeout{** WARNING: IEEEtran.bst: No hyphenation pattern has been}%
\typeout{** loaded for the language `#1'. Using the pattern for}%
\typeout{** the default language instead.}%
\else
\language=\csname l@#1\endcsname
\fi
#2}}
\providecommand{\BIBdecl}{\relax}
\BIBdecl

\bibitem{LW1993BPT}
E.~P. Lafortune and Y.~D. Willems, ``Bi-directional path tracing,'' in \emph{Proceedings of Third International Conference on Computational Graphics and Visualization Techniques (Compugraphics '93)}, Alvor, Portugal, December 1993, pp. 145--153.

\bibitem{veach1995bidirectional}
E.~Veach and L.~Guibas, ``Bidirectional estimators for light transport,'' in \emph{Photorealistic Rendering Techniques}.\hskip 1em plus 0.5em minus 0.4em\relax Berlin, Heidelberg: Springer, 1995, pp. 145--167.

\bibitem{hachisuka2008progressive}
T.~Hachisuka, S.~Ogaki, and H.~W. Jensen, ``Progressive photon mapping,'' \emph{ACM Transactions on Graphics (TOG)}, vol.~27, no.~5, p. 130, 2008.

\bibitem{jensen1996global}
H.~W. Jensen, ``Global illumination using photon maps,'' in \emph{Rendering Techniques' 96}.\hskip 1em plus 0.5em minus 0.4em\relax Vienna: Springer, 1996, pp. 21--30.

\bibitem{lin2020cppm}
\BIBentryALTinterwordspacing
Z.~Lin, S.~Li, X.~Zeng, C.~Zhang, J.~Jia, G.~Wang, and D.~Manocha, ``Cppm: Chi-squared progressive photon mapping,'' \emph{ACM Trans. Graph.}, vol.~39, no.~6, Nov. 2020. [Online]. Available: \url{https://doi.org/10.1145/3414685.3417822}
\BIBentrySTDinterwordspacing

\bibitem{Georgiev:2012:VCM}
\BIBentryALTinterwordspacing
I.~Georgiev, J.~K\v{r}iv\'{a}nek, T.~Davidovi\v{c}, and P.~Slusallek, ``Light transport simulation with vertex connection and merging,'' \emph{ACM Trans. Graph.}, vol.~31, no.~6, pp. 192:1--192:10, Nov. 2012. [Online]. Available: \url{http://doi.acm.org/10.1145/2366145.2366211}
\BIBentrySTDinterwordspacing

\bibitem{hachisuka2012ups}
\BIBentryALTinterwordspacing
T.~Hachisuka, J.~Pantaleoni, and H.~W. Jensen, ``A path space extension for robust light transport simulation,'' \emph{ACM Trans. Graph.}, vol.~31, no.~6, Nov. 2012. [Online]. Available: \url{https://doi.org/10.1145/2366145.2366210}
\BIBentrySTDinterwordspacing

\bibitem{kajiya1986rendering}
\BIBentryALTinterwordspacing
J.~T. Kajiya, ``The rendering equation,'' ser. SIGGRAPH '86.\hskip 1em plus 0.5em minus 0.4em\relax New York, NY, USA: Association for Computing Machinery, 1986, p. 143–150. [Online]. Available: \url{https://doi.org/10.1145/15922.15902}
\BIBentrySTDinterwordspacing

\bibitem{hachisuka2012state}
T.~Hachisuka, W.~Jarosz, G.~Bouchard, P.~Christensen, J.~R. Frisvad, W.~Jakob, H.~W. Jensen, M.~Kaschalk, C.~Knaus, A.~Selle \emph{et~al.}, ``State of the art in photon density estimation,'' in \emph{Acm Siggraph 2012 Courses}.\hskip 1em plus 0.5em minus 0.4em\relax ACM Siggraph, 2012, p.~6.

\bibitem{Kaplanyan2013appm}
\BIBentryALTinterwordspacing
A.~S. Kaplanyan and C.~Dachsbacher, ``Adaptive progressive photon mapping,'' \emph{ACM Trans. Graph.}, vol.~32, no.~2, Apr. 2013. [Online]. Available: \url{https://doi.org/10.1145/2451236.2451242}
\BIBentrySTDinterwordspacing

\bibitem{shilin2020deep}
S.~Zhu, Z.~Xu, H.~W. Jensen, H.~Su, and R.~Ramamoorthi, ``Deep kernel density estimation for photon mapping,'' \emph{Computer Graphics Forum}, vol.~39, no.~4, pp. 35--45, 2020.

\bibitem{toshiya2009stochastic}
T.~Hachisuka and H.~W. Jensen, ``Stochastic progressive photon mapping,'' \emph{ACM Transactions on Graphics (TOG)}, vol.~28, no.~5, p. 141, 2009.

\bibitem{vorba2011bidirectional}
J.~Vorba, ``Bidirectional photon mapping,'' in \emph{Proc. of the Central European Seminar on Computer Graphics (CESCG'11)}, 2011.

\bibitem{hachisuka2010progressive}
\BIBentryALTinterwordspacing
T.~Hachisuka, W.~Jarosz, and H.~W. Jensen, ``A progressive error estimation framework for photon density estimation,'' \emph{ACM Trans. Graph.}, vol.~29, no.~6, Dec. 2010. [Online]. Available: \url{https://doi.org/10.1145/1882261.1866170}
\BIBentrySTDinterwordspacing

\bibitem{knaus2011progressive}
C.~Knaus and M.~Zwicker, ``Progressive photon mapping: A probabilistic approach,'' \emph{ACM Transactions on Graphics (TOG)}, vol.~30, no.~3, p.~25, 2011.

\bibitem{qin2015unbiased}
H.~Qin, X.~Sun, Q.~Hou, B.~Guo, and K.~Zhou, ``Unbiased photon gathering for light transport simulation,'' \emph{ACM Transactions on Graphics (TOG)}, vol.~34, no.~6, pp. 1--14, 2015.

\bibitem{mcmcvcm}
\BIBentryALTinterwordspacing
M.~\v{S}ik, H.~Otsu, T.~Hachisuka, and J.~K\v{r}iv\'{a}nek, ``Robust light transport simulation via metropolised bidirectional estimators,'' \emph{ACM Trans. Graph.}, vol.~35, no.~6, pp. 245:1--245:12, Nov. 2016. [Online]. Available: \url{http://doi.acm.org/10.1145/2980179.2982411}
\BIBentrySTDinterwordspacing

\bibitem{calder1953statistical}
K.~Calder, ``Statistical inference,'' \emph{New York: Holt}, 1953.

\bibitem{everitt1998cambridge}
B.~Everitt, ``The cambridge dictionary of statistics,'' \emph{Cambridge University Press}, 1998.

\bibitem{william1952goodness}
W.~G. Cochran, ``The chi-square test of goodness of fit,'' \emph{The Annals of Mathematical Statistics}, vol.~23, no.~3, pp. 315--345, 1952.

\bibitem{miller1997beyond}
R.~G. Miller~Jr, \emph{Beyond ANOVA: basics of applied statistics}.\hskip 1em plus 0.5em minus 0.4em\relax CRC press, 1997.

\bibitem{fisher1921probable}
R.~A. Fisher, ``On the 'probable error' of a coefficient of correlation deduced from a small sample,'' \emph{Metron}, vol.~1, pp. 1--32, 1921.

\bibitem{kruskal1952use}
W.~H. Kruskal and W.~A. Wallis, ``Use of ranks in one-criterion variance analysis,'' \emph{Journal of the American statistical Association}, vol.~47, no. 260, pp. 583--621, 1952.

\bibitem{moore2021introduction}
D.~S. Moore, G.~P. McCabe, and B.~A. Craig, \emph{Introduction to the Practice of Statistics}.\hskip 1em plus 0.5em minus 0.4em\relax Macmillan, 2016.

\bibitem{blanca2017non}
M.~J. Blanca, R.~Alarc{\'o}n, J.~Arnau, R.~Bono, and R.~Bendayan, ``Non-normal data: Is anova still a valid option?'' \emph{Psicothema}, vol.~29, no.~4, pp. 552--557, 2017.

\bibitem{feir1974empirical}
B.~J. Feir-Walsh and L.~E. Toothaker, ``An empirical comparison of the anova f-test, normal scores test and kruskal-wallis test under violation of assumptions,'' \emph{Educational and Psychological Measurement}, vol.~34, no.~4, pp. 789--799, 1974.

\bibitem{BATHKE2004413}
\BIBentryALTinterwordspacing
A.~Bathke, ``The anova f test can still be used in some balanced designs with unequal variances and nonnormal data,'' \emph{Journal of Statistical Planning and Inference}, vol. 126, no.~2, pp. 413--422, 2004. [Online]. Available: \url{https://www.sciencedirect.com/science/article/pii/S0378375803002787}
\BIBentrySTDinterwordspacing

\bibitem{veach1997robust}
E.~Veach, \emph{Robust Monte Carlo methods for light transport simulation}.\hskip 1em plus 0.5em minus 0.4em\relax Stanford University PhD thesis, 1997, vol. 1610.

\bibitem{jan2014sample}
S.-L. Jan and G.~Shieh, ``Sample size determinations for welch's test in one-way heteroscedastic anova,'' \emph{British Journal of Mathematical and Statistical Psychology}, vol.~67, no.~1, pp. 72--93, 2014.

\bibitem{Mitsuba}
W.~Jakob, ``Mitsuba renderer,'' 2010, http://www.mitsuba-renderer.org.

\bibitem{Hachisuka2011Robust}
\BIBentryALTinterwordspacing
T.~Hachisuka and H.~W. Jensen, ``Robust adaptive photon tracing using photon path visibility,'' \emph{ACM Trans. Graph.}, vol.~30, no.~5, Oct. 2011. [Online]. Available: \url{https://doi.org/10.1145/2019627.2019633}
\BIBentrySTDinterwordspacing

\bibitem{Vorba2014On-Line}
\BIBentryALTinterwordspacing
J.~Vorba, O.~Karl\'{\i}k, M.~\v{S}ik, T.~Ritschel, and J.~K\v{r}iv\'{a}nek, ``On-line learning of parametric mixture models for light transport simulation,'' \emph{ACM Trans. Graph.}, vol.~33, no.~4, Jul. 2014. [Online]. Available: \url{https://doi.org/10.1145/2601097.2601203}
\BIBentrySTDinterwordspacing

\bibitem{muller2017practical}
T.~M{\"u}ller, M.~Gross, and J.~Nov{\'a}k, ``Practical path guiding for efficient light-transport simulation,'' in \emph{Computer Graphics Forum}, vol.~36, no.~4, 2017, pp. 91--100.

\bibitem{muller2019Neural}
\BIBentryALTinterwordspacing
T.~M\"{u}ller, B.~Mcwilliams, F.~Rousselle, M.~Gross, and J.~Nov\'{a}k, ``Neural importance sampling,'' \emph{ACM Trans. Graph.}, vol.~38, no.~5, Oct. 2019. [Online]. Available: \url{https://doi.org/10.1145/3341156}
\BIBentrySTDinterwordspacing

\bibitem{muller2020neural}
\BIBentryALTinterwordspacing
T.~M\"{u}ller, F.~Rousselle, A.~Keller, and J.~Nov\'{a}k, ``Neural control variates,'' \emph{ACM Trans. Graph.}, vol.~39, no.~6, Nov. 2020. [Online]. Available: \url{https://doi.org/10.1145/3414685.3417804}
\BIBentrySTDinterwordspacing

\bibitem{optimalImportanceSampling}
\BIBentryALTinterwordspacing
I.~Kondapaneni, P.~Vevoda, P.~Grittmann, T.~Sk\v{r}ivan, P.~Slusallek, and J.~K\v{r}iv\'{a}nek, ``Optimal multiple importance sampling,'' \emph{ACM Trans. Graph.}, vol.~38, no.~4, Jul. 2019. [Online]. Available: \url{https://doi.org/10.1145/3306346.3323009}
\BIBentrySTDinterwordspacing

\bibitem{grittmann2021correlated}
P.~Grittmann, I.~Georgiev, and P.~Slusallek, ``Correlation-aware multiple importance sampling for bidirectional rendering algorithms,'' \emph{Computer Graphics Forum}, vol.~40, no.~2, pp. 231--238, 2021.

\bibitem{su2022SPCBPT}
\BIBentryALTinterwordspacing
F.~Su, S.~Li, and G.~Wang, ``Spcbpt: Subspace-based probabilistic connections for bidirectional path tracing,'' \emph{ACM Trans. Graph.}, vol.~41, no.~4, jul 2022. [Online]. Available: \url{https://doi.org/10.1145/3528223.3530183}
\BIBentrySTDinterwordspacing

\bibitem{stochasticsampling1986}
\BIBentryALTinterwordspacing
R.~L. Cook, ``Stochastic sampling in computer graphics,'' \emph{ACM Trans. Graph.}, vol.~5, no.~1, p. 51–72, jan 1986. [Online]. Available: \url{https://doi.org/10.1145/7529.8927}
\BIBentrySTDinterwordspacing

\bibitem{jaume2003criteria}
J.~Rigau, M.~Feixas, and M.~Sbert, ``{Refinement Criteria Based on f-Divergences},'' in \emph{Eurographics Workshop on Rendering}.\hskip 1em plus 0.5em minus 0.4em\relax The Eurographics Association, 2003.

\end{thebibliography}

\end{document}